\documentclass[12pt]{article}
\usepackage{amsmath,amssymb,amsfonts,color,graphicx,cite,color}
\usepackage{latexsym}
\usepackage{epsfig,psfrag,rotating,soul}
\usepackage{rotfloat}
\input paperdef

\evensidemargin \oddsidemargin
\allowdisplaybreaks

\begin{document}
\thispagestyle{empty}

\def\thefootnote{\fnsymbol{footnote}}

\begin{flushright}
MPP--2014--321 
\end{flushright}

\vspace{0.5cm}

\begin{center}

\begin{large}
\textbf{Higgs masses and Electroweak Precision Observables}
\\[2ex]
\textbf{in the Lepton-Flavor-Violating MSSM}
\end{large}

\vspace{1cm}

{\sc
M.E.~G{\'o}mez$^{1}$%
\footnote{email: mario.gomez@dfa.uhu.es}%
, T.~Hahn$^2$%
\footnote{email: hahn@feynarts.de}%
, S.~Heinemeyer$^{3}$%
\footnote{email: Sven.Heinemeyer@cern.ch}%
,~and M.~Rehman$^{3}$%
\footnote{email: rehman@ifca.unican.es}%
\footnote{MultiDark Scholar}
 
}

\vspace*{.7cm}

{\sl
$^1$Department of Applied Physics, University of Huelva, 21071 Huelva, Spain

\vspace*{0.1cm}

$^2$Max-Planck-Institut f\"ur Physik,
F\"ohringer Ring 6, \\
D--80805 M\"unchen, Germany

\vspace*{0.1cm}
$^3$Instituto de F\'isica de Cantabria (CSIC-UC), Santander, Spain

}

\end{center}

\vspace*{0.1cm}

\begin{abstract}
\noindent
We study the effects of Lepton Flavor Violation (LFV) in the scalar 
lepton sector of the MSSM on precision observables such as the $W$-boson 
mass and the effective weak leptonic mixing angle, and on the Higgs-boson 
mass predictions.  The slepton mass matrices are parameterized in a 
model-independent way by a complete set of dimensionless parameters
which we constrain through LFV decay processes and the precision 
observables.  We find regions where both conditions are similarly 
constraining.  The necessary prerequisites for the calculation have 
been added to FeynArts and FormCalc and are thus publicly available for 
further studies.  The obtained results are available in FeynHiggs.
\end{abstract}

\def\thefootnote{\arabic{footnote}}
\setcounter{page}{0}
\setcounter{footnote}{0}

\newpage


\section{Introduction}

Lepton Flavor Violating (LFV) processes provide one of the most 
interesting probes to physics beyond the Standard Model (SM) of particle 
physics.  All SM interactions preserve lepton flavor number and 
therefore a measurement of any (charged) LFV process would be an 
unambiguous signal of physics beyond the SM and provide interesting 
information on the involved flavor mixing, as well as on the underlying 
origin for this mixing (for a review see \citere{Kuno:1999jp}, for 
instance).

The data from past and ongoing neutrino oscillation experiments, as well 
as from cosmology and astrophysics, have confirmed that neutrinos have 
different non-zero masses and that the three neutrino flavors $\nu_e$, 
$\nu_\mu$, $\nu_\tau$ mix to form three mass eigenstates.  This implies 
non-conservation of lepton flavor, clearly beyond the SM.  Thus, 
lepton-flavor-violating processes are expected in the lepton sector just 
as quark-flavor-violating processes arise in the quark sector.

Within the Minimal Supersymetric Standard Model (MSSM)~\cite{mssm}, LFV 
can occur in the scalar lepton sector.  The most general way to 
introduce slepton flavor mixing within the MSSM is through the 
off-diagonal soft-SUSY-breaking parameters (both mass parameters and 
trilinear couplings) in the slepton sector.  The off-diagonality in the 
slepton mass matrix reflects the misalignment (in flavor space) between 
lepton and slepton mass matrices, which cannot be diagonalized 
simultaneously.  This misalignment can have various origins; for 
instance, off-diagonal slepton mass matrix entries can be generated by 
Renormalization Group Equations running from high energies, where heavy 
right-handed neutrinos are assumed to be active, down to low energies 
where LFV processes can occur~\cite{Hall:1985dx,Borzumati:1986qx}.

In this work we do not investigate the possible dynamical origin of this 
lepton--slepton misalignment, nor particular predictions for 
off-diagonal slepton soft-SUSY-breaking mass terms in specific SUSY 
models, but instead parameterize the slepton mass matrix and explore the 
phenomenological implications of LFV on various observables.

Specifically, we write the off-diagonal slepton mass matrix elements in 
terms of a complete set of generic dimensionless parameters 
$\delta_{\{12,13,23\}}^{\{LL,LR,RL,RR\}}$, where $L,R$ refers to the 
left-/right-handed SUSY partner of the corresponding leptonic degree of 
freedom and $1,2,3$ are the involved generation indices, and explore the 
sensitivity of several precision observables to the $\deABij$'s, 
extending a program carried out for flavor violation in the scalar quark 
sector~\cite{arana}.

Besides direct searches, which have not turned up evidence for any 
additional particles so far, SUSY can also be probed through its effects 
on precision observables via virtual particles, see \citere{PomssmRep} 
for a review.  Electroweak precision observables (EWPO) like the 
$W$-boson mass or the effective weak leptonic mixing angle have been 
measured to a very high precision, and the anticipated improved 
precision in current and future experiments for these observables makes 
them very sensitive to physics beyond the SM.

Besides EWPO we also explore the effects of LFV on the MSSM Higgs 
sector, again extending existing analyses on flavor violation in the 
scalar quark sector~\cite{delrhoNMFV,arana}.  The MSSM Higgs sector 
consist of two Higgs doublets and predicts five physical Higgs bosons, 
the light and heavy $\cp$-even $h$ and $H$, the $\cp$-odd $A$, and the 
charged Higgs boson $H^\pm$.  At tree level the Higgs sector is 
described with the help of two parameters: the mass of the $A$~boson, 
$\MA$, and $\tb := v_2/v_1$, the ratio of the two vacuum expectation 
values.  After the spectacular discovery of a Higgs particle at the LHC, 
the precision of the measured mass value is already below the GeV 
level~\cite{ATLAS:2013mma,CMS:yva}, and at a future ILC, a precision 
even below $\sim 50\mev$ is anticipated~\cite{dbd}.  We evaluate the 
effects of LFV on the predictions of the masses of the light and heavy 
$\cp$-even Higgs bosons, $\Mh$ and $\MH$, as well as on the charged 
Higgs-boson mass $\MHp$.  Based on the evaluations in the scalar quark 
sector~\cite{arana}, theoretical uncertainties from LFV effects on the 
evaluation of the Higgs-boson masses are substantially larger than the 
future experimental accuracy could be expected, motivating the 
analytical calculation of these corrections.

For our calculations we prepared (and thoroughly tested) an add-on model 
file for FeynArts~\cite{feynarts,famssm} which adds LFV effects to the 
existing MSSM model file.  No renormalization as in \citere{mssmct} is 
included yet (and also not necessary for the present work since the SM 
is lepton-flavor conserving and hence there is no tree-level 
contribution).  The FormCalc~\cite{formcalc} driver files were also 
modified accordingly.  We checked that the LFV Feynman rules yield 
finite results for all our calculations.  The results derived with this 
setup, the Higgs-boson masses as well as the EWPO, were added to 
FeynHiggs 2.10.2.

This paper is organized as follows: First we review the main features of 
the MSSM with general slepton flavor mixing and set the relevant 
notation for the $\deABij$'s in ~\refse{sec:nmfv}.  The selection of 
specific MSSM scenarios as well as their experimental restrictions from 
LFV processes is presented in \refse{sec:scenariosLFV}.  The numerical 
analysis is given in \refse{sec:results}, showing for the first time the 
LFV effects on the MSSM Higgs boson masses and on the EWPO.  
\refse{sec:conclusions} summarizes our conclusions.


\def\vspc{\vrule width 0pt height 2.5ex depth 1.5ex}
\def\noent{\vspc~\cdot~}
\def\diag{\mathop{\mathrm{diag}}}
\def\Re{\mathop{\mathrm{Re}}}

\section{Calculational Basis}
\label{sec:nmfv}

We work in MSSM scenarios with general flavor mixing in the sleptons.  
Within these MSSM-FV scenarios, lepton flavor violation is induced by 
the PMNS matrix of the neutrino sector and transmitted by the small 
neutrino Yukawa couplings which we ignore here.  Flavor mixing in the 
slepton mass matrix is the main generator of LFV.  In the following we 
give a brief overview about the relevant sectors of the MSSM with LFV.


\subsection{Scalar lepton sector with LFV}
\label{sec:sfermions}

For the slepton sector of the MSSM including LFV contributions we use 
the same notation as \citere{Arana-Catania:2013nha}.  The most general 
hypothesis for flavor mixing in the slepton sector assumes a 
non-diagonal mass matrix for both charged sleptons and sneutrinos.  For 
the charged sleptons this is a $6\times 6$ mass matrix since there are 
six electroweak interaction eigenstates, $\til_{L,R}$ with $\ell = e, 
\mu, \tau$, while for the sneutrinos the matrix is only $3 \times 3$ 
corresponding to the three states $\tilde\nu_L$ with $\nu = \nu_e, 
\nu_\mu, \nu_\tau$.

The non-diagonal entries in the $6 \times 6$ general matrix for charged 
sleptons can be described in a model-independent way in terms of a set 
of dimensionless parameters $\deABij$ ($A,B=L,R$; $i,j=1,2,3$, $i \neq 
j$), where $L,R$ refer to the left-/right-handed SUSY partners of the 
corresponding leptonic degrees of freedom, and the indices $i,j$ run 
over the three generations.  These scenarios with general sfermion 
flavor mixing lead generally to larger LFV rates than in the so-called 
Minimal Flavor Violation Scenarios, where the mixing is induced 
exclusively by the Yukawa coupling of the corresponding fermion sector.  
This is true for both squarks and sleptons but it is obviously of 
special interest in the slepton case due to the extremely small size of 
the lepton Yukawa couplings, suppressing LFV processes from this origin. 
Hence in the present case of slepton mixing we assume that the 
$\deABij$'s provide the sole source of LFV processes with potentially 
measurable rates.

The non-diagonal $6\times 6$ slepton mass matrix, which we order here as 
$(\SelL, \SmuL, \StauL, \SelR, \SmuR, \StauR)$, is usually decomposed 
into left- and right-handed $3\times 3$ blocks $M^2_{\til,AB}$ as
\begin{equation}
\label{eq:slep-6x6}
\mathcal{M}_{\til}^2 = \begin{pmatrix}
M^2_{\til,LL} & M^2_{\til,LR} \\[.3em]
M_{\til,LR}^{2\,\dagger} & M^2_{\til,RR}
\end{pmatrix},
\end{equation} 
where
\begin{equation}
\label{eq:slep-matrix}
\begin{aligned}
(M^2_{\til,LL})_{ij}
&= (m^2_{\tilde L})_{ij} + \left( m_{\ell_i}^2 +
   (-\tfrac 12 + \sw^2) \MZ^2 \cos 2\beta \right) \delta_{ij}\,, \\
(M^2_{\til,RR})_{ij}
&= (m^2_{\tilde E})_{ij} + \left( m_{\ell_i}^2 -
   \sw^2 \MZ^2 \cos 2\beta \right) \delta_{ij}\,, \\
(M^2_{\til,LR})_{ij}
&= v_1\mathcal{A}_{ij}^\ell - m_{\ell_i}\mu\tb\,\delta_{ij}\,,
\end{aligned}
\end{equation}
with flavor indexes $i,j = 1,2,3$, $\sw = \sqrt{1 - \cw^2}$ with $\cw = 
\MW/\MZ$, lepton masses $(m_{\ell_i}) = (m_e, m_\mu, m_\tau)$, and 
Higgsino mass parameter $\mu$.  The off-diagonal elements arise 
exclusively from the soft SUSY-breaking parameters: the doublet mass 
parameters $m^2_{\tilde L}$, the singlet mass parameters $m^2_{\tilde 
E}$, and the trilinear couplings $\mathcal{A}^\ell$, which are all 
$3\times 3$ matrices in flavor space.

The sneutrino mass matrix contains only a single $3\times 3$ block 
(ordered as $(\tinu_{eL}, \tinu_{\mu L}, \tinu_{\tau L})$) to start 
with since the singlet components are absent:
\begin{equation}
\label{eq:sneu-3x3}
(\mathcal{M}^2_{\tinu})_{ij} = (M^2_{\tinu,LL})_{ij} =
(m^2_{\tilde L})_{ij} + \tfrac 12\MZ^2\cos 2\beta \delta_{ij}\,.
\end{equation} 
Note that, due to $SU(2)_L$ gauge invariance, the same doublet mass 
parameters $m^2_{\tilde L}$ enter the slepton and sneutrino $LL$ mass 
matrices.

If neutrino masses and neutrino flavor mixings (oscillations) were taken 
into account, the soft-SUSY-breaking parameters for the sneutrinos would 
differ from the ones for charged sleptons by a rotation with the PMNS 
matrix.  Taking the neutrino masses and oscillations into account in the 
SM leads to LFV effects that are extremelly small; for instance, in 
$\mu \to e \gamma$ they are of \order{10^{-47}} in case of Dirac 
neutrinos with mass around 1~eV and maximal 
mixing~\cite{Kuno:1999jp,DiracNu,MajoranaNu}, and of \order{10^{-40}} in 
case of Majorana neutrinos~\cite{Kuno:1999jp,MajoranaNu}.  Consequently 
we do not expect large effects from the inclusion of neutrino mass 
effects here.

The dimensionless parameters $\deABij$ allow for a unified description 
of the off-diagonal soft-SUSY-breaking parameters to which they are 
related as follows:
\begin{align}
m^2_{\tilde L} &= \begin{pmatrix}
\vspc
m^2_{\tilde L_1} &
	\delta_{12}^{LL} m_{\tilde L_1} m_{\tilde L_2} &
		\delta_{13}^{LL} m_{\tilde L_1} m_{\tilde L_3} \\
\vspc
\delta_{21}^{LL} m_{\tilde L_2}m_{\tilde L_1} &
	m^2_{\tilde L_2} &
		\delta_{23}^{LL} m_{\tilde L_2} m_{\tilde L_3} \\
\vspc
\delta_{31}^{LL} m_{\tilde L_3} m_{\tilde L_1} &
	\delta_{32}^{LL} m_{\tilde L_3} m_{\tilde L_2} &
		m^2_{\tilde L_3}
\end{pmatrix}, \\
m^2_{\tilde E} &= \begin{pmatrix}
\vspc
m^2_{\tilde E_1} &
	\delta_{12}^{RR} m_{\tilde E_1} m_{\tilde E_2} &
		\delta_{13}^{RR} m_{\tilde E_1} m_{\tilde E_3} \\
\vspc
\delta_{21}^{RR} m_{\tilde E_2} m_{\tilde E_1} &
	m^2_{\tilde E_2} &
		\delta_{23}^{RR} m_{\tilde E_2}m_{\tilde E_3} \\
\vspc
\delta_{31}^{RR} m_{\tilde E_3} m_{\tilde E_1} &
	\delta_{32}^{RR} m_{\tilde E_3} m_{\tilde E_2} &
		m^2_{\tilde E_3}
\end{pmatrix}, \\
v_1 \mathcal{A}^\ell &= \begin{pmatrix}
\vspc
m_e A_e &
	\delta_{12}^{LR} m_{\tilde L_1}m_{\tilde E_2} & 
		\delta_{13}^{LR} m_{\tilde L_1}m_{\tilde E_3} \\
\vspc
\delta_{21}^{LR} m_{\tilde L_2}m_{\tilde E_1} &
	m_\mu A_\mu &
		\delta_{23}^{LR} m_{\tilde L_2} m_{\tilde E_3} \\
\vspc
\delta_{31}^{LR} m_{\tilde L_3} m_{\tilde E_1} &
	\delta_{32}^{LR} m_{\tilde L_3} m_{\tilde E_2} &
		m_\tau A_\tau
\end{pmatrix}.
\end{align}
This parameterization is purely phenomenological and does not rely on
any specific assumptions on the origin of the soft-SUSY-breaking
parameters.

The next step is to rotate the sleptons and sneutrinos from the 
electroweak interaction basis into the physical mass eigenstate basis,
\begin{equation}
\begin{pmatrix}
\til_1 \\ \til_2 \\ \til_3 \\ \til_4 \\ \til_5 \\ \til_6
\end{pmatrix} = R^{\til}\begin{pmatrix}
\SelL \\ \SmuL \\ \StauL \\ \SelR \\ \SmuR \\ \StauR
\end{pmatrix},
\qquad
\begin{pmatrix}
\tinu_1 \\ \tinu_2  \\ \tinu_3
\end{pmatrix} = R^{\tinu}\begin{pmatrix}
\tinu_{e L} \\ \tinu_{\mu L} \\ \tinu_{\tau L}
\end{pmatrix},
\end{equation} 
where $R^{\til}$ and $R^{\tinu}$ are the unitary matrices resulting from 
diagonalizing the mass matrices,
\begin{equation}
\begin{aligned}
R^{\til}\,\mathcal{M}^2_{\til}\,R^{\til\dagger}
&= \diag\{m_{\til_1}^2, m_{\til_2}^2, m_{\til_3}^2,
          m_{\til_4}^2, m_{\til_5}^2, m_{\til_6}^2\}\,, \\
R^{\tinu}\,\mathcal{M}^2_{\tinu}\,R^{\tinu\dagger}
&= \diag\{m_{\tinu_1}^2, m_{\tinu_2}^2, m_{\tinu_3}^2\}\,.
\end{aligned}
\end{equation}


\subsection{Higgs masses and mixing}
\label{sec:mhiggs}

In this section we shortly review the relevant features of the MSSM 
Higgs sector\footnote{%
  We restrict ourselves to the case of real parameters.  For the 
  case of complex parameters see \citeres{mhcMSSMlong,mhcMSSM2L} and 
  references therein.}
at tree-level.  Unlike the SM, the MSSM requires two Higgs doublets.  
The Higgs potential~\cite{hhg}
\begin{equation}
\label{eq:higgspot}
\begin{aligned}
V &= m_1^2 |\cHe|^2 + m_2^2 |\cHz|^2 
      - m_{12}^2 (\epsilon_{ab} \cHe^a\cHz^b + \hc) + {} \\
&\qquad \frac 18 (g_1^2 + g_2^2) \left[ |\cHe|^2 - |\cHz|^2 \right]^2
    + \frac 12 g_2^2|\cHe^{\dag} \cHz|^2\,,
\end{aligned}
\end{equation}
contains $m_1$, $m_2$, $m_{12}$ as soft-SUSY-breaking parameters;
$g_2, g_1$ are the $SU(2)$ and $U(1)$ gauge couplings, and 
$\epsilon$ is the spinor metric with $\epsilon_{12} = -1$.

The doublet fields $H_1$ and $H_2$ are decomposed as
\begin{equation}
\begin{aligned}
\cHe &= \begin{pmatrix} \cHe^0 \\[0.5ex] \cHe^- \end{pmatrix}
      = \begin{pmatrix} v_1 + \ed{\wz}(\phi_1^0 - i\chi_1^0) \\[0.5ex]
          -\phi_1^- \end{pmatrix}, \\
\cHz &= \begin{pmatrix} \cHz^+ \\[0.5ex] \cHz^0 \end{pmatrix}
      = \begin{pmatrix} \phi_2^+ \\[0.5ex]
          v_2 + \ed{\wz}(\phi_2^0 + i\chi_2^0) \end{pmatrix}.
\end{aligned}
\end{equation}
The Higgs potential is thus characterized at tree level by only two 
independent parameters: $\Tb = v_2/v_1$ and $M_A^2 = -m_{12}^2 (\Tb + 
\CTb)$, where $M_A$ is the mass of the $\cp$-odd Higgs boson~$A$.

The bilinear part of the Higgs potential is diagonalized by orthogonal
transformations
\begin{align}
\label{eq:hHdiag}
\begin{pmatrix} H \\[0.5ex] h \end{pmatrix}
&= \begin{pmatrix}
   \Ca & \Sa \\[0.5ex]
   -\Sa & \Ca
   \end{pmatrix} \begin{pmatrix}
   \phi_1^0 \\[0.5ex] \phi_2^0
   \end{pmatrix},  \\
\label{eq:AGdiag}
\begin{pmatrix} G \\[0.5ex] A \end{pmatrix}
&= \begin{pmatrix}
   \Cb & \Sbe \\[0.5ex] 
   -\Sbe & \Cb
   \end{pmatrix} \begin{pmatrix}
   \chi_1^0 \\[0.5ex] \chi_2^0
   \end{pmatrix}, \\
\label{eq:Hpmdiag}
\begin{pmatrix} G^{\pm} \\[0.5ex] H^\pm \end{pmatrix}
&= \begin{pmatrix}
   \Cb & \Sbe \\[0.5ex]
   -\Sbe & \Cb
   \end{pmatrix} \begin{pmatrix}
   \phi_1^\pm \\[0.5ex] \phi_2^\pm
   \end{pmatrix}.
\end{align}
where the tree-level mixing angle $\al$ is given by
\begin{equation}
\al = \arctan\KKL 
  \frac{-(\MA^2 + \MZ^2) \Sbe \Cb}
       {\MZ^2 \CQb + \MA^2 \SQb - m^2_{h,{\rm tree}}} \KKR,
\quad
-\frac{\pi}{2} < \al < 0\,.
\end{equation}
The Higgs spectrum is thus:
\begin{align*}
\mbox{2 neutral bosons},\, {\cal CP} = +1 &:\quad h,\ H, \\
\mbox{1 neutral boson},\, {\cal CP} = -1  &:\quad A, \\
\mbox{2 charged bosons}                   &:\quad H^+,\ H^-, \\
\mbox{3 unphysical Goldstone bosons}      &:\quad G,\ G^+,\ G^-.
\end{align*}
At tree level the neutral $\cp$-even Higgs-boson masses are determined 
from
\begin{equation}
\label{eq:higgsmassmatrixtree}
M_{\text{Higgs}}^{2,\text{tree}} = \begin{pmatrix}
\MA^2 \SQb + \MZ^2 \CQb & -(\MA^2 + \MZ^2) \Sbe \Cb \\
-(\MA^2 + \MZ^2) \Sbe \Cb & \MA^2 \CQb + \MZ^2 \SQb
\end{pmatrix}
\overset{\rlap{\large $\circlearrowleft$}\,\,%
  \raise 1.2bp\hbox{\scriptsize $\alpha$}~}{\longrightarrow}
\begin{pmatrix}
m_{H,\text{tree}}^2 & 0 \\ 0 &  m_{h,\text{tree}}^2
\end{pmatrix}.
\end{equation}
which yields
\begin{equation}
(m_{H,h}^2)_{\text{tree}} = \frac 12\left[ \MA^2 + \MZ^2 \pm
  \sqrt{(\MA^2+\MZ^2)^2 - 4\MZ^2\MA^2\cos^2 2\be} \right]
\end{equation}
and the charged Higgs-boson mass is given by
\begin{equation}
m^2_{H^\pm,\text{tree}} = \MA^2 + \MW^2\,.
\end{equation}


\subsection{Calculation of higher-order corrections in the Higgs sector}
\label{sec:higgs-ho}

We briefly review the procedure of \citeres{mhcMSSMlong,mhiggsf1lC} 
for the computation of one-loop corrections to the Higgs-boson masses.
The parameters appearing in the Higgs potential, \refeq{eq:higgspot}, 
are renormalized as follows:
\begin{equation}
\label{eq:rMSSM:PhysParamRenorm}
\begin{aligned}
\MZ^2 &\to \MZ^2 + \dMZsq\,,	& \tadh &\to \tadh + \dtadh\,, \\ 
\MW^2 &\to \MW^2 + \dMWsq\,,	& \tadH &\to \tadH + \dtadH\,, \\ 
M_{\text{Higgs}}^2 &\to M_{\text{Higgs}}^2 + \de M_{\text{Higgs}}^2\,,
\quad & 
	\tanb &\to \tanb (1 + \dtanb)\,.
\end{aligned}
\end{equation}
$M_{\text{Higgs}}^2$ denotes the tree-level Higgs-boson mass matrix of 
\refeq{eq:higgsmassmatrixtree}, and $\tadh$ and $\tadH$ are the 
tree-level tadpoles, i.e.\ the terms linear in $h$ and $H$ in the Higgs 
potential.

In the $\cp$-even sector the mass and field renormalization can be 
set up symmetrically,
\begin{align}
\label{eq:rMSSM:higgsfeldren}
\delta M_{\rm Higgs}^2 = \begin{pmatrix}
  \dmhsq  & \dmhHsq \\[.5em]
  \dmhHsq & \dmHsq  
\end{pmatrix},
\qquad
\begin{pmatrix} h \\[.5em] H \end{pmatrix} \to \begin{pmatrix}
  1 + \tfrac 12 \dZ{hh} & \tfrac 12 \dZ{hH} \\[.5em]
  \tfrac 12 \dZ{hH} & 1 + \tfrac 12 \dZ{HH}
\end{pmatrix}\begin{pmatrix}
  h \\[.5em] H
\end{pmatrix}.
\end{align}
The renormalized self-energies $\hSi(p^2)$ are expressed through the 
unrenormalized self-energies $\Si(p^2)$, the field renormalization 
constants, and the mass counter-terms as follows:
\begin{equation}
\label{rMSSM:renses_higgssector}
\begin{aligned}
\ser{hh}(p^2) &= \se{hh}(p^2) + \dZ{hh} (p^2-\mhtree^2) - \dmhsq\,, \\
\ser{hH}(p^2) &= \se{hH}(p^2) + \dZ{hH}
                 (p^2-\tfrac{1}{2}(\mhtree^2+\mHtree^2)) - \dmhHsq\,, \\ 
\ser{HH}(p^2)  &= \se{HH}(p^2) + \dZ{HH} (p^2-\mHtree^2) - \dmHsq\,.
\end{aligned}
\end{equation}
Inserting the renormalization transformation into the Higgs mass terms 
gives the following Higgs-mass counter-terms:
\begin{equation}
\begin{aligned}
\dmhsq &= \de\MA^2 \cos^2(\al-\be) + \delta \MZ^2 \sin^2(\al+\be) + {} \\
&\qquad \tfrac e{2\MZ\sw\cw} (\dtadH \cos(\al-\be) \sin^2(\al-\be) + 
  \dtadh \sin(\al-\be) (1+\cos^2(\al-\be))) + {} \\ 
&\qquad \dtanb\sinb\cosb (\MA^2 \sin 2 (\al-\be) +
  \MZ^2 \sin 2 (\al+\be))\,, \\[1ex]
\dmhHsq &= \tfrac 12 (\de\MA^2 \sin 2(\al-\be) - \dMZsq \sin 2(\al+\be)) + {} \\ 
&\qquad \tfrac e{2\MZ\sw\cw} (\dtadH \sin^3(\al-\be) -
  \dtadh \cos^3(\al-\be)) - {} \\ 
&\qquad \dtanb \sinb \cosb (\MA^2 \cos 2 (\al-\be) +
  \MZ^2 \cos 2 (\al+\be))\,, \\[1ex]
\dmHsq &= \de\MA^2 \sin^2(\al-\be) + \dMZsq \cos^2(\al+\be) - {} \\
&\qquad \tfrac e{2 \MZ \sw \cw} (\dtadH \cos(\al-\be)
  (1+\sin^2(\al-\be)) + \dtadh \sin(\al-\be) \cos^2(\al-\be)) - {} \\ 
&\qquad \dtanb \sinb \cosb (\MA^2 \sin 2 (\al-\be) +
  \MZ^2 \sin 2 (\al+\be))\,.
\end{aligned}
\end{equation}
We give the Higgs doublets one renormalization constant each,
\begin{equation}
  \cHe \to (1 + \tfrac 12 \dZ{\cHe}) \cHe\,, \quad
  \cHz \to (1 + \tfrac 12 \dZ{\cHz}) \cHz\,,
\end{equation}
which leads to the field renormalization constants
\begin{equation}
\begin{aligned}
\dZ{hh} &= \sinasq \dZ{\cHe} + \cosasq \dZ{\cHz}\,, \\[.2em]
\dZ{hH} &= \sina \cosa (\dZ{\cHz} - \dZ{\cHe})\,, \\[.2em]
\dZ{HH} &= \cosasq \dZ{\cHe} + \sinasq \dZ{\cHz}\,.
\end{aligned}
\end{equation}
The counter-term for $\tb$ can be expressed in terms of the vacuum
expectation values as
\begin{equation}
\de\tb = \frac 12\KL\dZ{\cHz} - \dZ{\cHe}\KR +
\frac{\de v_2}{v_2} - \frac{\de v_1}{v_1}\,,
\end{equation}
where the $\de v_i$ are the renormalization constants of the $v_i$:
\begin{equation}
v_1 \to \KL 1 + \dZ{\cHe} \KR \KL v_1 + \de v_1 \KR, \quad
v_2 \to \KL 1 + \dZ{\cHz} \KR \KL v_2 + \de v_2 \KR.
\end{equation}
It can be shown that the divergent parts of $\de v_1/v_1$ and $\de 
v_2/v_2$ are equal~\cite{mhiggsf1l,mhiggsf1lC}, thus we set $\de v_2/v_2 
- \de v_1/v_1$ to zero.

In the charged Higgs sector, the renormalized self-energy is written 
similarly as
\begin{equation}
\hat\Sigma_{H^- H^+}(p^2) = \Sigma_{H^- H^+}(p^2) +
  \delta Z_{H^- H^+}(p^2 - m^2_{H^\pm,\text{tree}}) -
  \delta m_{H^\pm}^2,
\end{equation}
where
\begin{align}
\delta m_{H^\pm}^2 &= \delta M_A^2 + \delta M_W^2\,, \\
\delta Z_{H^- H^+} &= \sin^2\beta\,\dZ{\cHe} + \cos^2\beta\,\dZ{\cHz}\,.
\end{align}
We apply on-shell conditions for the masses
\begin{equation}
\dMZsq = \re\se{ZZ}^T(\MZ^2)\,, \quad
\dMWsq = \re\se{WW}^T(\MW^2)\,, \quad
\de\MA^2 = \re\se{AA}(\MA^2)\,. 
\end{equation}
Since the tadpole coefficients are chosen to vanish in all orders,
their counter-terms follow from $T_{\{h,H\}} + \de T_{\{h,H\}} = 0$: 
\begin{equation}
\dtadh = -\tadh, \quad
\dtadH = -\tadH\,.
\end{equation}
\drbar\ renormalization is the most convenient choice for the remaining 
renormalization constants
\begin{equation}
\begin{aligned}
\dZ{\cHe} &= \dZ{\cHe}^{\drbarm}
     = -\KKL \re \Sip_{HH \; |\al = 0} \KKR^{\rm div}, \\[.5em]
\dZ{\cHz} &= \dZ{\cHz}^{\drbarm}
= -\KKL\re\Sip_{hh \; |\al = 0} \KKR^{\text{div}}, \\[.5em]
\dtanb &= \frac 12 (\dZ{\cHz} - \dZ{\cHe}) = \dtanb^{\drbarm}\,.
\end{aligned}
\end{equation}
We choose a renormalization scale of $\mudim = \mt$ in all numerical 
evaluations. 

Finally, in the Feynman-diagrammatic approach we are following here, the 
higher-order-corrected $\cp$-even Higgs-boson masses are derived by 
finding the poles of the $(h,H)$-propagator matrix.  The inverse of this 
matrix is
\begin{equation}
\label{eq:higgsmassmatrixnondiag}
\bigl(\Delta_{\text{Higgs}}\bigr)^{-1}
= -\mathrm{i} \begin{pmatrix}
p^2 - \mHtree^2 + \hSi_{HH}(p^2) & \hSi_{hH}(p^2) \\
\hSi_{hH}(p^2) & p^2 -  \mhtree^2 + \hSi_{hh}(p^2)
\end{pmatrix}.
\end{equation}
Determining its poles is thus equivalent to solving the equation
\begin{equation}
\label{eq:proppole}
\left[p^2 - \mhtree^2 + \hSi_{hh}(p^2) \right]
\left[p^2 - \mHtree^2 + \hSi_{HH}(p^2) \right] -
\left[\hSi_{hH}(p^2)\right]^2 = 0\,.
\end{equation}
The corrected charged Higgs mass is analogously derived as the position
of the pole of the charged-Higgs propagator,
\begin{equation}
p^2 - m^2_{H^\pm,\text{tree}} + \hat{\Sigma}_{H^- H^+}(p^2) = 0\,.
\end{equation}

\bigskip

We calculated the LFV contribution originating from the mixing in the 
slepton sector in a model-independent approach to the Higgs-boson 
masses.  The present experimental uncertainty at the LHC for $\Mh$, the 
mass of the light neutral Higgs boson, is about $350 
\mev$~\cite{ATLAS:2013mma,CMS:yva}.  This can possibly be reduced by 
about 50\% at the LHC and below the level of $\sim 50 \mev$ at the 
ILC~\cite{dbd}.  Similarly, for the masses of the heavy neutral Higgs 
$\MH$ and charged Higgs boson $\MHp$, an uncertainity at the $1\%$ level 
could be expected at the LHC~\cite{cmsHiggs}.  This sets the goal for 
the theoretical uncertainty, which should be reduced to the same (or 
higher) level of accuracy.

\begin{figure}[htb!]
\centerline{\includegraphics{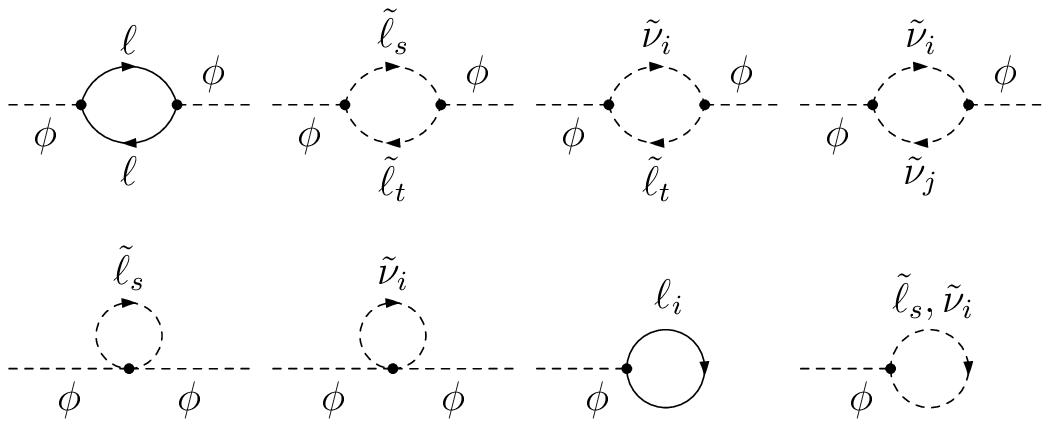}}
\caption[]{\label{fig:FeynDiagHH}Generic Feynman diagrams for the 
Higgs-boson self-energies and tadpoles.  $\phi$ denotes any of the Higgs 
bosons, $h$, $H$, $A$ or $H^\pm$; $\ell$ stands for $e,\mu,\tau$; 
$\til_x$ are the six mass eigenstates of charged sleptons, and $\tinu_x$ 
are the three sneutrino states $\tinu_e$, $\tinu_\mu$, $\tinu_\tau$.}
\end{figure}

The generic Feynman diagrams for the one-loop Higgs-boson self-energies 
relevant for our work are shown in \reffi{fig:FeynDiagHH}.  The diagrams 
were generated with FeynArts and further evaluated using FormCalc, see 
\refse{sec:feynhiggs}.

\subsection{Calculation of EWPO}
\label{sec:EWPO-calc}

EWPO that are known with an accuracy at the per-mille level or better 
have the potential to allow a discrimination between quantum effects of 
the SM and SUSY models, see \citere{PomssmRep} for a review.  Examples 
are the $W$-boson mass $\MW$ and the $Z$-boson observables, such as the 
effective leptonic weak mixing angle $\sweff$, whose present 
experimental uncertainities are \cite{LEPEWWG}
\begin{equation}
\label{eq:EWPO-today}
\de\MW^{\text{exp,today}} \sim 15 \mev, \quad
\de\sweff^{\text{exp,today}} \sim 15 \times 10^{-5}\,, 
\end{equation}
The experimental uncertanity will further be reduced~\cite{Baak:2013fwa} 
to
\begin{equation}
\label{eq:EWPO-future}
\de\MW^{\text{exp,future}} \sim 4\mev, \quad
\de\sweff^{\text{exp,future}} \sim 1.3 \times 10^{-5}\,,
\end{equation}
at the ILC and at the GigaZ option of the ILC, respectively. 

The $W$-boson mass can be evaluated from
\begin{equation}
\MW^2 \KL 1 - \frac{\MW^2}{\MZ^2} \KR = 
\frac{\pi \al}{\wz \Gmu} (1 + \De r)
\end{equation}
where $\al$ is the fine-structure constant and $\Gmu$ the Fermi 
constant.  This relation arises from comparing the prediction for muon 
decay with the experimentally precisely known Fermi constant.  The 
one-loop contributions to $\De r$ can be written as
\begin{equation}
\De r = \De\al - \frac{\cw^2}{\sw^2}\De\rho + (\De r)_{\text{rem}},
\end{equation}
where $\De\al$ is the shift in the fine-structure constant due to the 
light fermions of the SM, $\De\al \propto \log(\MZ/m_f)$, and $\De\rho$ 
is the leading contribution to the $\rho$ parameter~\cite{rho} from 
(certain) fermion and sfermion loops.  The remainder part $(\De 
r)_{\text{rem}}$ contains in particular the contributions from the 
Higgs sector.

The effective leptonic weak mixing angle at the $Z$-boson resonance, 
$\sweff$, is defined through the vector and axial-vector couplings 
($g_{\text{V}}^\ell$ and $g_{\text{A}}^\ell$) of leptons ($\ell$) to the 
$Z$~boson, measured at the $Z$-boson pole.  If this vertex is written as 
$i\bar \ell\ga^\mu (g_{\text{V}}^\ell - g_{\text{A}}^\ell \ga_5) \ell 
Z_\mu$ then
\begin{equation}
\sweff = \frac 14 \KL 1 -
  \Re\frac{g_{\text{V}}^\ell}{g_{\text{A}}^\ell}\KR\,.
\end{equation}
At tree level this coincides with the sine of the weak mixing angle, 
$\sin^2\theta_{\text{W}} = 1 - \MW^2/\MZ^2$, in the on-shell scheme. 
Loop corrections enter through higher-order contributions to 
$g_{\text{V}}^\ell$ and $g_{\text{A}}^\ell$.

Both of these (pseudo-)observables are affected by shifts in the 
quantity $\De\rho$ according to
\begin{equation}
\label{eq:precobs}
\De\MW \approx \frac{\MW}{2}\frac{\cw^2}{\cw^2 - \sw^2} \De\rho\,, \quad
\De\sweff \approx - \frac{\cw^2 \sw^2}{\cw^2 - \sw^2} \De\rho\,.
\end{equation}
The quantity $\De\rho$ is defined by the relation
\begin{equation}
\De\rho = \frac{\Si_Z^{\text{T}}(0)}{\MZ^2} -
          \frac{\Si_W^{\text{T}}(0)}{\MW^2}
\end{equation} 
with the unrenormalized transverse parts of the $Z$- and $W$-boson 
self-energies at zero momentum, $\Si_{Z,W}^{\text{T}}(0)$.  It 
represents the leading universal corrections to the electroweak 
precision observables induced by mass splitting between partners in 
isospin doublets~\cite{rho}. Consequently, it is sensitive to the 
mass-splitting effects induced by non-minimal flavour mixing.

Beyond the $\De\rho$ approximation, the shifts in $\MW$ and $\sweff$ 
originate from the complete sfermion contributions to the quantity $\De 
r$ and to other combinations of the various vector-boson self-energies.  
It has been numerically verified that $\De\rho$ yields an excellent 
approximation for the full calculation in the case of NMFV 
effects~\cite{PomssmRep,delrhoNMFV}, however.

\bigskip

\begin{figure}[htb!]
\centerline{\includegraphics{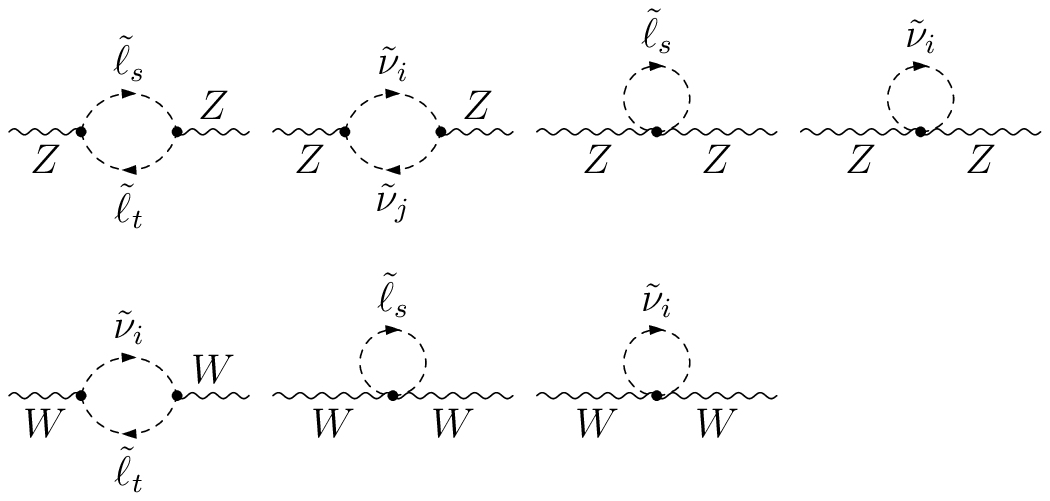}}
\caption{\label{fig:FeynDiagWZ}Generic Feynman diagrams for the $W$- and 
$Z$-boson self-energies containing sleptons in loops.  The six mass 
eigenstates of charged sleptons are denoted by $\til_x$, and $\tinu_x$ 
stands for the three sneutrino states $\tinu_e$, $\tinu_\mu$, 
$\tinu_\tau$.}
\end{figure}

We calculated the LFV contribution to the above-mentioned observables 
entering the $Z$- and $W$-boson self-energies at the one-loop level 
through the $\rho$-parameter.  The generic Feynman diagrams contributing 
to our calculation are shown in \reffi{fig:FeynDiagWZ}.  The diagrams 
were generated with FeynArts and further evaluated using FormCalc, see 
\refse{sec:feynhiggs}.  The resulting evaluation of $\De\rho$ has been 
made publicly available in \fh.  Using \refeq{eq:precobs} the shifts in 
$\MW$ and $\sweff$ induced by LFV have been evaluated, see 
\refse{sec:results}.

\subsection{Changes in FeynArts, FormCalc, and FeynHiggs}
\label{sec:feynhiggs}

\newcommand\MSQ[1]{M_{\tilde Q,#1}}
\newcommand\MSU[1]{M_{\tilde U,#1}}
\newcommand\MSD[1]{M_{\tilde D,#1}}
\newcommand\MSL[1]{M_{\tilde L,#1}}
\newcommand\MSR[1]{M_{\tilde R,#1}}
\newcommand\MSX[1]{M_{\tilde X,#1}}
\newcommand\LL{\mathrm{LL}}
\newcommand\LR{\mathrm{LR}}
\newcommand\RL{\mathrm{RL}}
\newcommand\RR{\mathrm{RR}}
\newcommand\uct{\{{u},{c},{t}\}}
\newcommand\dsb{\{{d},{s},{b}\}}
\newcommand\ru{{u}}
\newcommand\rd{{d}}
\newcommand\deu[3]{(\delta_{\tinu})_{#1} &:= \del{#2}{#3}}
\newcommand\ded[3]{(\delta_{\til})_{#1} &:= \del{#2}{#3}}

FeynArts\cite{feynarts} and FormCalc\cite{formcalc} provide a high level 
of automation for perturbative calculations up to one loop.  This is 
particularly important for models with a large particle content such as 
the MSSM \cite{famssm}.  Here we briefly describe the recent extension 
of the implementation of the MSSM in these packages to include LFV.  
Details on the previous inclusion of NMFV can be found in 
\citeres{feynarts,interplay}.  This involves firstly the modification of 
the slepton couplings in the existing FeynArts model file for the MSSM 
and secondly the corresponding initialization routines for the slepton 
masses and mixings, i.e.\ the $6\times 6$ and $3\times3$ diagonalization 
of the mass matrices in FormCalc.

\subsubsection{FeynArts Model File}

FeynArts' add-on model file \texttt{FV.mod} applies algebraic 
substitutions to the Feynman rules of \texttt{MSSM.mod} to upgrade 
minimal to non-minimal flavor mixing in the sfermion sector.  The 
original version modified only the squark sector, i.e.\ NMFV, and needed 
to be generalized to include LFV.  We solved this by allowing the user 
to choose which sfermion types to introduce non-minimal mixing for 
through the variable \texttt{\$FV} (set before model initialization, of 
course).  For example,
\begin{verbatim}
  $FV = {11, 12, 13, 14};
  InsertFields[..., Model -> {MSSM, FV}]
\end{verbatim}
sets non-minimal mixing for all four sfermion types, with 11 = $\tinu$, 
12 = $\til$, 13 = $\tiu$, and 14 = $\tid$ as usual in \texttt{MSSM.mod}.  
For compatibility with the old NMFV-only version, the default is 
\verb|$FV = {13, 14}|.

\texttt{FV.mod} introduces the following new quantities:
\begin{center}
\begin{tabular}{ll}
\texttt{UASf[$s_1$,$s_2$,$t$]} &
	the slepton mixing matrix $R$, where \\
&	$s_1, s_2 = 1\dots 6$, \\
&	$t = 1\,(\tinu\/), 2\,(\til\/), 3\,(\tiu\/), 4\,(\tid\/)$, \\[1ex]
\texttt{MASf[$s$,$t$]} &
	the slepton masses, where \\
&	$s = 1\dots 6$, \\
&	$t = 1\,(\tinu\/), 2\,(\til\/), 3\,(\tiu\/), 4\,(\tid\/)$.
\end{tabular}
\end{center}
Entries $4\dots 6$ are unused for the sneutrino.


\subsubsection{Model Initialization in FormCalc}

The initialization of the generalized slepton-mixing parameters 
\texttt{MASf} and \texttt{UASf} is already built into FormCalc's regular 
MSSM model-initialization file \texttt{model\_mssm.F} but not turned on 
by default.  It must be enabled by adjusting the \texttt{FV} 
preprocessor flag in \texttt{run.F}:
\begin{verbatim}
   #define FV 2
\end{verbatim}
where 2 is the lowest sfermion type $t$ for which flavor violation is 
enabled, i.e.\ $\til$.

The flavor-violating parameters $\deABij$ are represented in FormCalc 
by the \texttt{deltaSf} matrix:
\begin{center}
\begin{tabular}{ll}
\texttt{double complex deltaSf($s_1$,$s_2$,$t$)} &
	the matrix $(\delta_t)_{s_1s_2}$, where \\
&	$s_1, s_2 = 1\dots 6$ ($1\dots 3$ for $\tinu$), \\
&	$t = 2\,(\til\/), 3\,(\tiu\/), 4\,(\tid\/)$.
\end{tabular}
\end{center}
Since $\delta$ is an Hermitian matrix, only the entries above the
diagonal are considered.  The $\deABij$ are located at the following 
places in the matrix $\delta$:
$$
\left(\begin{array}{ccc|ccc}
\noent & \delta^{LL}_{12} & \delta^{LL}_{13} &
\noent & \delta^{LR}_{12} & \delta^{LR}_{13} \\
\noent & \noent & \delta^{LL}_{23} &
\delta^{RL*}_{12} & \noent & \delta^{LR}_{23} \\
\noent & \noent & \noent &
\delta^{RL*}_{13} & \delta^{RL*}_{23} & \noent \\ \hline
\noent & \noent & \noent &
\noent & \delta^{RR}_{12} & \delta^{RR}_{13} \\
\noent & \noent & \noent &
\noent & \noent & \delta^{RR}_{23} \\
\noent & \noent & \noent &
\noent & \noent & \noent
\end{array}\right)
$$
The trilinear couplings $A_f$ acquire non-zero off-diagonal entries in 
the presence of LFV through the relations
\begin{equation}
m_{f,i} (A_f)_{ij} = (M_{\tif,LR}^2)_{ij}\,,
\quad
i, j = 1\dots 3\, ,
\end{equation}
see \refeq{eq:slep-matrix}.  These off-diagonal trilinear couplings (and 
hence the $\delta$'s) appear directly in the Higgs--slepton--slepton 
couplings, whereas all other effects are mediated through the masses 
and mixings.

The described changes are contained in FeynArts 3.9 and FormCalc 8.4, 
which are publicly available from \texttt{feynarts.de}.


\subsubsection{Inclusion of LFV into FeynHiggs}

As discussed above, the new corrections to the (renormalized) 
Higgs-boson self-energies (and thus to the Higgs-boson masses), as well 
as to $\De\rho$ (and thus to $\MW$ and $\sweff$) have been included in 
\fh~\cite{feynhiggs,mhiggslong,mhiggsAEC,mhcMSSMlong,Mh-logresum}.

The corrections are activated by setting one or more of the $\deABij$ to 
non-zero values.  All $\deABij$ that are not set are assumed to be zero. 
The non-zero value can be set in three ways:
\begin{itemize}
\item
by including them in the input file, e.g.

\texttt{deltaLLL23~~~~0.1}

where the general format of the identifier is

\texttt{delta$F$$XY$$ij$, $F$ = L,E,Q,U,D, $XY$ = LL,LR,RL,RR, $ij$ = 12,23,13}

\item
by calling the subroutine \texttt{FHSetLFV(\ldots)} from your 
Fortran/C/C++ code.

\item
by calling the routine \texttt{FHSetLFV[\ldots]} from your Mathematica
code.

\end{itemize}
The detailed invocation of \texttt{FHSetLFV} is given in the 
corresponding man page included in the \fh\ distribution.  The LFV 
corrections are included starting from \fh\ version 2.10.2,
available from \texttt{feynhiggs.de}.

\section{Selection of Input Parameters}
\label{sec:scenariosLFV}

\subsection{MSSM scenarios} 

For the following numerical analysis we chose the MSSM parameter sets of 
\citere{Arana-Catania:2013nha}.  This framework contains six specific 
points S1\dots S6 in the MSSM parameter space, all of which are well 
compatible with present data, including recent LHC searches and the 
measurements of the muon anomalous magnetic moment.  The values of the 
various MSSM parameters as well as the values of the predicted MSSM mass 
spectra are summarized in \refta{tab:spectra}.  They were evaluated with 
the program 
\fh~\cite{feynhiggs,mhiggslong,mhiggsAEC,mhcMSSMlong,Mh-logresum}.

For simplicity, and to reduce the number of independent MSSM input 
parameters, we assume equal soft masses for the sleptons of the first 
and second generations (similarly for the squarks), and for the left and 
right slepton sectors (similarly for the squarks).  We choose equal 
trilinear couplings for the stop and sbottom squarks and for the 
sleptons consider only the stau trilinear coupling; the others are set 
to zero.  We assume an approximate GUT relation for the gaugino 
soft-SUSY-breaking parameters.  The pseudoscalar Higgs mass $\MA$ and 
the $\mu$ parameter are taken as independent input parameters.  In 
summary, the six points S1\dots S6 are defined in terms of the 
following subset of ten input MSSM parameters:
\begin{align*}
&m_{\tilde L_1} = m_{\tilde L_2}\,, &
&m_{\tilde L_3}\,, &
&(\text{with~} m_{\tilde L_i} = m_{\tilde E_i},\ i = 1,2,3) \\
&m_{\tilde Q_1} = m_{\tilde Q_2} &
&m_{\tilde Q_3}\,, &
&(\text{with~} m_{\tilde Q_i} = m_{\tilde U_i} = m_{\tilde D_i},\ i = 1,2,3) \\
&A_t = A_b\,, &
&A_\tau\,, \\
&M_2 = 2 M_1 = M_3/4\,, &
&\mu\,, \\
&\MA\,, &
&\tb\,.
\end{align*}

\begin{table}[h!]
\caption{\label{tab:spectra}Selected points in the MSSM parameter space 
(upper part) and their corresponding spectra (lower part).  All 
dimensionful quantities are in GeV.}
\centerline{\begin{tabular}{|c|c|c|c|c|c|c|}
\hline
 & S1 & S2 & S3 & S4 & S5 & S6 \\\hline
$m_{\tilde L_{1,2}}$& 500 & 750 & 1000 & 800 & 500 &  1500 \\
$m_{\tilde L_{3}}$ & 500 & 750 & 1000 & 500 & 500 &  1500 \\
$M_2$ & 500 & 500 & 500 & 500 & 750 &  300 \\
$A_\tau$ & 500 & 750 & 1000 & 500 & 0 & 1500  \\
$\mu$ & 400 & 400 & 400 & 400 & 800 &  300 \\
$\tb$ & 20 & 30 & 50 & 40 & 10 & 40  \\
$\MA$ & 500 & 1000 & 1000 & 1000 & 1000 & 1500  \\
$m_{\tilde Q_{1,2}}$ & 2000 & 2000 & 2000 & 2000 & 2500 & 1500  \\
$m_{\tilde Q_{3}}$ & 2000 & 2000 & 2000 & 500 & 2500 & 1500  \\
$A_t$ & 2300 & 2300 & 2300 & 1000 & 2500 &  1500 \\\hline
$m_{\til_{1\dots 6}}$ & 489--515 & 738--765 & 984--1018 & 474--802  & 488--516 & 1494--1507  \\
$m_{\tinu_{1\dots 3}}$ & 496 & 747 & 998 & 496--797 & 496 &  1499 \\
$m_{{\tilde\chi}_{1,2}^\pm}$ & 375--531 & 376--530 & 377--530 & 377--530  & 710--844 & 247--363  \\
$m_{{\tilde\chi}^0_{1\dots 4}}$ & 244--531 & 245--531 & 245--530 & 245--530  & 373--844 & 145--363  \\
$M_h$ & 126.6 & 127.0 & 127.3 & 123.1 & 123.8 & 125.1  \\
$M_H$ & 500 & 1000 & 999 & 1001 & 1000 & 1499  \\
$M_A$ & 500 & 1000 & 1000 & 1000 & 1000 & 1500  \\
$M_{H^\pm}$ & 507 & 1003 & 1003 & 1005 & 1003 & 1502  \\
$m_{\tilde u_{1\dots 6}}$& 1909--2100 & 1909--2100 & 1908--2100 & 336--2000 & 2423--2585 & 1423--1589  \\
$m_{\tilde d_{1\dots 6}}$ & 1997--2004 & 1994--2007 & 1990--2011 & 474--2001 & 2498--2503 &  1492--1509 \\
$m_{\tilde g}$ & 2000 & 2000 & 2000 & 2000 & 3000 &  1200 \\
\hline
\end{tabular}}
\end{table}

The specific values of these ten MSSM parameters in \refta{tab:spectra} 
are chosen to provide different patterns in the various sparticle 
masses, but all leading to rather heavy spectra and thus naturally in 
agreement with the absence of SUSY signals at the LHC.  In particular, 
all points lead to rather heavy squarks and gluinos above $1200\gev$ and 
heavy sleptons above $500\gev$ (where the LHC limits would also permit 
substantially lighter sleptons).  The values of $\MA$ within the 
interval $(500,1500)\gev$, $\tb$ within the interval $(10,50)$ and a 
large $A_t$ within $(1000,2500)\gev$ are fixed such that a light Higgs 
boson $h$ within the LHC-favoured range $(123,127)\gev$ is obtained.

The large values of $\MA\geqslant 500$ GeV place the Higgs sector of our 
scenarios in the so-called decoupling regime\cite{Haber:1989xc}, where 
the couplings of $h$ to gauge bosons and fermions are close to the SM 
Higgs couplings, and the heavy $H$ couples like the pseudoscalar $A$, 
and all heavy Higgs bosons are close in mass.  With increasing $\MA$, 
the heavy Higgs bosons tend to decouple from low-energy physics and the 
light $h$ behaves like $H_{\text{SM}}$.  This type of MSSM Higgs sector 
seems to be in good agreement with recent LHC data\cite{LHCHiggs}.  We 
checked with the code HiggsBounds~\cite{higgsbounds} that this is indeed 
the case (although S3 is right `at the border').

Particularly, the absence of gluinos at the LHC so far forbids too low 
$M_3$ and, through the assumed GUT relation, also a too low $M_2$.  This 
is reflected by our choice of $M_2$ and $\mu$ which give gaugino masses 
compatible with present LHC bounds.  Finally, we required that all our 
points lead to a prediction of the anomalous magnetic moment of the muon 
in the MSSM that can fill the present discrepancy between the Standard 
Model prediction and the experimental value.


\subsection{Selection of \boldmath{$\deABij$} mixings}
\label{sec:limits}

Finally, we need to set the range of values for the explored 
$\deABij$'s.  We use the constraints of \citere{Arana-Catania:2013nha},
calculated from the following LFV processes:
\begin{enumerate}
\item Radiative LFV decays: $\mu\to e\gamma$, $\tau\to e\gamma$, and
$\tau\to \mu\gamma$.  These are sensitive to the $\deABij$'s via the
$(\ell_i \ell_j\gamma)_{\text{1-loop}}$ vertices with a real photon. 

\item Leptonic LFV decays: $\mu\to 3 e$, $\tau\to 3 e$, and $\tau\to 3 
\mu$.  These are sensitive to the $\deABij$'s via the 
$(\ell_i\ell_j\gamma)_{\text{1-loop}}$ vertices with a virtual photon, 
via the $(\ell_i\ell_j Z)_{\text{1-loop}}$ vertices with a virtual $Z$, 
and via the $(\ell_i\ell_j h)_{\text{1-loop}}$, $(\ell_i\ell_j 
H)_{\text{1-loop}}$ and $(\ell_i\ell_j A)_{\text{1-loop}}$ vertices with 
virtual Higgs bosons.

\item Semileptonic LFV tau decays: $\tau\to \mu\eta$ and $\tau\to 
e\eta$.  These are sensitive to the $\deABij$'s via the $(\tau\ell 
A)_{\text{1-loop}}$ vertex with a virtual $A$ and the $(\tau\ell 
Z)_{\text{1-loop}}$ vertex with a virtual $Z$, where $\ell = \mu, e$, 
respectively.

\item Conversion of $\mu$ into $e$ in heavy nuclei: These are sensitive 
to the $\deABij$'s via the $(\mu e\gamma)_{\text{1-loop}}$ vertex with a 
virtual photon, the $(\mu e Z)_{\text{1-loop}}$ vertex with a virtual 
$Z$, and the $(\mu e h)_{\text{1-loop}}$ and $(\mu e H)_{\text{1-loop}}$ 
vertices with a virtual Higgs boson.
\end{enumerate}
 
\begin{table}[h!]
\caption{\label{tab:boundsSpoints}Present upper bounds on the slepton 
mixing parameters $|\delta^{AB}_{ij}|$ for the MSSM points S1\dots S6 
defined in \refta{tab:spectra}.  The bounds for $|\delta^{RL}_{ij}|$ are 
similar to those of $|\delta^{LR}_{ij}|$.}
\centerline{\begin{tabular}{|c|c|c|c|c|c|c|}
\hline
 &  S1 &  S2 &  S3 &  S4 &  S5 & S6  
 \\ \hline
 & & & & & & \\
$|\delta^{LL}_{12}|_{\rm max}$ & $10 \times 10^{-5}$ & $7.5\times 10^{-5}$ &   $5 \times 10^{-5}$& $6 \times 10^{-5}$ & $42\times 10^{-5}$  &  $8\times 10^{-5}$  \\ 
& & & & & & \\
\hline
& & & & & & \\
$|\delta^{LR}_{12}|_{\rm max}$ & $2\times 10^{-6}$ & $3\times 10^{-6}$ &
$4\times 10^{-6}$  & $3\times 10^{-6}$ & $2\times 10^{-6}$  & $1.2\times 10^{-5}$   \\ 
& & & & & & \\
\hline
& & & & & & \\
$|\delta^{RR}_{12}|_{\rm max}$ & $1.5 \times 10^{-3}$& $1.2 \times 10^{-3}$ & 
$1.1 \times 10^{-3}$ & $1 \times 10^{-3}$ & $2 \times 10^{-3}$ & $5.2 \times 10^{-3}$   \\ 
& & & & & & \\
\hline
& & & & & & \\
$|\delta^{LL}_{13}|_{\rm max} $ &  $5 \times 10^{-2}$ & $5 \times 10^{-2}$ & 
$3 \times 10^{-2}$ &  $3 \times 10^{-2}$& $23 \times 10^{-2}$ & $5 \times 10^{-2}$   \\ 
& & & & & & \\
\hline
& & & & & & \\
 $|\delta^{LR}_{13}|_{\rm max}$& $2\times 10^{-2}$  & $3\times 10^{-2}$ & $4\times 10^{-2}$ & $2.5\times 10^{-2}$ & $2\times 10^{-2}$ & $11\times 10^{-2}$   \\ 
& & & & & & \\
 \hline
 & & & & & & \\
$|\delta^{RR}_{13}|_{\rm max}$ & $5.4\times 10^{-1}$  & $5\times 10^{-1}$ & 
 $4.8\times 10^{-1}$ &$5.3\times 10^{-1}$  & $7.7\times 10^{-1}$ & $7.7\times 10^{-1}$ 
  \\ 
 & & & & & & \\ 
  \hline
 & & & & & & \\ 
$|\delta^{LL}_{23}|_{\rm max}$ & $6\times 10^{-2}$  & $6\times 10^{-2}$ & 
 $4\times 10^{-2}$& $4\times 10^{-2}$ & $27\times 10^{-2}$ & $6\times 10^{-2}$ 
  \\ 
 & & & & & & \\ 
  \hline
 & & & & & & \\ 
$|\delta^{LR}_{23}|_{\rm max}$ & $2\times 10^{-2}$   & $3\times 10^{-2}$ & 
$4\times 10^{-2}$ & $3\times 10^{-2}$ & $2\times 10^{-2}$ & $12\times 10^{-2}$ 
  \\ 
 & & & & & & \\ 
  \hline
 & & & & & & \\ 
$|\delta^{RR}_{23}|_{\rm max}$ & $5.7\times 10^{-1}$  & $5.2\times 10^{-1}$ & 
 $5\times 10^{-1}$& $5.6\times 10^{-1}$ & $8.3\times 10^{-1}$ & $8\times 10^{-1}$ 
  \\ 
 & & & & & & \\  
  \hline
\end{tabular}}
\end{table}

Applying the most recent constraints from the LFV processes listed above 
yields up-to-date limits on the $\deABij$~\cite{Arana-Catania:2013nha}. 
Using these upper bounds on $\delta^{AB}_{ij}$ given in 
\refta{tab:boundsSpoints}, we calculate the corrections to the Higgs 
boson masses and the EWPO.  For each explored non-vanishing delta, the 
corresponding physical sfermion masses and mixings, as well as the EWPO 
and Higgs masses were numerically computed with FeynHiggs 2.10.2, which 
includes the analytical results of our calculations.


\section{Results and discussion} 
\label{sec:results}

We implemented the full one-loop results including all LFV mixing terms 
for the $W$-, $Z$-, and Higgs-boson self-energies in \fh\ 2.10.2.  The 
analytical results are lengthy and not shown here.  For the numerical 
study we analyzed all 12 $\deABij$ for the MSSM scenarios defined in 
\refta{tab:spectra}.  For a better view of the LFV effects we shall plot
only the differences
\begin{align}
\De\rho^{\text{LFV}} &= \De\rho-\De\rho^{\text{MSSM}}\,, \\
\de\MW^{\text{LFV}} &= \MW-\MW^{\text{MSSM}}\,, \\
\de\sweff^{\text{LFV}} &= \sweff-\sweff^{\text{MSSM}}\,,
\end{align}
where $\De\rho^{\text{MSSM}}$, $\MW^{\text{MSSM}}$, and 
$\sweff^{\text{MSSM}}$ are the values with $\deABij = 0$ (the latter two 
evaluated with the help of \refeq{eq:precobs}).  Furthermore we use
\begin{align}
\De\Mh^{\text{LFV}} &= \Mh - \Mh^{\text{MSSM}}\,, \\
\De\MH^{\text{LFV}} &= \MH - \MH^{\text{MSSM}}\,, \\
\De\MHp^{\text{LFV}} &= \MHp - \MHp^{\text{MSSM}}\,,
\end{align}
where again $\Mh^{\text{MSSM}}$, $\MH^{\text{MSSM}}$ and 
$\MHp^{\text{MSSM}}$ are the values for $\deABij = 0$.  The SM results 
for $\MW$ and $\sweff$ are $\MW=80.361 \gev$ and $\sweff=0.23152$ as 
evaluated with \fh\ (using the approximation formulas given in 
\citeres{MWSMapprox,sw2effSMapprox}.  The numerical values of $\De\rho$, 
$\MW$, $\sweff$, $\Mh$, $\MH$ and $\MHp$ in the MSSM with all $\deABij = 
0$ are summarized in \refta{tab:absolutevalues}.

\begin{table}[h!]
\begin{tabular}{|c|c|c|c|c|c|c|}
\hline
 &  S1 &  S2 &  S3 &  S4 &  S5 & S6  
 \\ \hline
 & & & & & & \\
$\De\rho$ & $2.66 \times 10^{-5}$ & $1.72\times 10^{-5}$ &   $1.39 \times 10^{-5}$& $2.35 \times 10^{-4}$ & $2.36\times 10^{-5}$  &  $2.14\times 10^{-5}$  \\ 
& & & & & & \\
\hline
& & & & & & \\
$\MW$ & $ 80.362$ & $ 80.362 $ & $80.361$  & $80.375$ & $80.364$  & $80.363$   \\ 
& & & & & & \\
\hline
& & & & & & \\
$\sweff$ & $0.23151$& $0.23152$ & $0.23152$ & $0.23143$ & $0.23150$ & $0.23151$   \\ 
& & & & & & \\
\hline
& & & & & & \\
$ M_{h} $ &  $126.257$ & $126.629$ & $126.916$ &  $123.205$& $123.220$ & $124.695$   \\ 
& & & & & & \\
\hline
& & & & & & \\
 $M_{H} $ & $500.187$  & $999.580$ & $999.206$ & $1001.428$ & $1000.239$ & $1499.365$   \\ 
& & & & & & \\
 \hline
 & & & & & & \\
$M_{H^{\pm}} $ & $506.888$  & $1003.182$ & $1003.005$ &$1005.605$  & $1003.454$ & $1501.553$ 
  \\ 
 & & & & & & \\ 
 \hline
\end{tabular}
\caption{The values of  $\De\rho$, $\MW$, $\sweff$, 
$\Mh$, $\MH$ and $\MHp$ for the selected S1-S6 MSSM points defined
in \refta{tab:spectra} (i.e.\ with all $\deABij = 0$). Mass values are
in~GeV.}  
\label{tab:absolutevalues}
\end{table}

\begin{figure}[ht!]
\begin{center}
\psfig{file=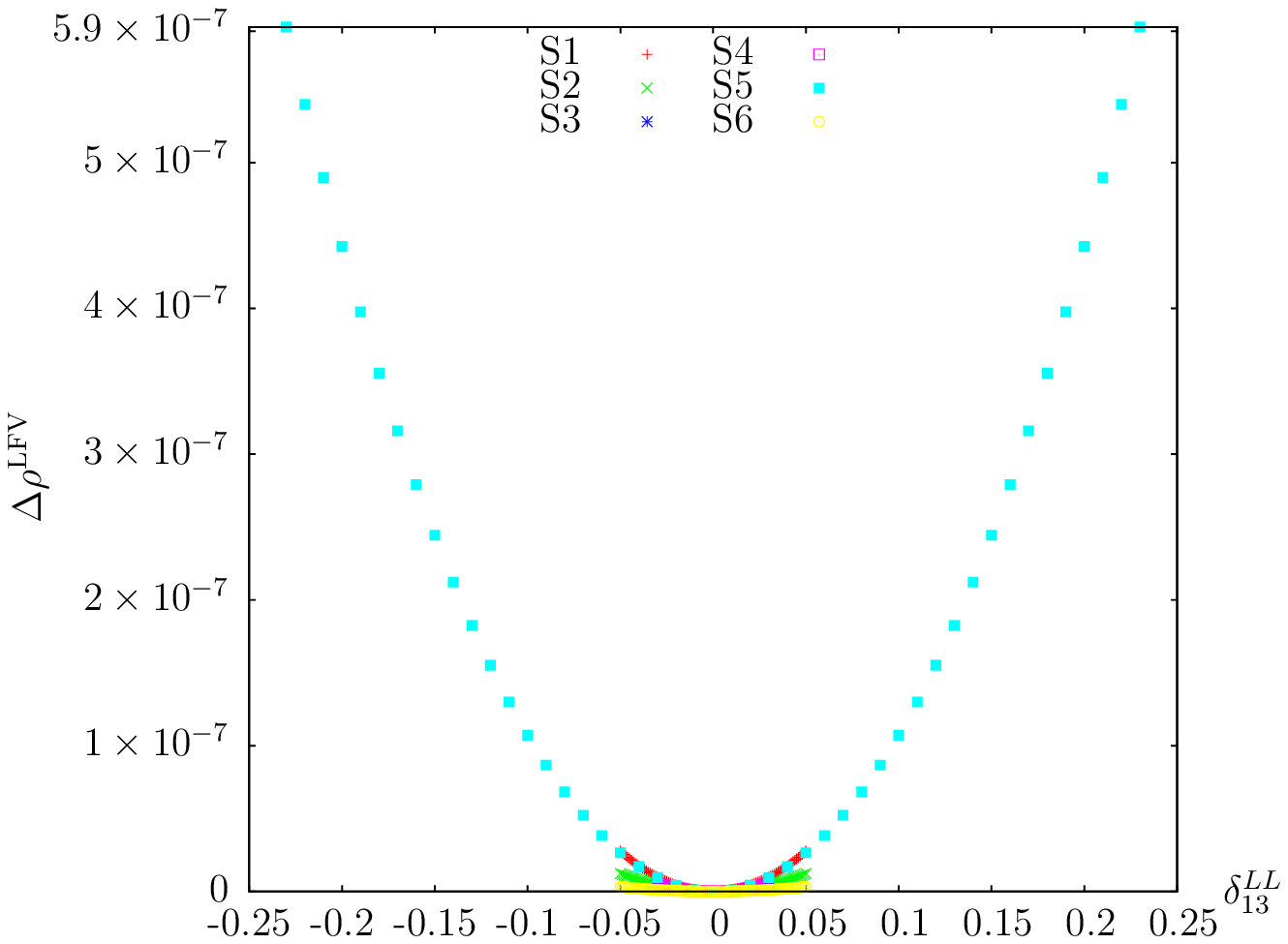  ,scale=0.57,angle=0,clip=}
\psfig{file=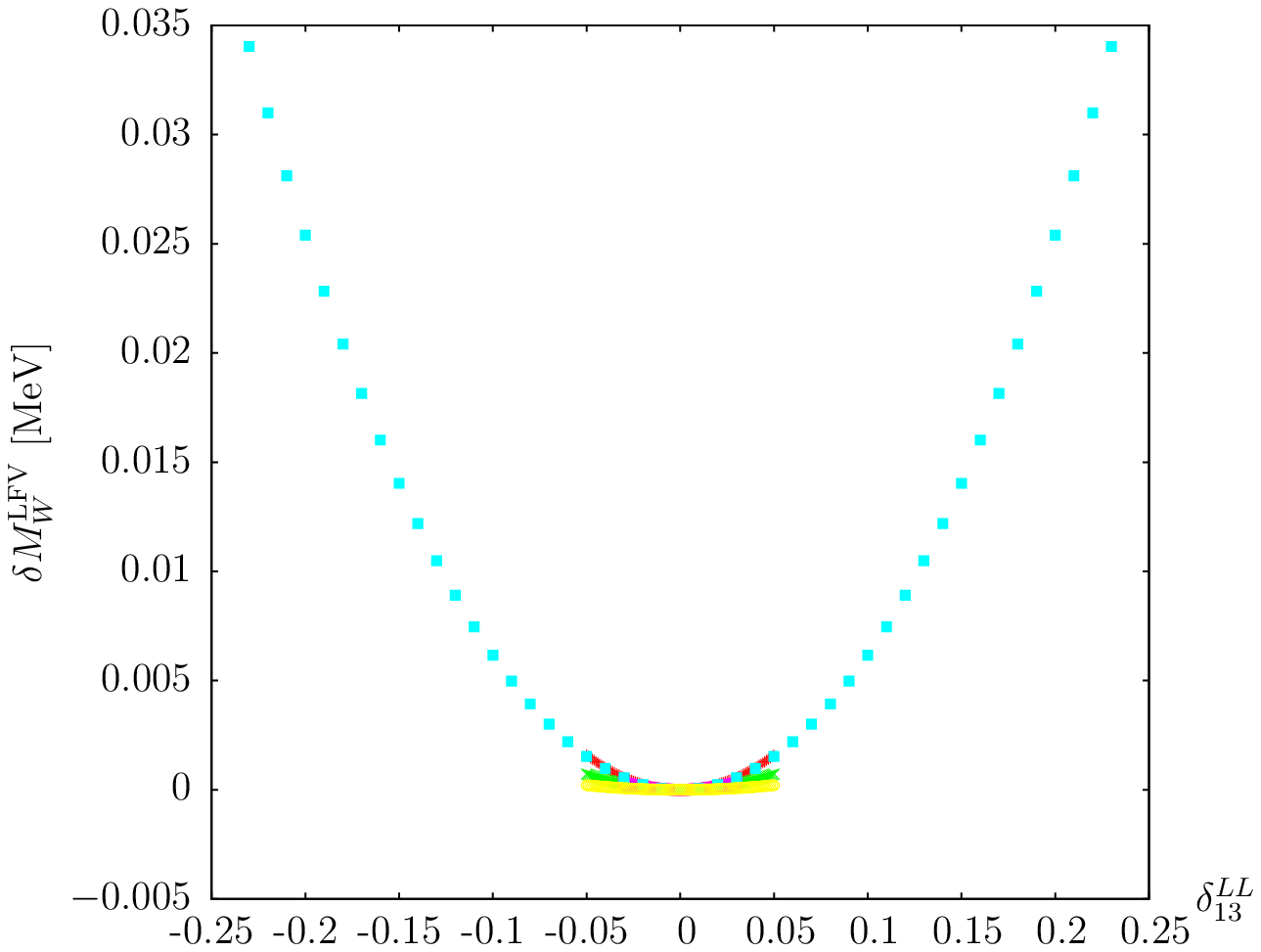  ,scale=0.57,angle=0,clip=}\\
\vspace{0.5cm}
\psfig{file=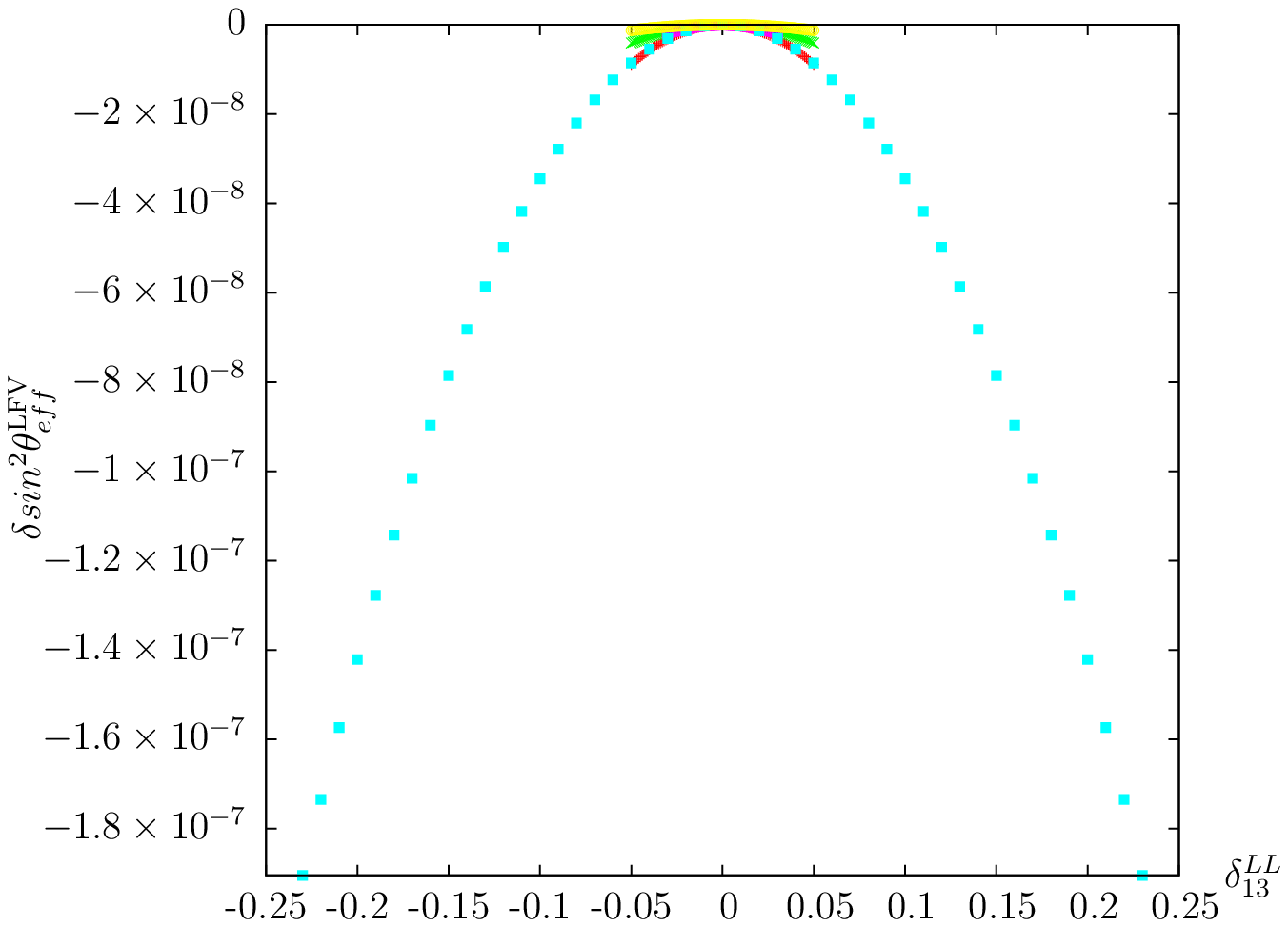 ,scale=0.57,angle=0,clip=}
\psfig{file=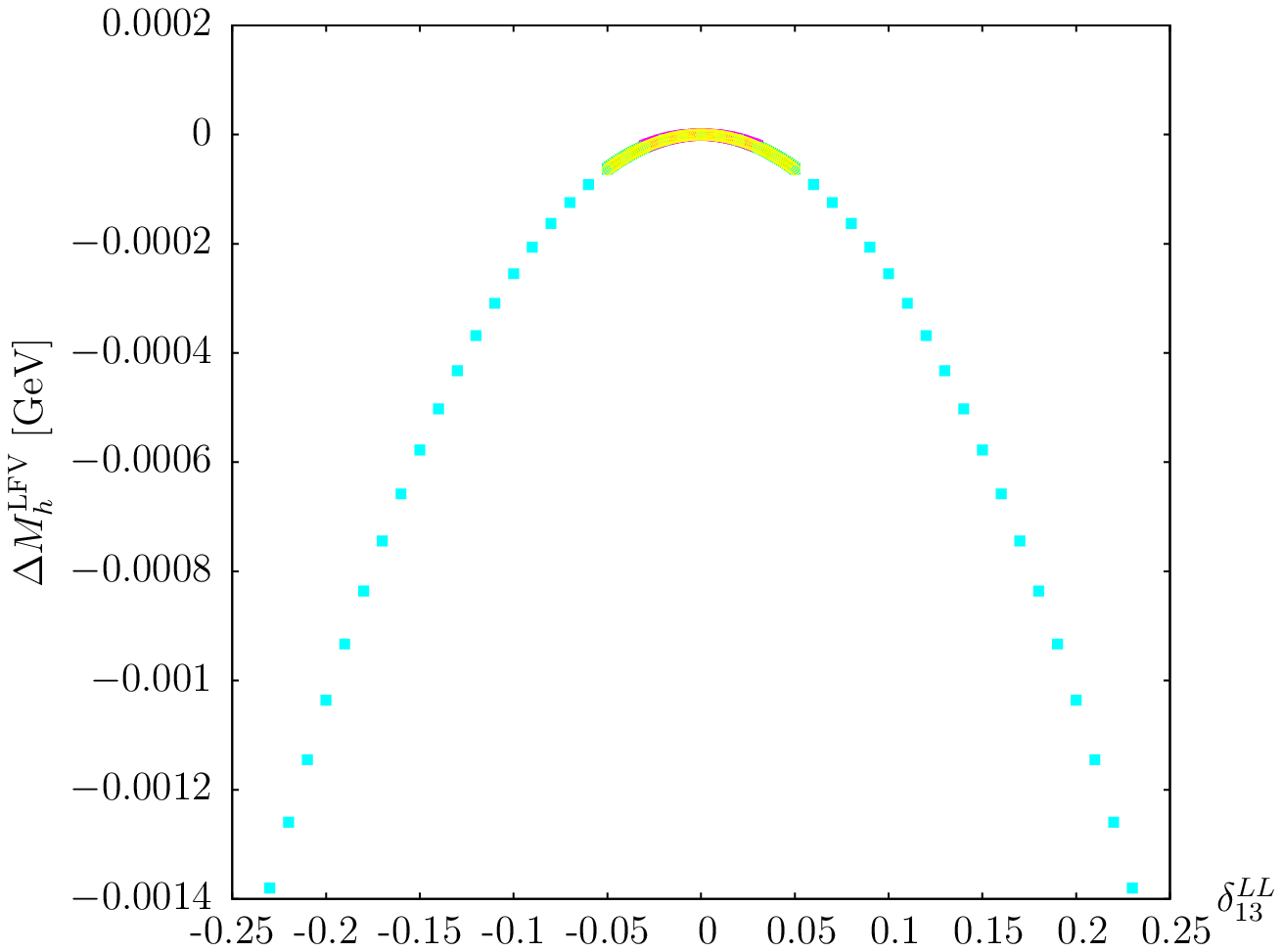   ,scale=0.57,angle=0,clip=}\\
\vspace{0.5cm}
\psfig{file=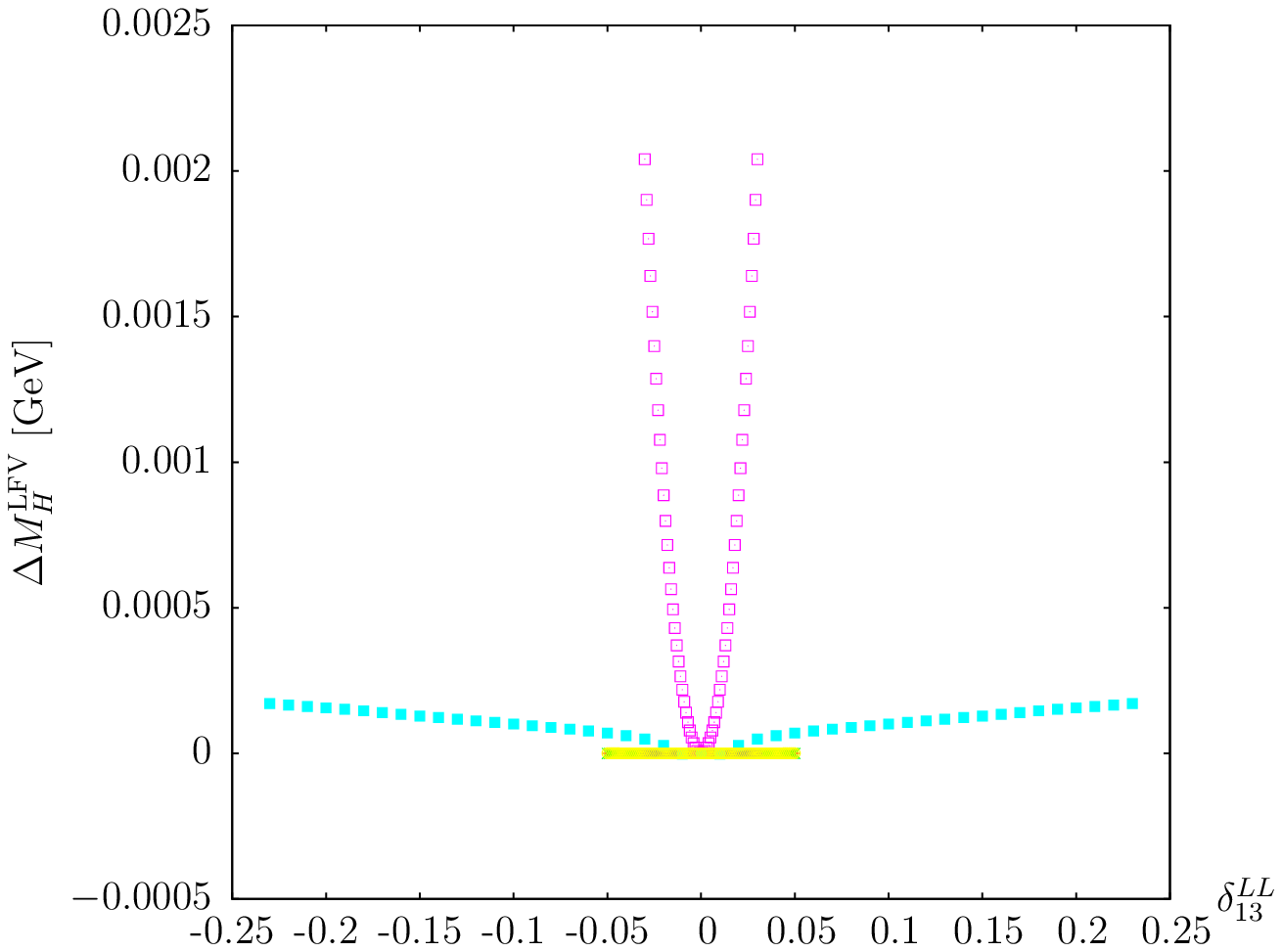  ,scale=0.57,angle=0,clip=}
\psfig{file=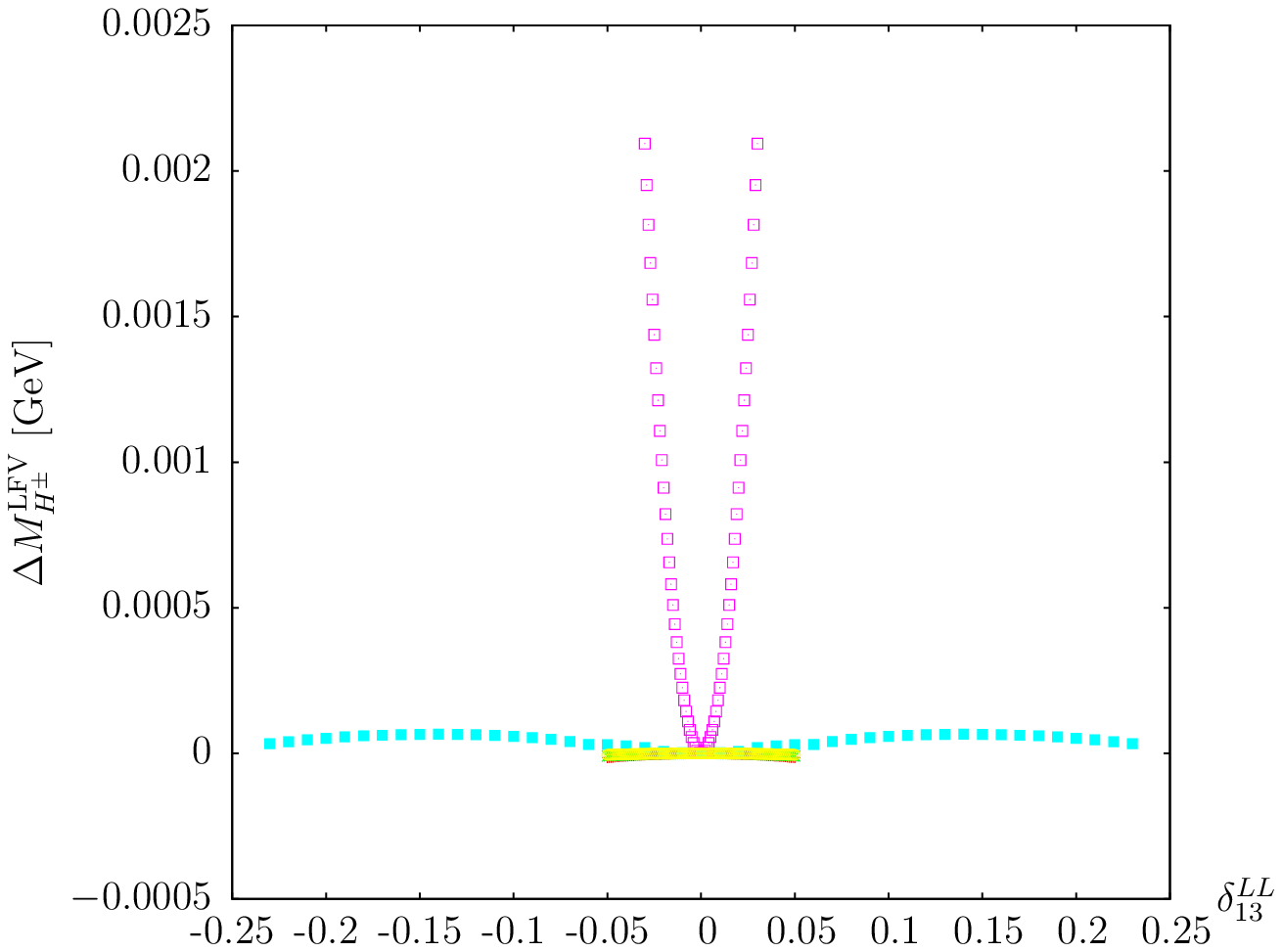  ,scale=0.57,angle=0,clip=}\\

\end{center}
\caption{EWPO and Higgs masses as a function of slepton
  mixing $\delta^{LL}_{13}$ for the six points defined in the \refta{tab:spectra}.}  
\label{figdLL13}
\end{figure}

\begin{figure}[ht!]
\begin{center}
\psfig{file=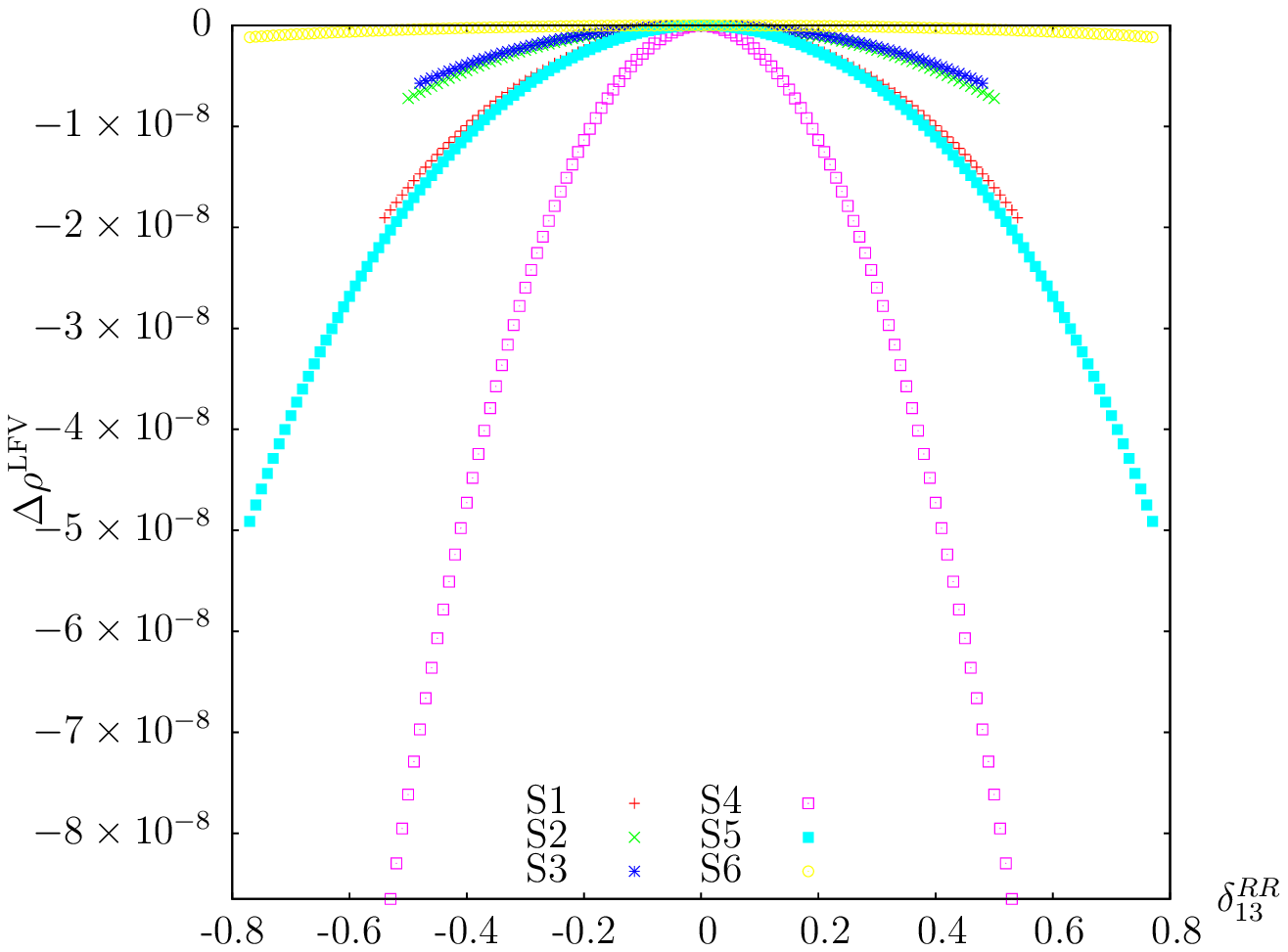  ,scale=0.57,angle=0,clip=}
\psfig{file=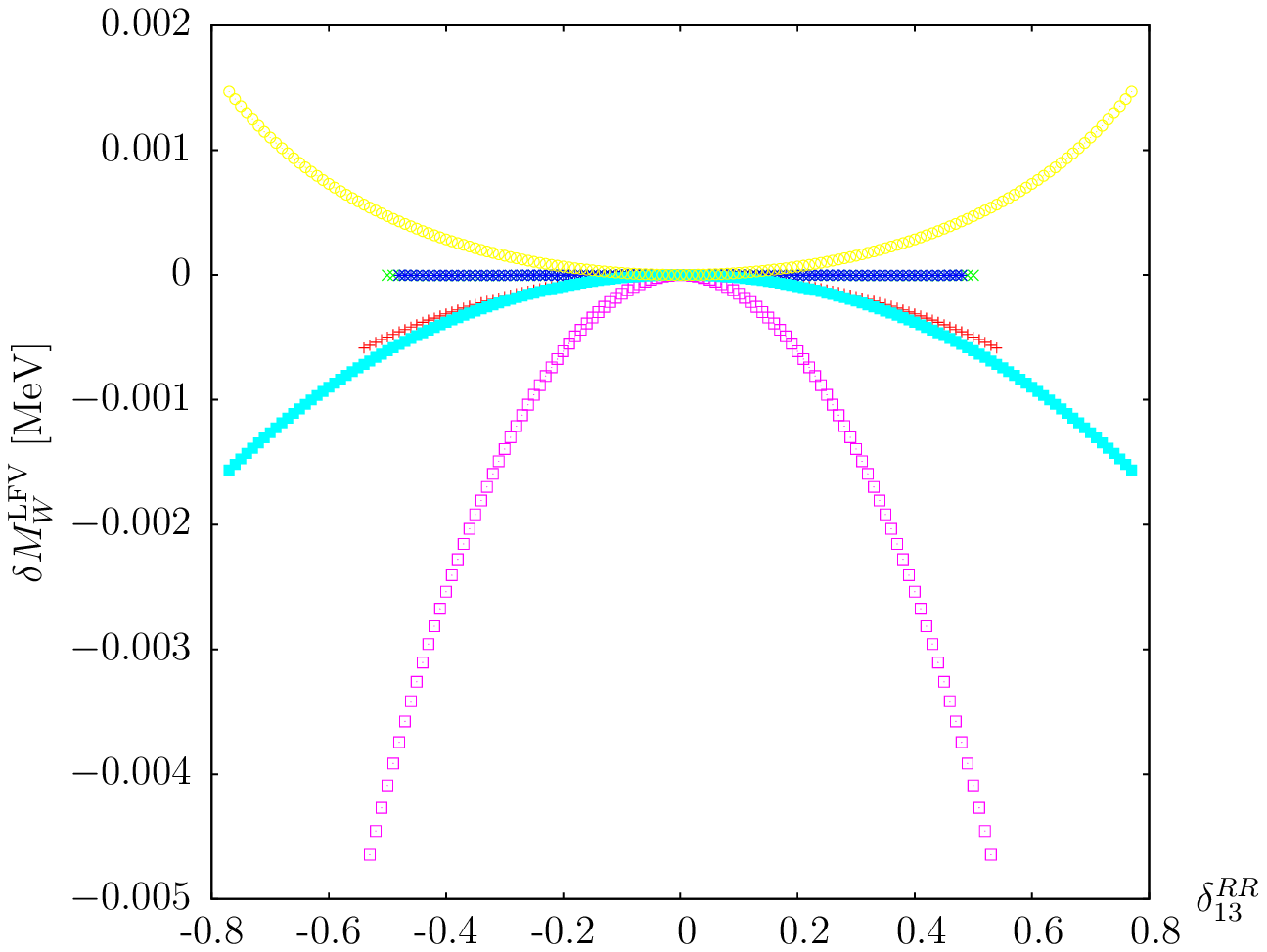  ,scale=0.57,angle=0,clip=}\\
\vspace{0.5cm}
\psfig{file=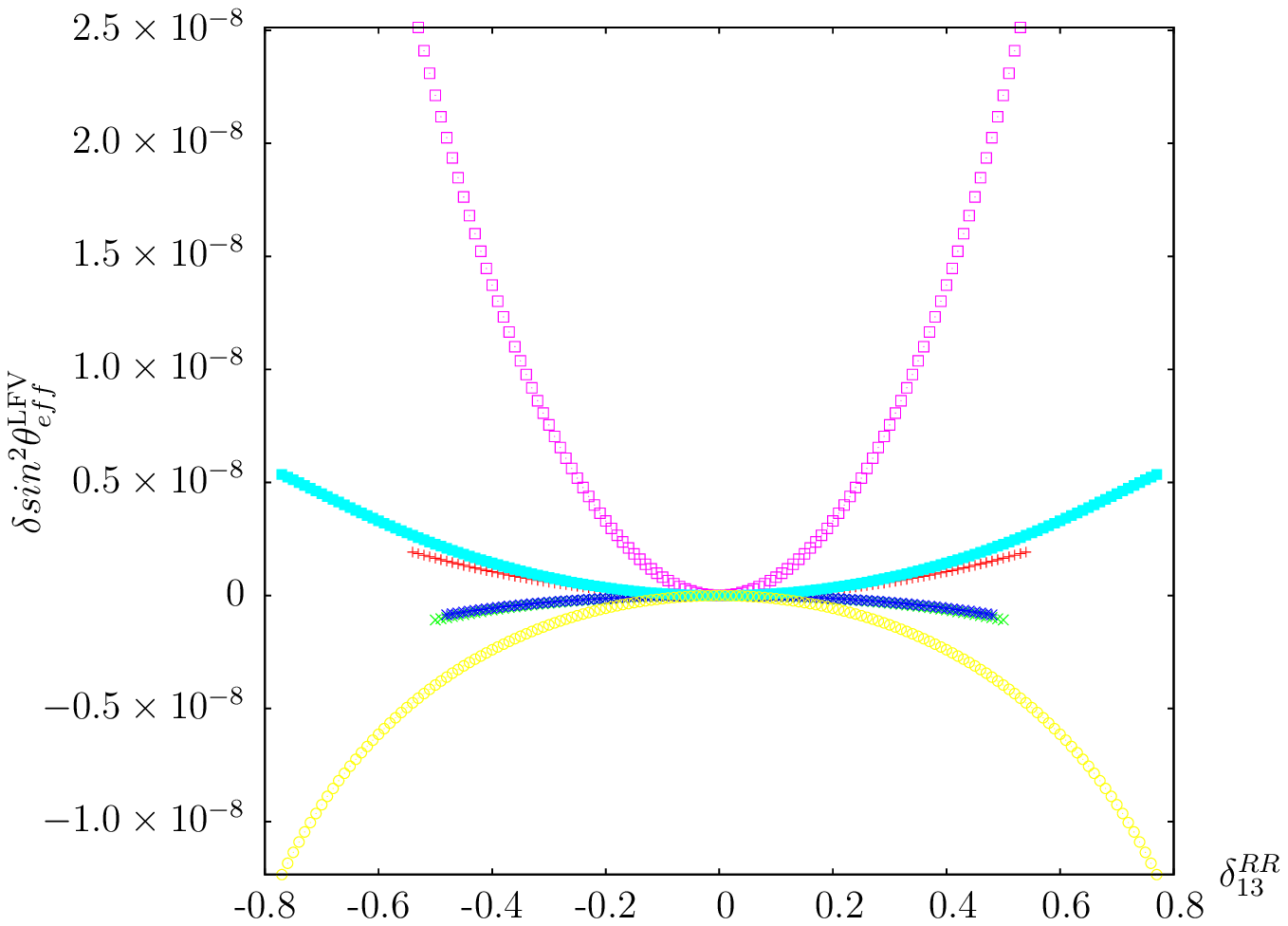 ,scale=0.56,angle=0,clip=}
\psfig{file=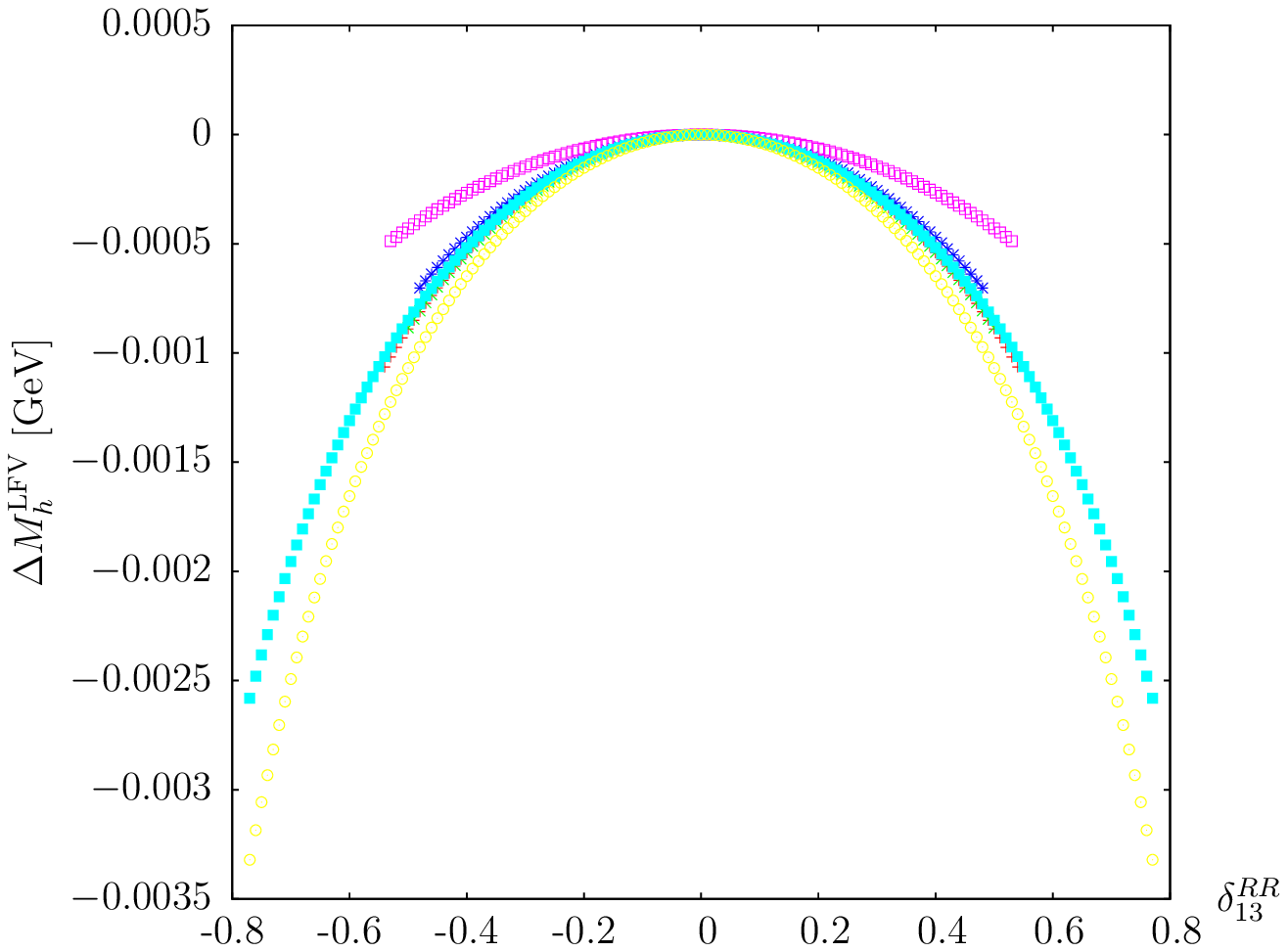   ,scale=0.56,angle=0,clip=}\\
\vspace{0.5cm}
\psfig{file=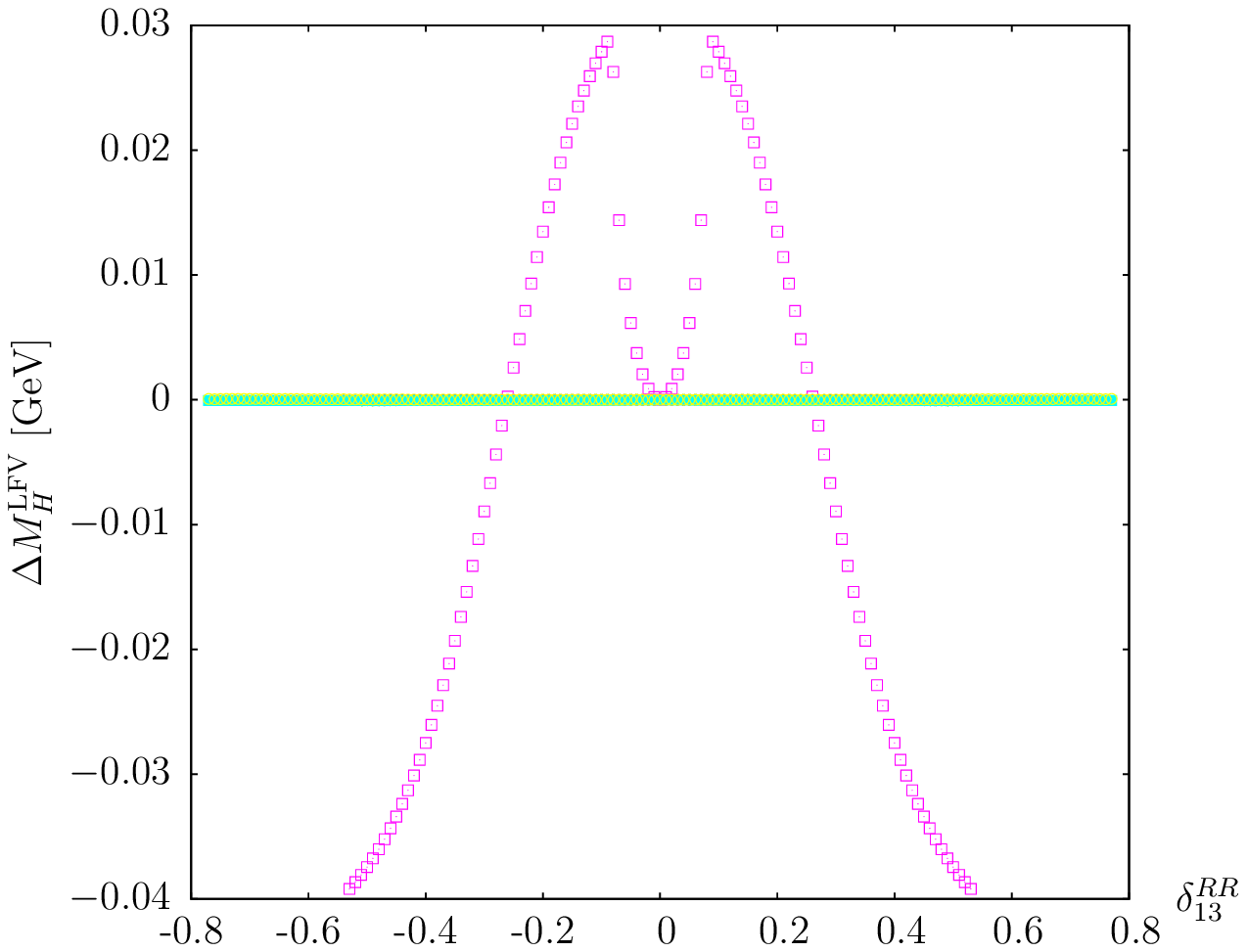  ,scale=0.57,angle=0,clip=}
\psfig{file=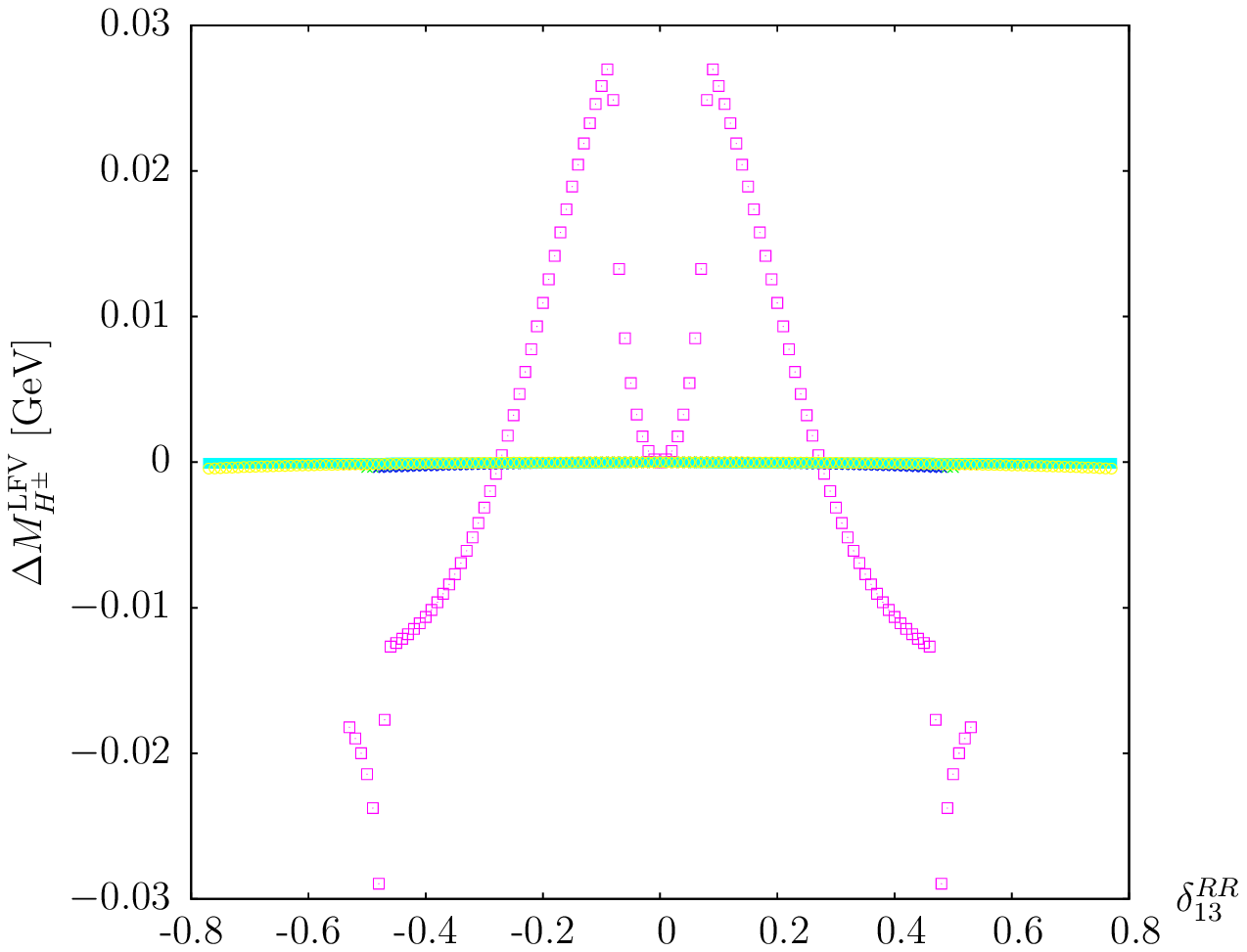  ,scale=0.57,angle=0,clip=}\\

\end{center}
\caption{EWPO and Higgs masses as a function of slepton
  mixing $\delta^{RR}_{13}$ for the six points defined in the \refta{tab:spectra}.}  
\label{figdRR13}
\end{figure}

\begin{figure}[ht!]
\begin{center}
\psfig{file=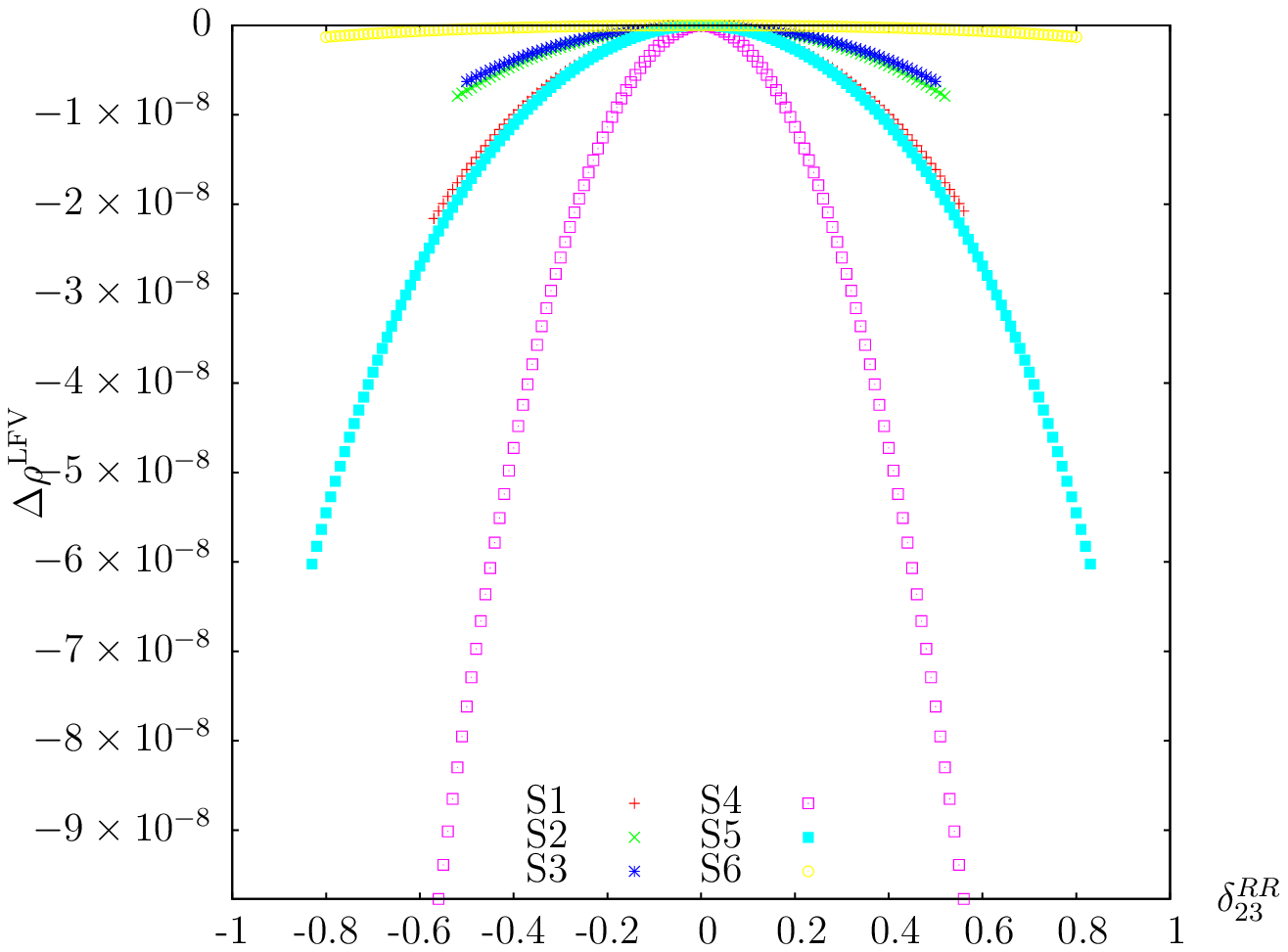  ,scale=0.57}
\psfig{file=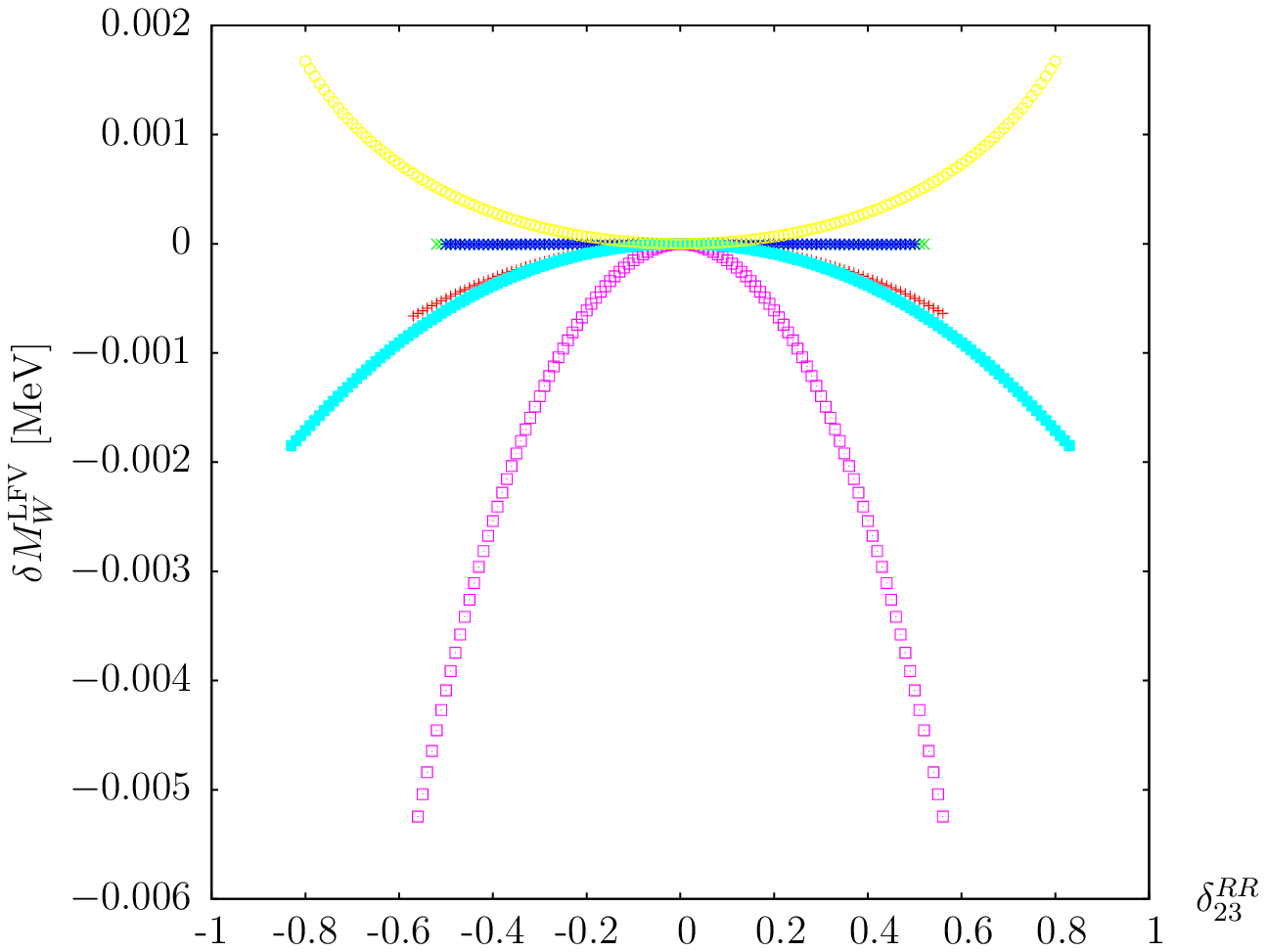  ,scale=0.57}\\
\vspace{0.5cm}
\psfig{file=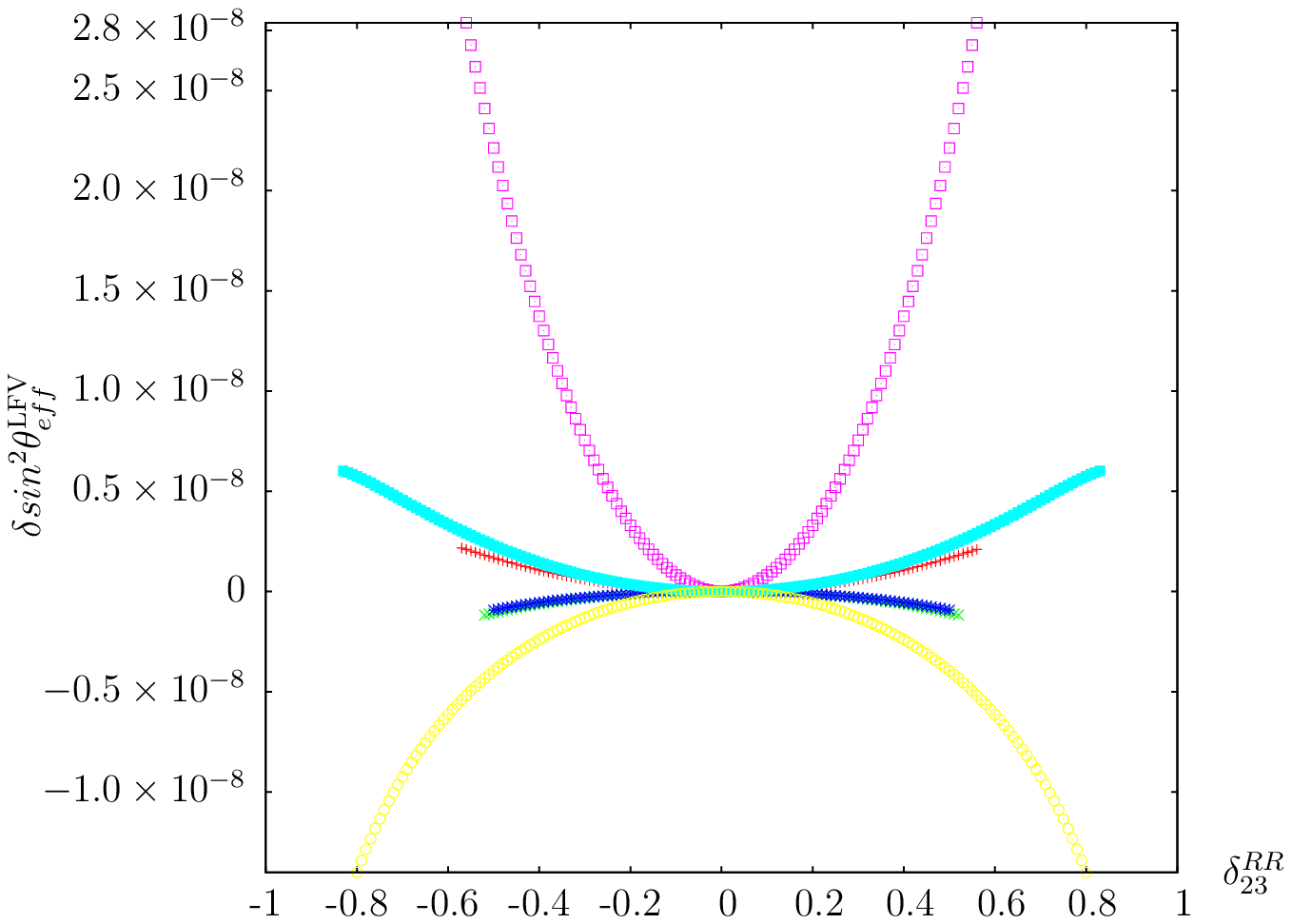 ,scale=0.56}
\psfig{file=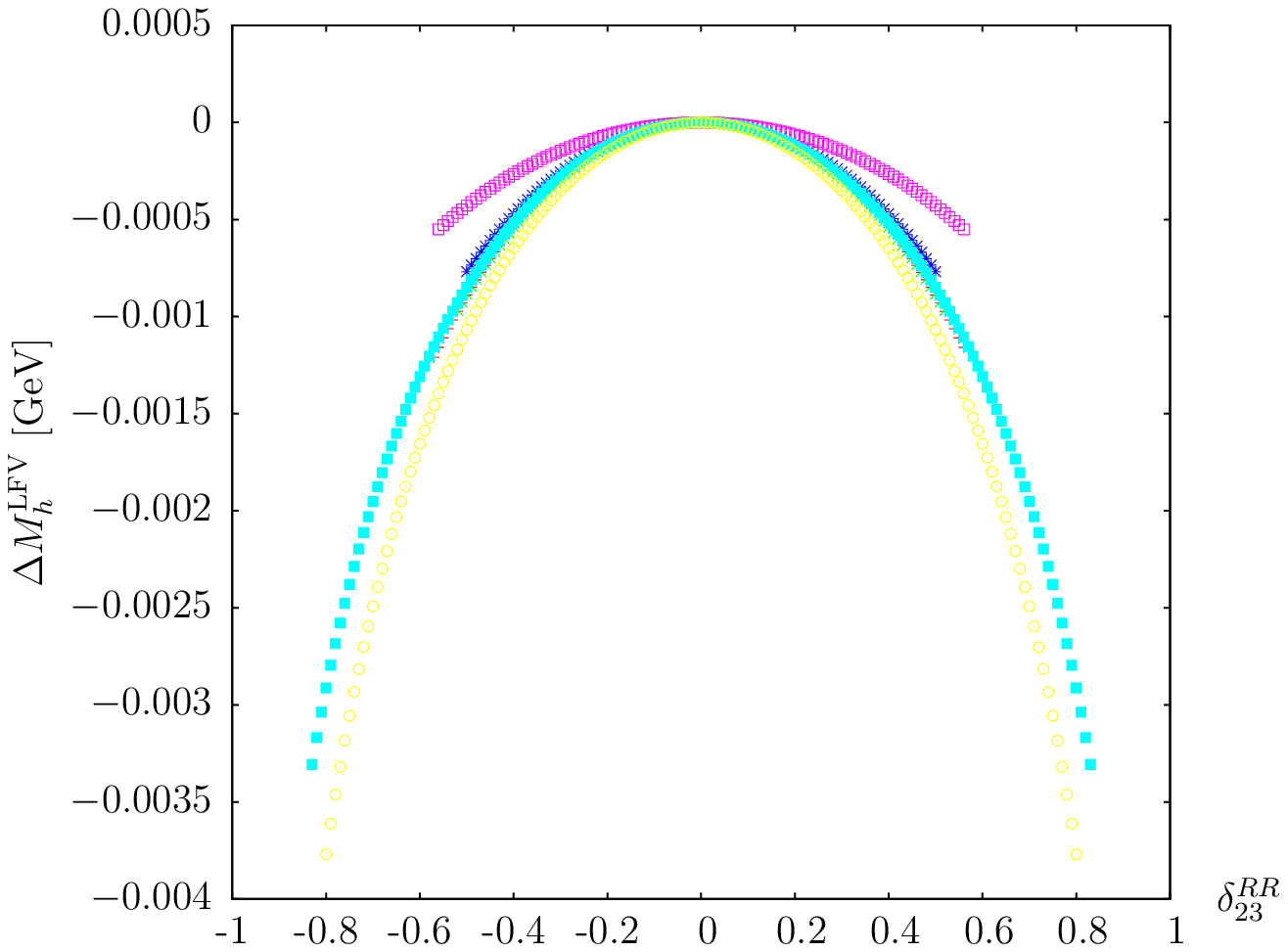   ,scale=0.56}\\
\vspace{0.5cm}
\psfig{file=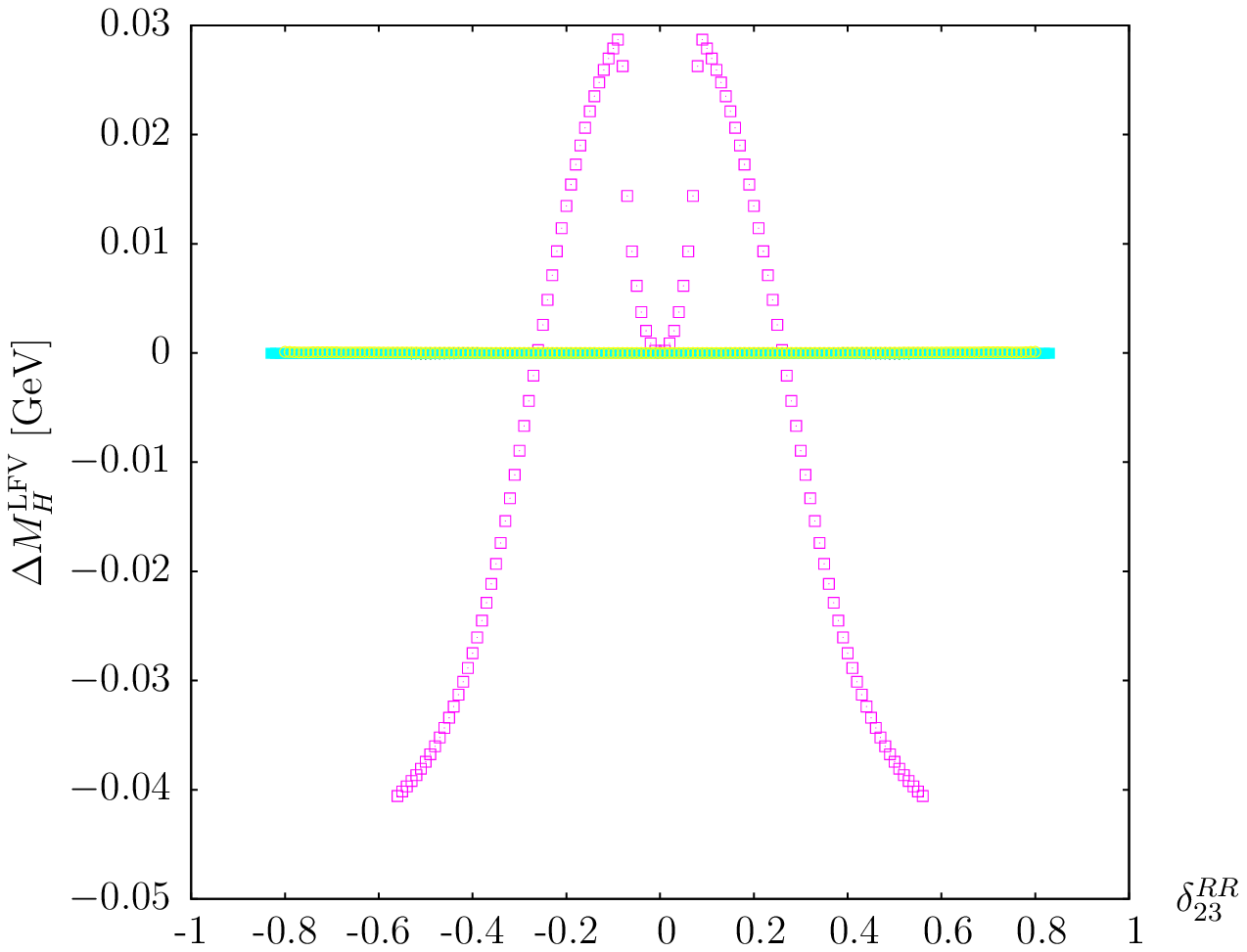  ,scale=0.57}
\psfig{file=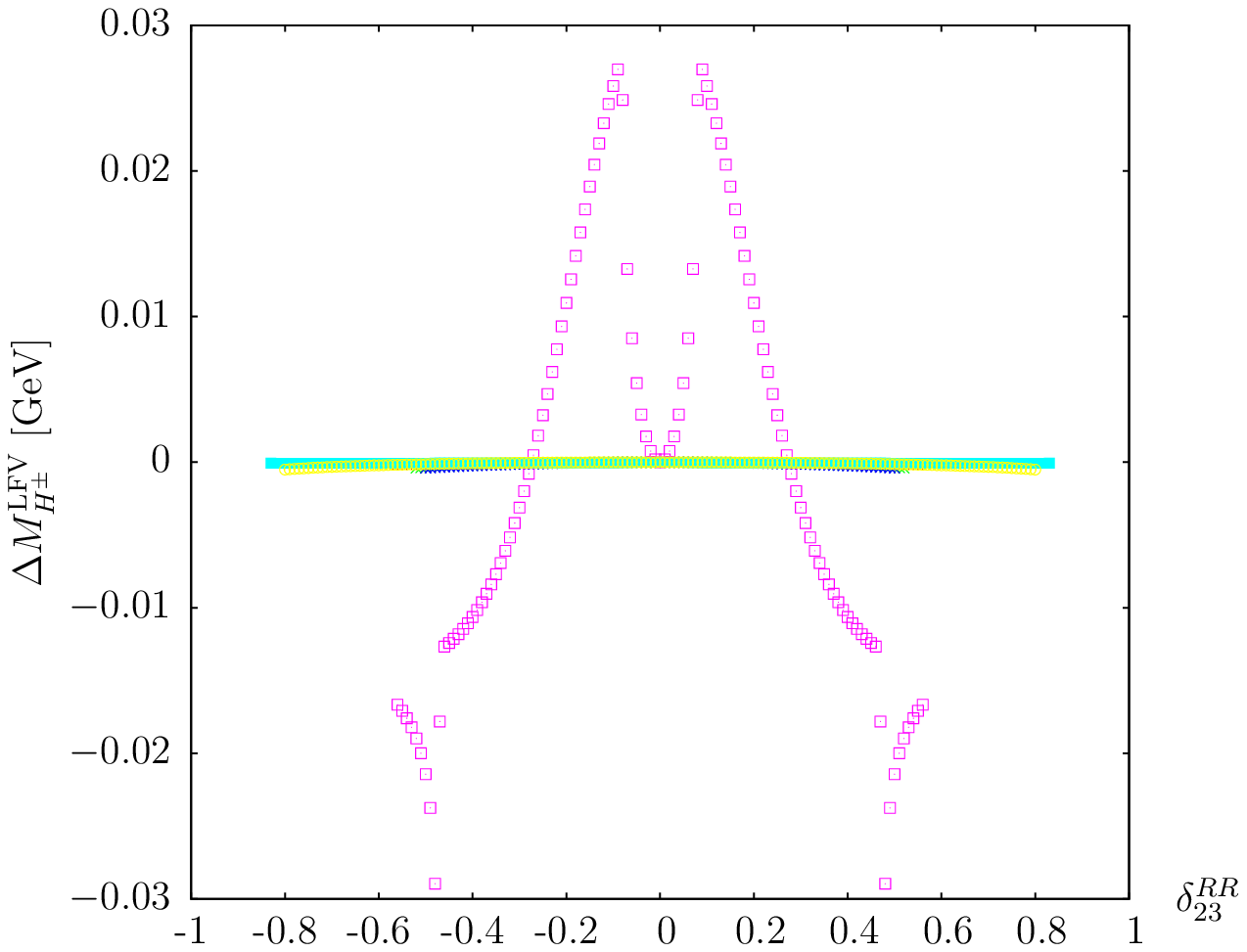  ,scale=0.57}\\

\end{center}
\caption{EWPO and Higgs masses as a function of slepton
  mixing $\delta^{RR}_{23}$ for the six points defined in the \refta{tab:spectra}.}  
\label{figdRR23}
\end{figure}

Our numerical results are shown in Figs.~\ref{figdLL13}--\ref{figdRR23}. 
The six plots in each figure are ordered as follows.
Upper left: $\De\rho^{\text{LFV}}$, 
upper right: $\de\MW^{\text{LFV}}$, 
middle left: $\de\sweff^{\text{LFV}}$, 
middle right: $\De \Mh^{\text{LFV}}$, 
lower left: $\De \MH^{\text{LFV}}$, 
and lower right: $\De \MHp^{\text{LFV}}$, 
as a function of $\de^{LL}_{13}$ (Fig.~\ref{figdLL13}), 
$\de^{LL}_{23}$ (Fig.~\ref{figdLL23}),
$\de^{LR}_{13}$ (Fig.~\ref{figdLR13}), 
$\de^{LR}_{23}$ (Fig.~\ref{figdLR23}), 
$\de^{RL}_{13}$ (Fig.~\ref{figdRL13}),
$\de^{RL}_{23}$ (Fig.~\ref{figdRL23}), 
$\de^{RR}_{13}$ (Fig.~\ref{figdRR13}) 
and $\de^{RR}_{23}$ (Fig.\ref{figdRR23}).
The legends are shown only in the first plot of each figure. 
We do not show results for LFV effects involving only the first and
second generation.  While they are included for completeness in our
analytical results, they are expected to have a negligible effect on the
observables considered here.  The latter is confirmed by the numerical
analysis presented in the next subsections.


\subsection{EWPO}
\label{sec:ewpo}

\begin{figure}[ht!]
\begin{center}
\psfig{file=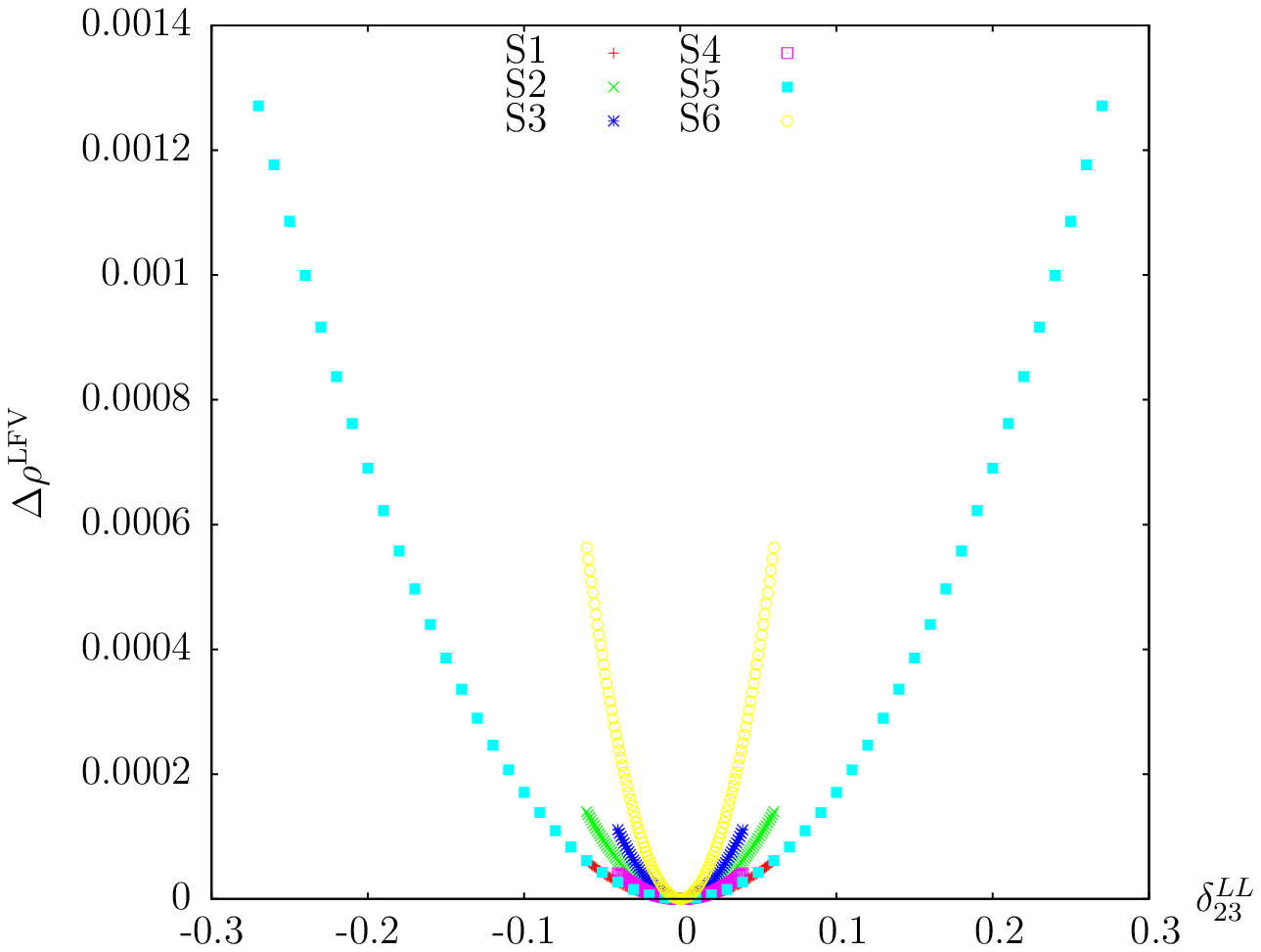  ,scale=0.57,angle=0,clip=}
\psfig{file=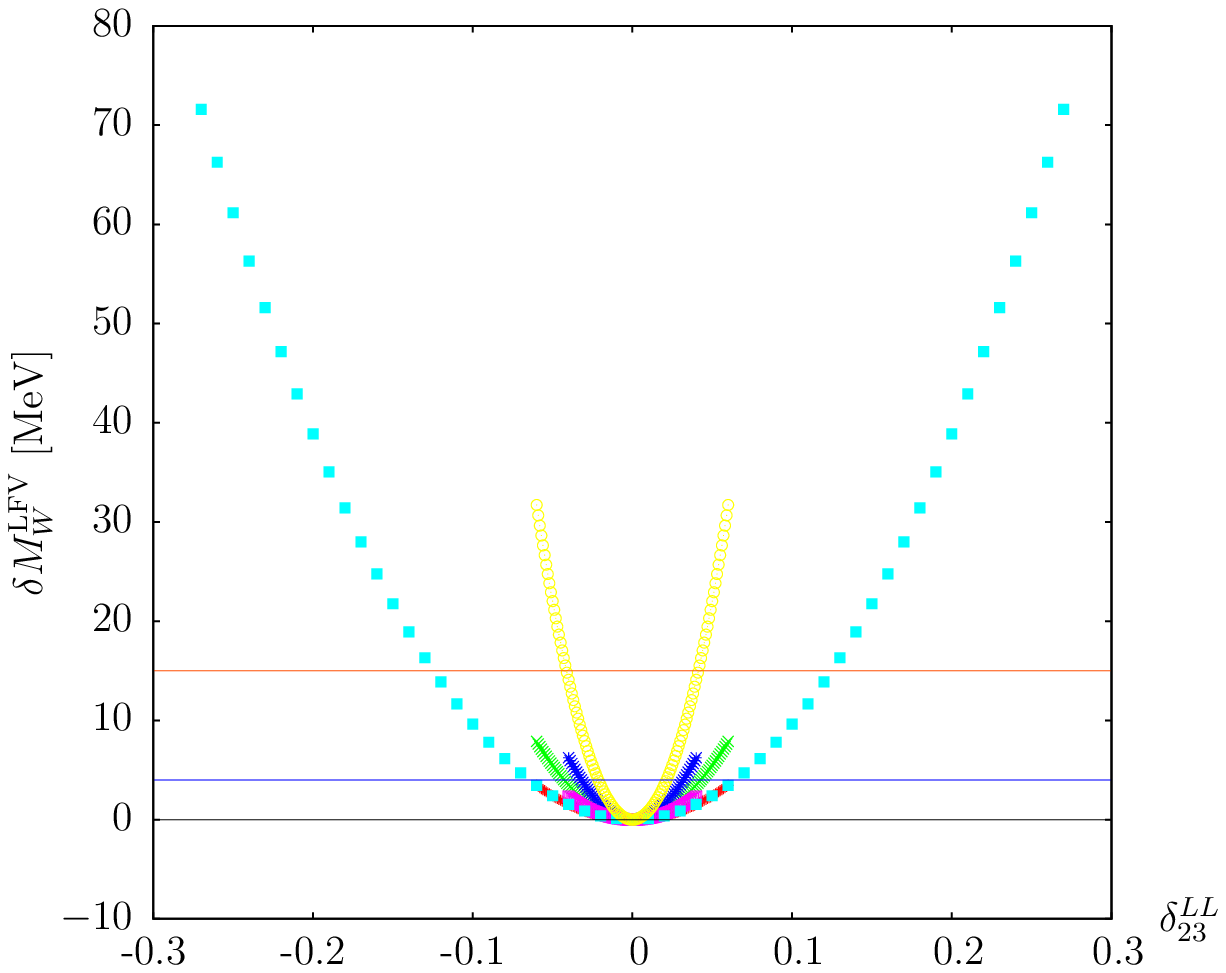  ,scale=0.57,angle=0,clip=}\\
\vspace{0.5cm}
\psfig{file=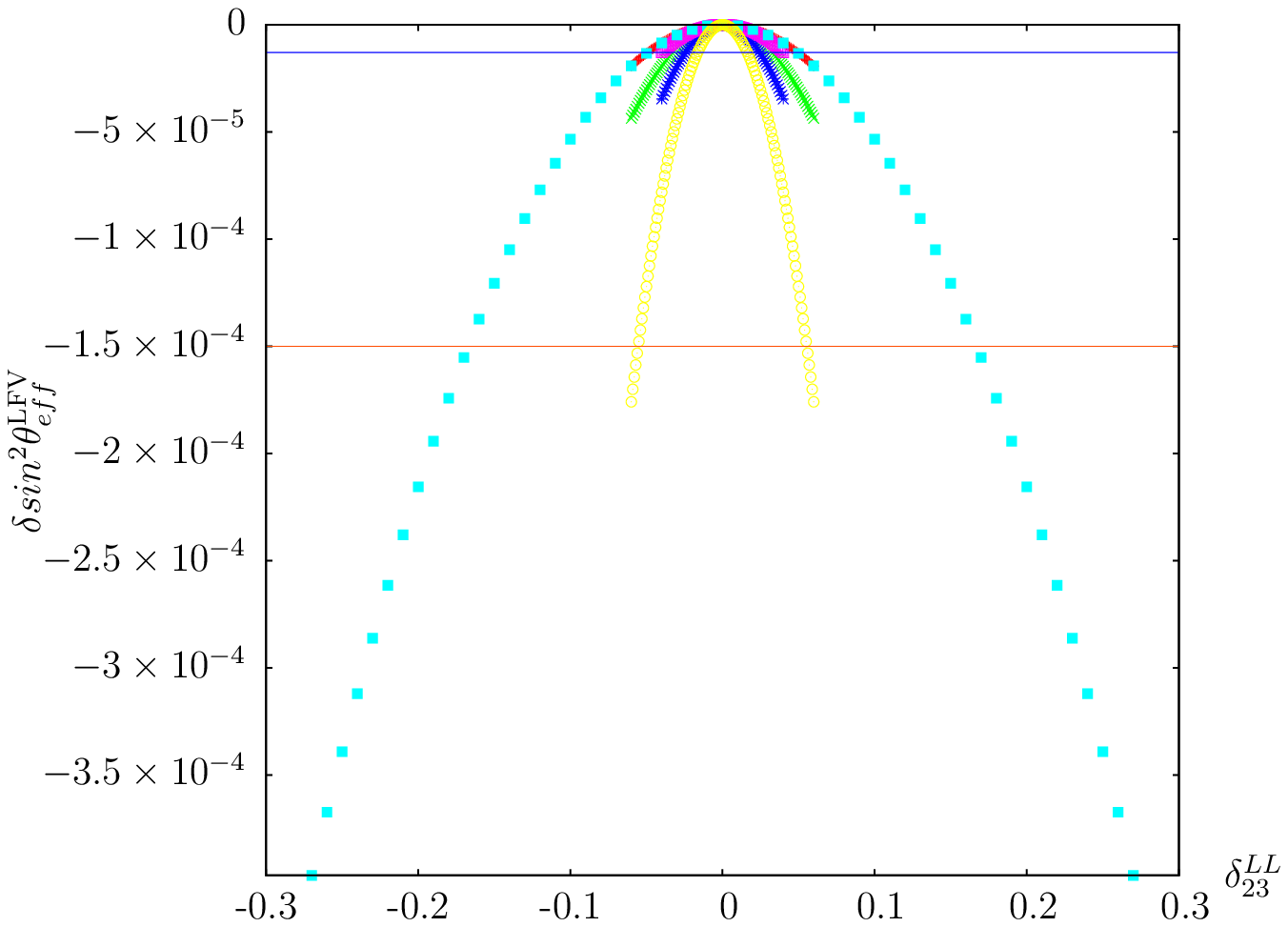 ,scale=0.56,angle=0,clip=}
\psfig{file=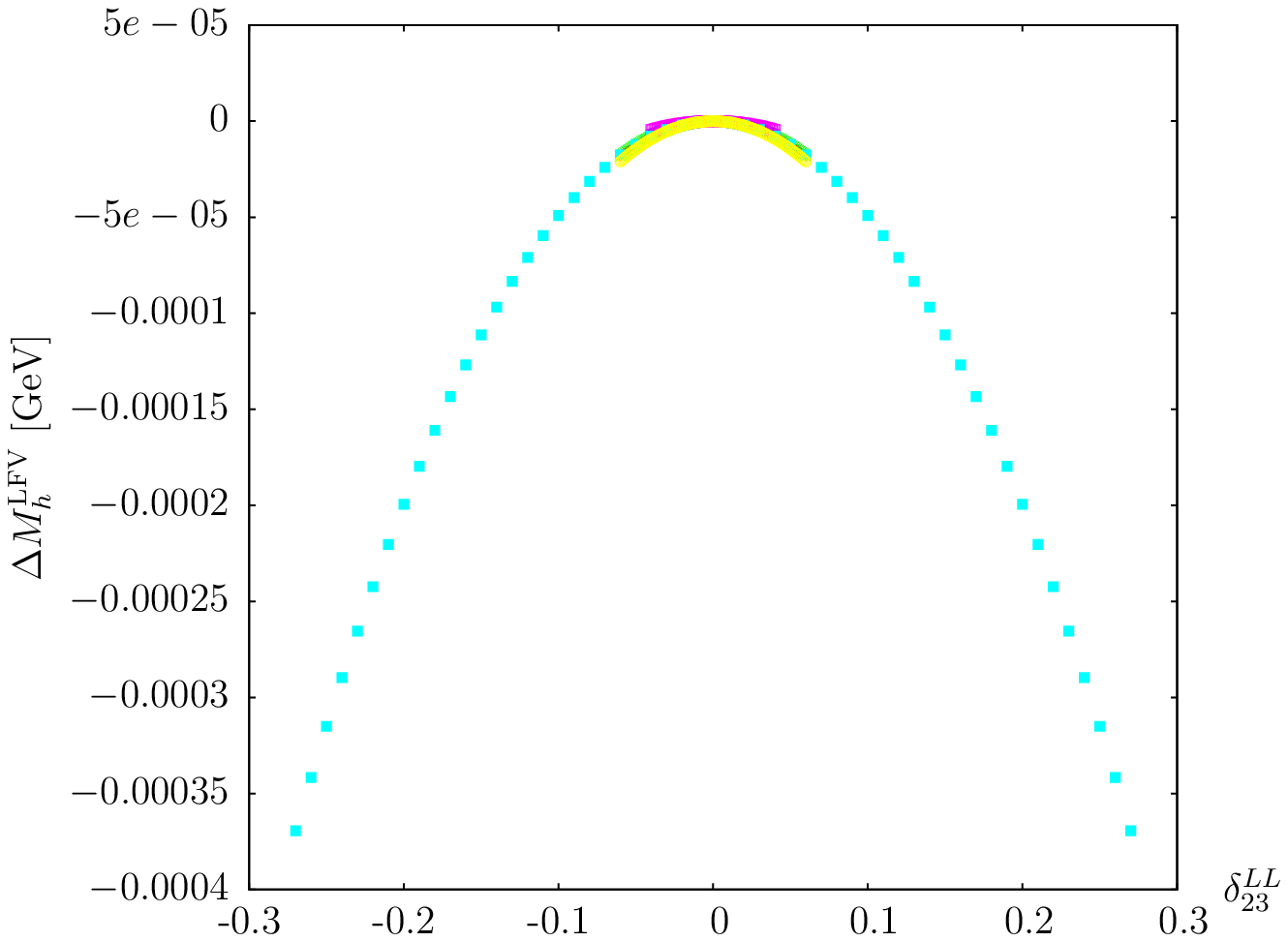   ,scale=0.56,angle=0,clip=}\\
\vspace{0.5cm}
\psfig{file=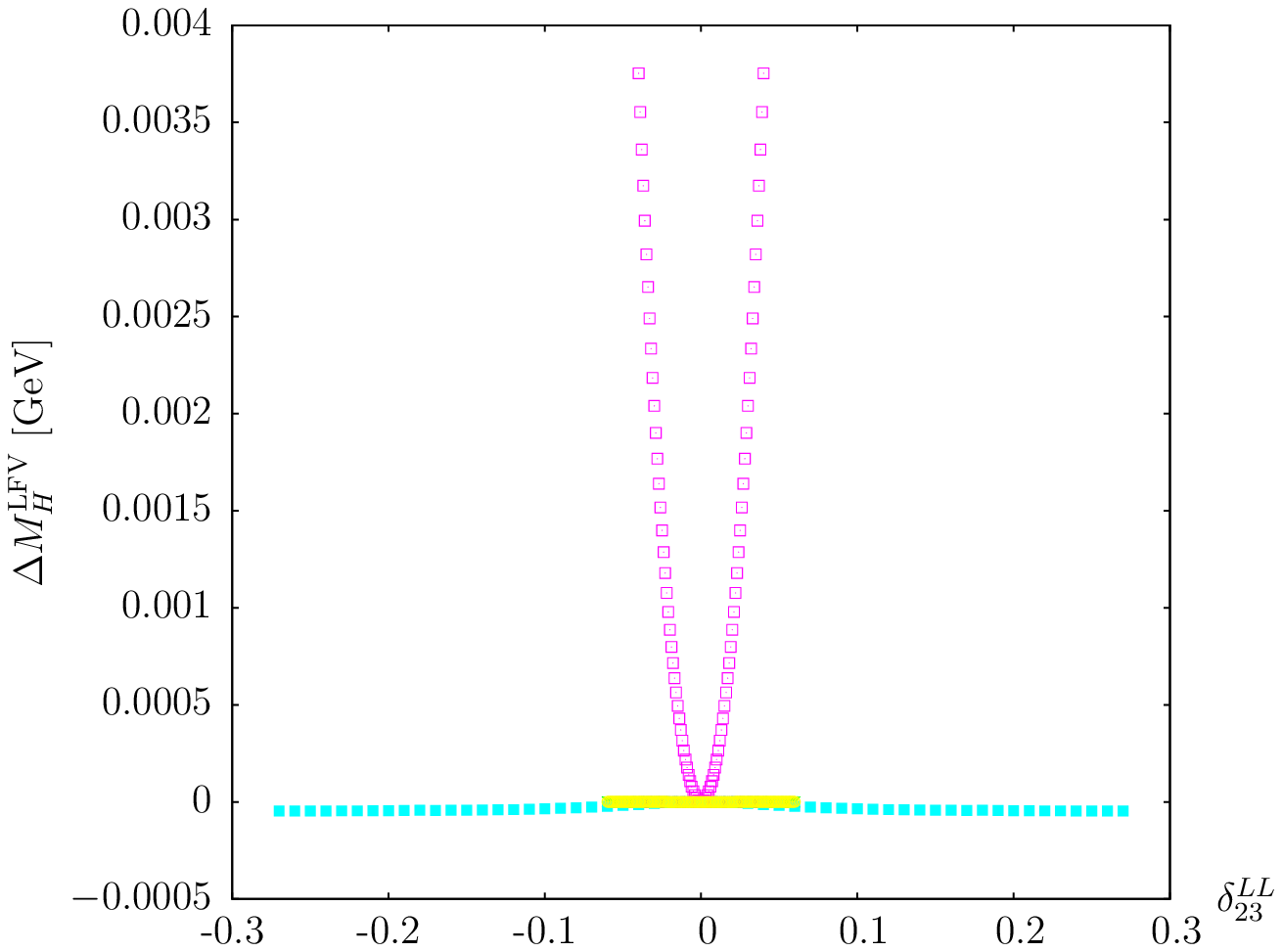  ,scale=0.57,angle=0,clip=}
\psfig{file=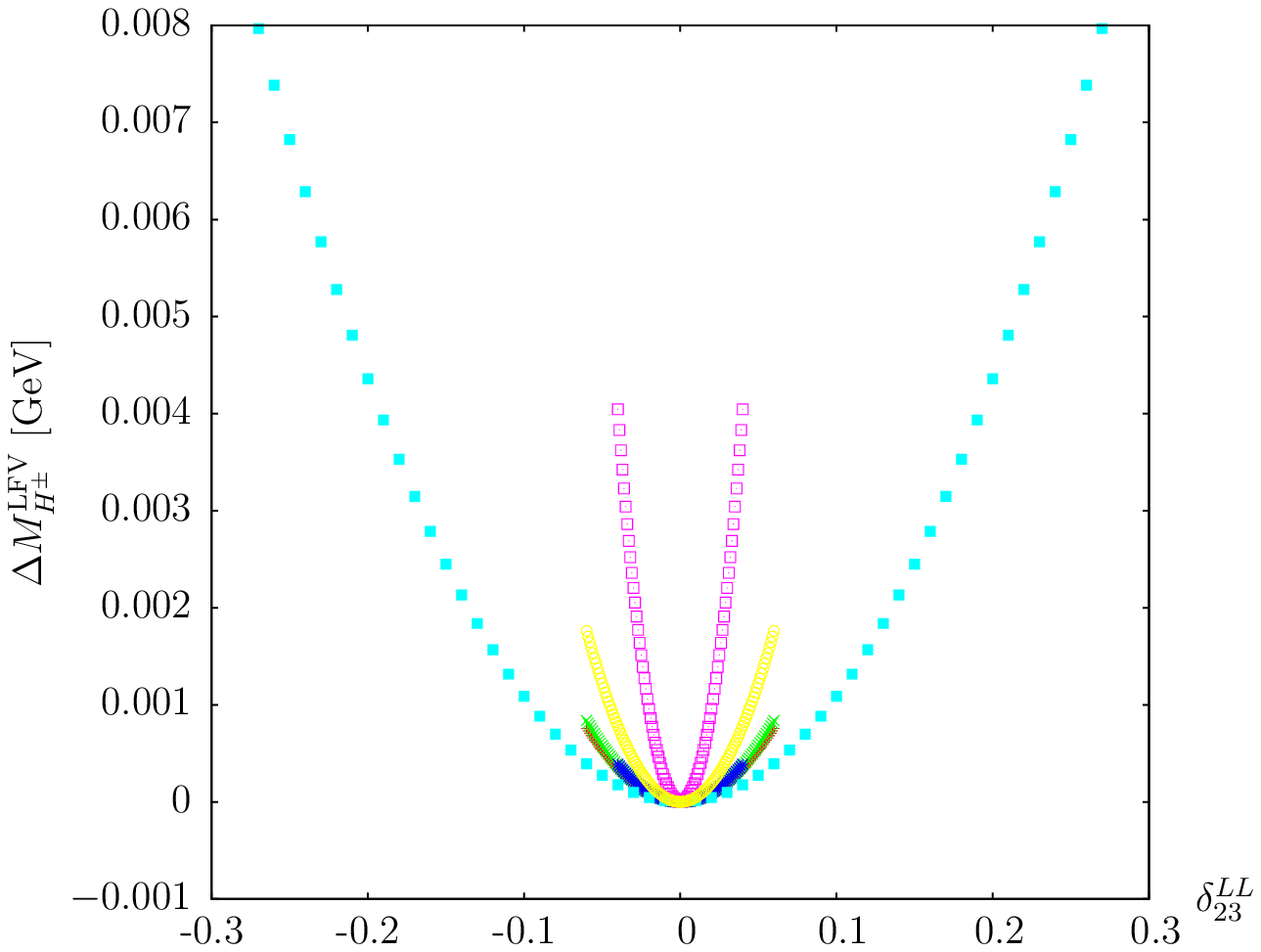  ,scale=0.57,angle=0,clip=}\\

\end{center}
\caption{EWPO and Higgs masses as a function of slepton
  mixing $\delta^{LL}_{23}$ for the six points defined in the \refta{tab:spectra}.
Solid red (blue) line shows the present (future) experimental uncertainty.}  
\label{figdLL23}
\end{figure}

We start with the investigation of the LFV effects on the electroweak 
precision observables. Experimental bounds on the $\de^{AB}_{12}$ are 
very strict (see \refta{tab:boundsSpoints}) and hence it does not 
contribute sizably.  The bounds on the other $\de^{AB}_{ij}$'s are less 
strict but in most cases we still do not get appreciable contributions 
to the EWPO (but now can quantify their corresponding sizes).  The only 
significant contribution comes from $\de^{LL}_{23}$.  The upper left 
plot in \reffi{figdLL23} shows our results for $\De\rho$ as functions of 
$\de^{LL}_{23}$ under the presently allowed experimental range given in 
\refta{tab:boundsSpoints}.  Depending on the choice of scenario (S1\dots 
S6), values of \order{10^{-3}} can be reached.  The largest values are 
found in S5, where the values of $\del{LL}{23}$ of up to $\pm 0.3$ are 
permitted.  For the same value of $\del{LL}{23}$ we find the largest 
contributions in S6, which possesses the relatively largest values of 
soft-SUSY-breaking parameters in the slepton sector.  This indicates 
that in general large contributions to the EWPO are possible as soon as 
heavy sleptons are involved.  Conversely, while such heavy sleptons are 
in general difficult to detect directly at the LHC or the ILC, their 
presence could be visible in case of large LFV contributions via a shift 
in the EWPO.

Turning to the (pseudo-)observables $\MW$ and $\sweff$, respectively 
shown in the upper right and middle left plot of \reffi{figdLL23}, we 
can compare the size of the LFV contributions to the current and future 
anticipated accuracies in these observables.  The black line in both 
plots indicates the result for $\del{LL}{23} = 0$.  The red line shows 
the current level of accuracy, \refeq{eq:EWPO-today}, while the blue 
line indicates the future ILC/GigaZ precision, \refeq{eq:EWPO-future}.  
We refrain from putting the absolute values of these observables since 
their values strongly depend on the choice of the stop/sbottom sector 
(see \citere{PomssmRep} and references therein), which is independent on 
the slepton sector under investigation here.  While the current level of 
accuracy only has the potential to restrict $\del{LL}{23}$ in S5 and S6, 
the future accuracy (particularly in $\sweff$) can set stringent bounds 
in all six scenarios.

The overall conclusion for the EWPO is that, while $\del{LL}{23}$ is the 
most difficult one to restrict using `conventional' LFV observables (see 
\refse{sec:limits}), it has (by far) the strongest impact on EWPO.  
Depending on the stop/sbottom sector, new bounds beyond the 
`conventional' LFV observables can be obtained even with the current 
precision, and still better with the (anticipated) future accuracies.


\subsection{Higgs masses}
\label{sec:Mh}

We now turn to the effects of the LFV contribtions on the prediction of 
the neutral $\cp$-even and the charged MSSM Higgs-boson masses.  As 
discussed in \refse{sec:higgs-ho}, the theoretical accuracy should reach 
a precision of $\sim 50 \mev$ in the case of $\Mh$ and about $\sim 1\%$ 
in the case of the heavy Higgs bosons.  The calculation of $\Mh$ in the 
presence of non-minimal flavor violation (NMFV) in the squark sector 
\cite{arana} indicated that corrections as large as \order{10 \gev} are 
possible (for the NMFV $\deABij$ in agreement with all other precision 
data).  Similar or even larger corrections were found for the heavy 
Higgs bosons, in particular for the charged Higgs boson.  Large 
corrections were associated especially with non-zero values of 
$\del{LR,RL}{23}$.

Even though the corrections from the slepton sector are naturally much 
smaller than from the squark sector, the LFV contributions could be 
expected to exceed future and possibly even current experimental 
uncertainties.  Indeed, the estimated theoretical uncertainties for 
the LFV contributions of at least \order{100\mev} for $\Mh$ and 
\order{10\gev} for $\MHp$ were at the level of or exceeding the future 
anticipated accuracies.  Thus, the LFV had to be evaluated and analyzed 
in order to reach the required level of precision.

The Higgs-boson masses are shown in the middle right plot ($\Mh$), the 
lower left ($\MH$) and the lower right plot ($\MHp$) of each figure.  As 
expected from the NMFV analysis in the squark sector~\cite{arana}, the 
largest effects are found for $\del{LR,RL}{23}$, but similarly for 
$\del{LR,RL}{13}$, indicating that only the electroweak, not the Yukawa 
couplings, play a relevant role in these corrections.  Contrary to 
expectations, the corrections to $\Mh$ \emph{always} stay below the 
level of a few~MeV.  Though this result obviates the above-mentioned 
uncertainty of \order{100 \mev}, these contributions are too small to 
yield a sizable numerical effect.

\begin{figure}[ht!]
\begin{center}
\psfig{file=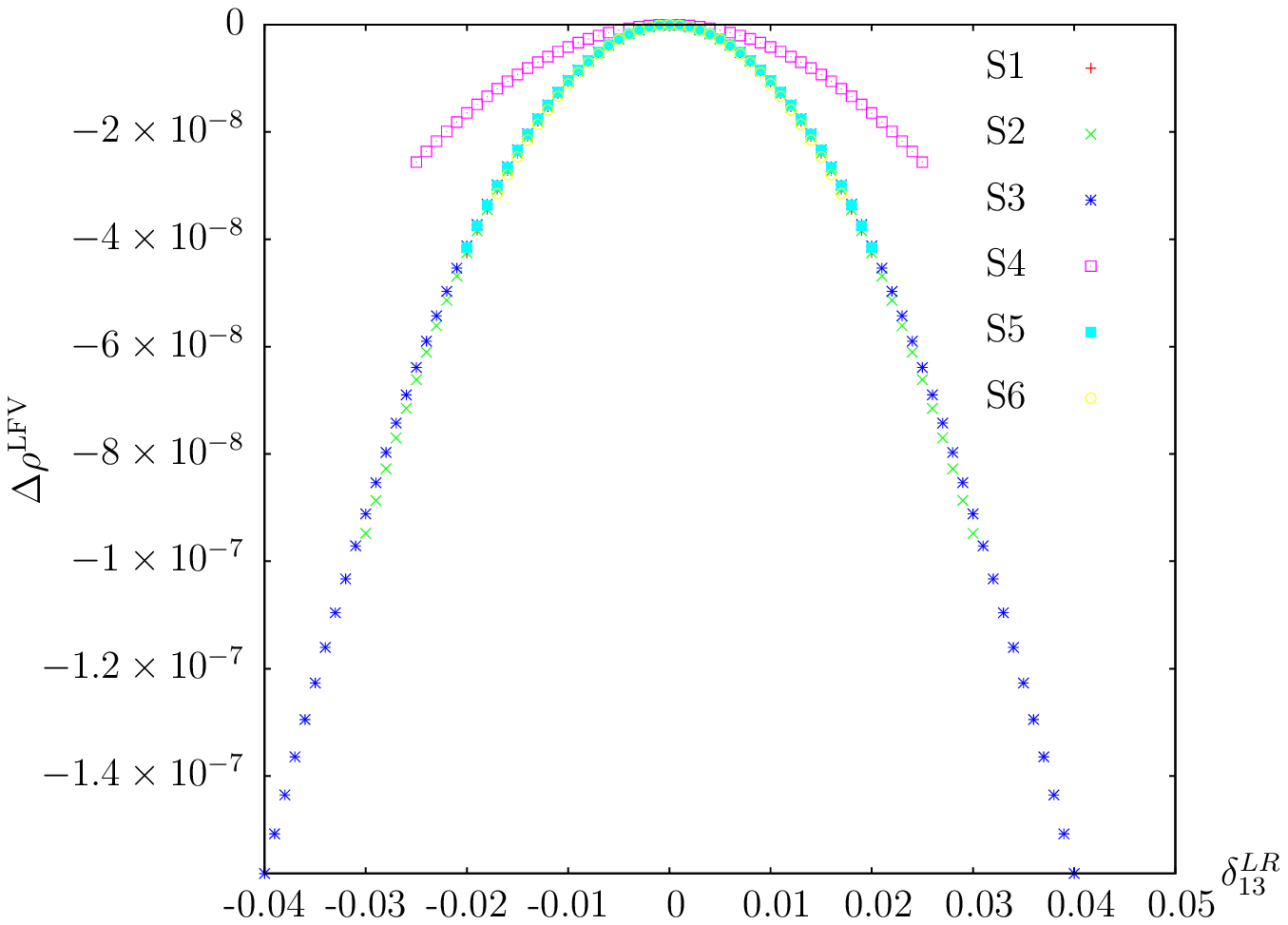  ,scale=0.57,angle=0,clip=}
\psfig{file=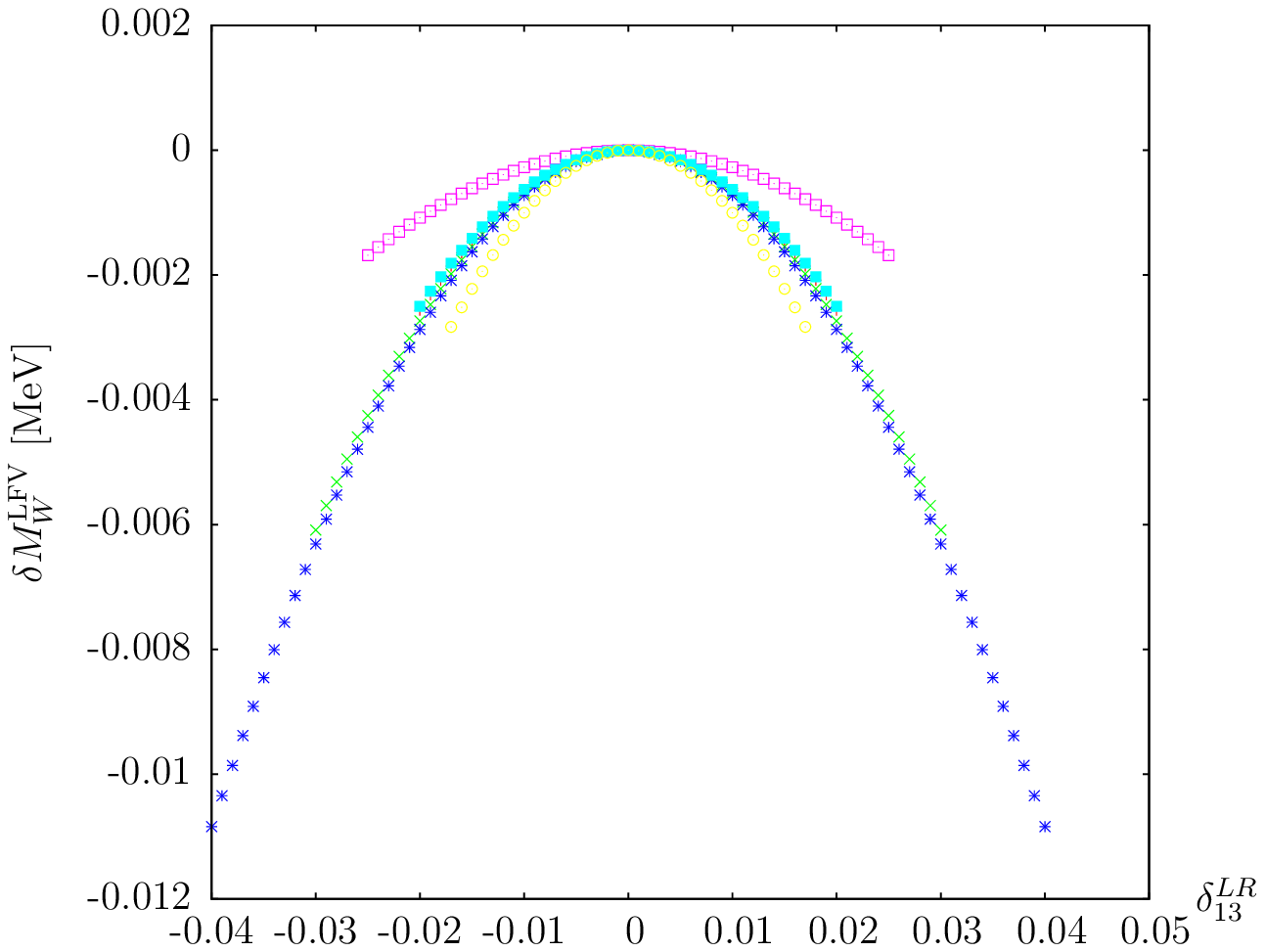  ,scale=0.57,angle=0,clip=}\\
\vspace{0.5cm}
\psfig{file=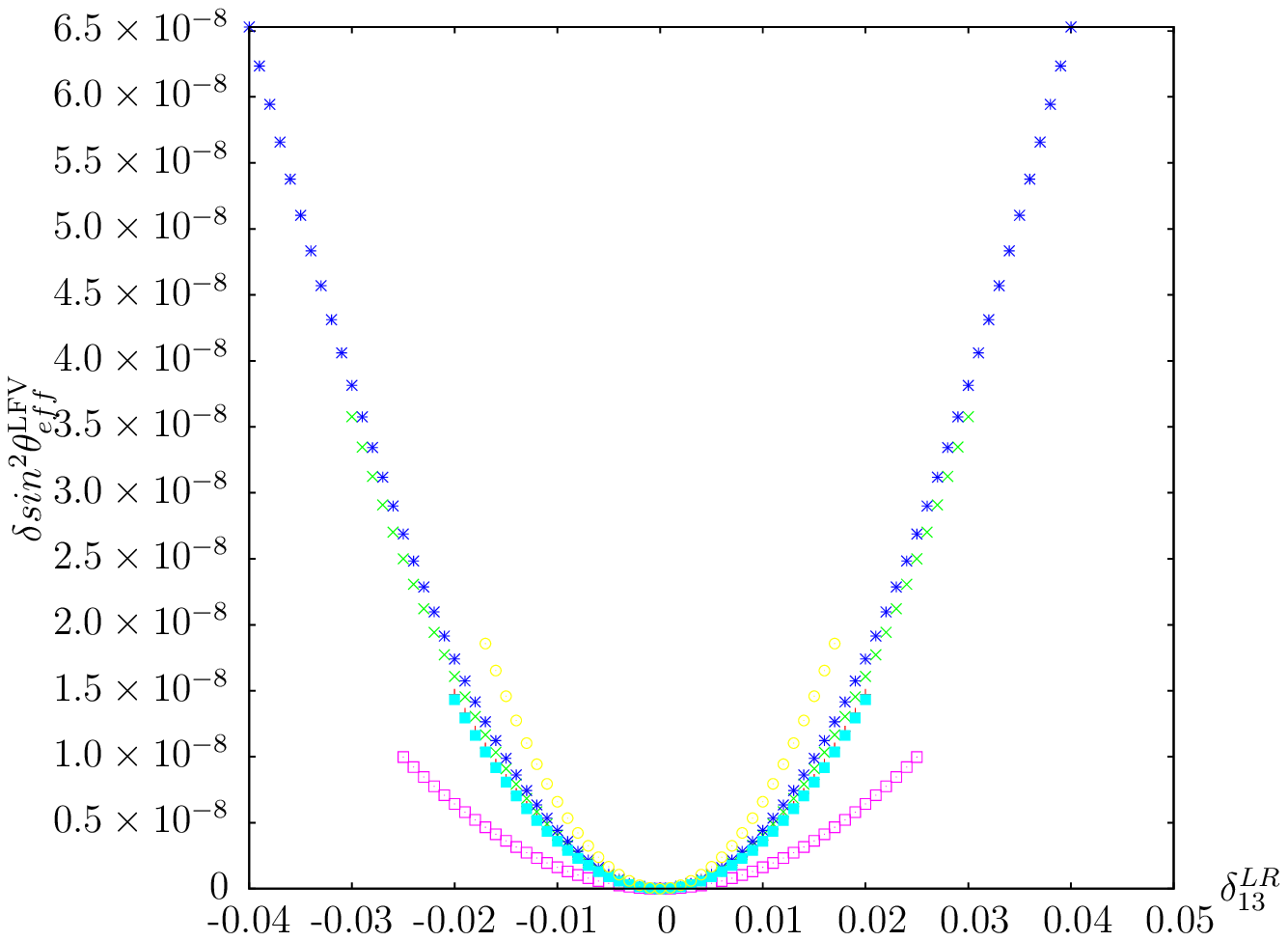 ,scale=0.57,angle=0,clip=}
\psfig{file=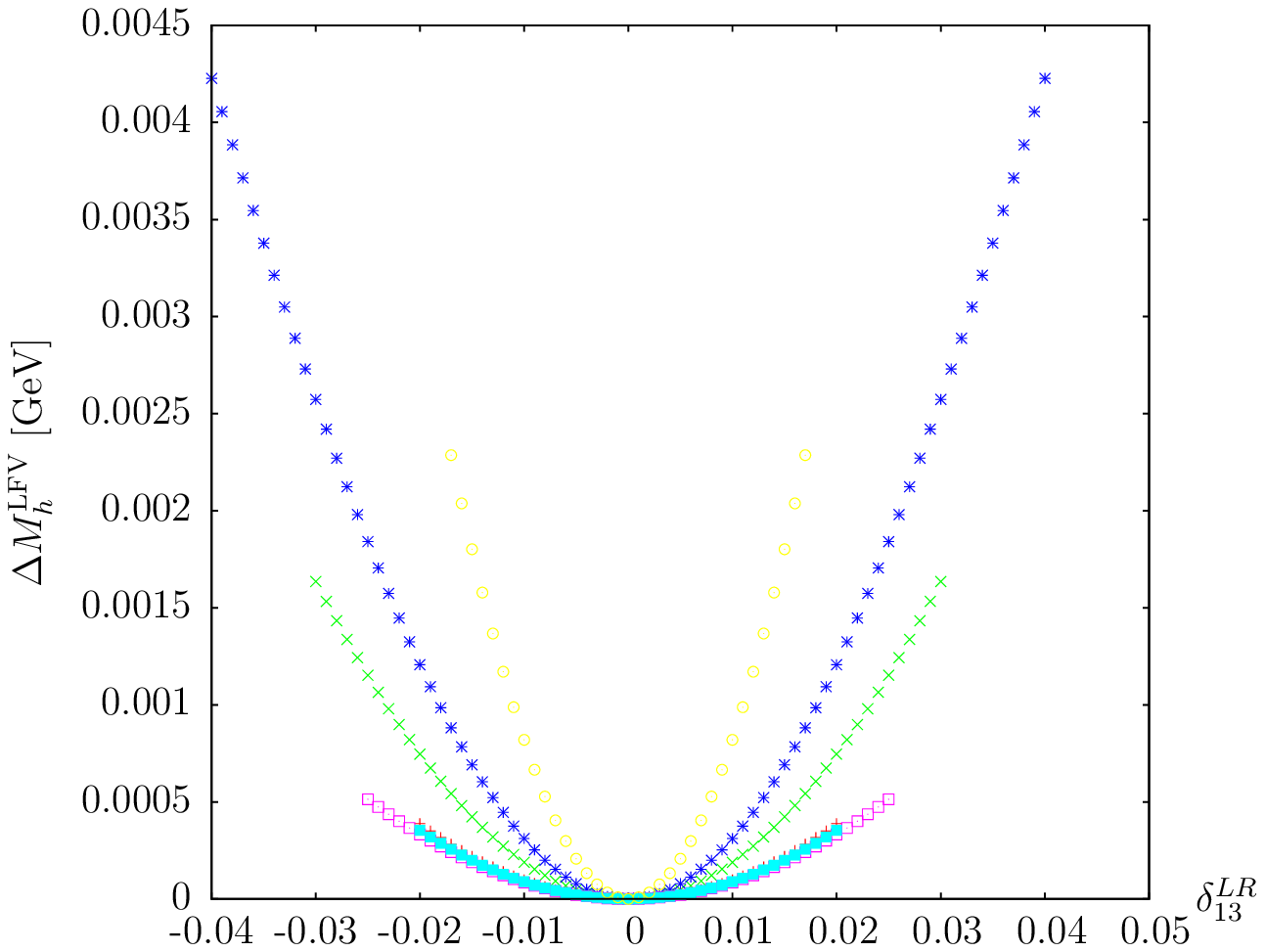   ,scale=0.57,angle=0,clip=}\\
\vspace{0.5cm}
\psfig{file=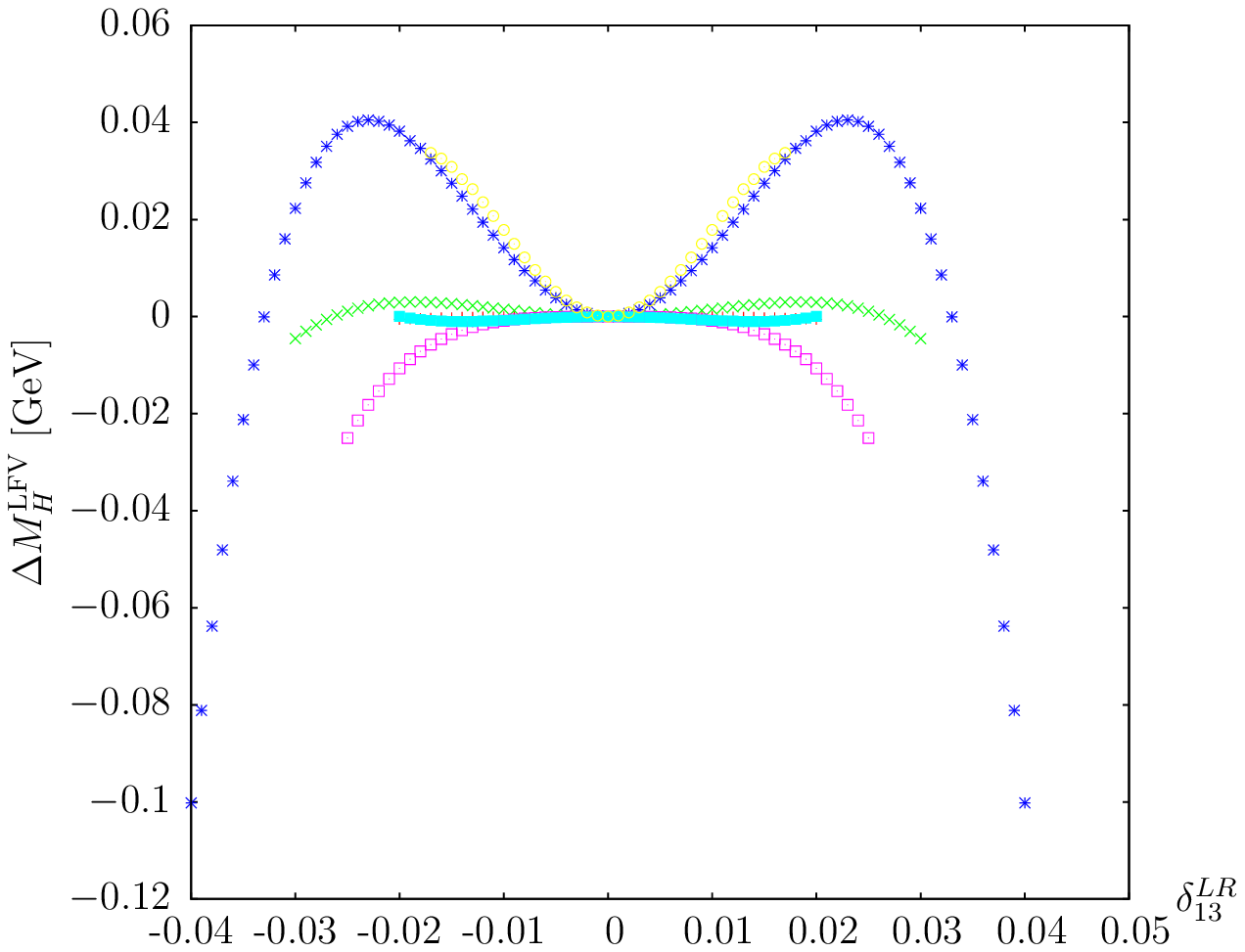  ,scale=0.57,angle=0,clip=}
\psfig{file=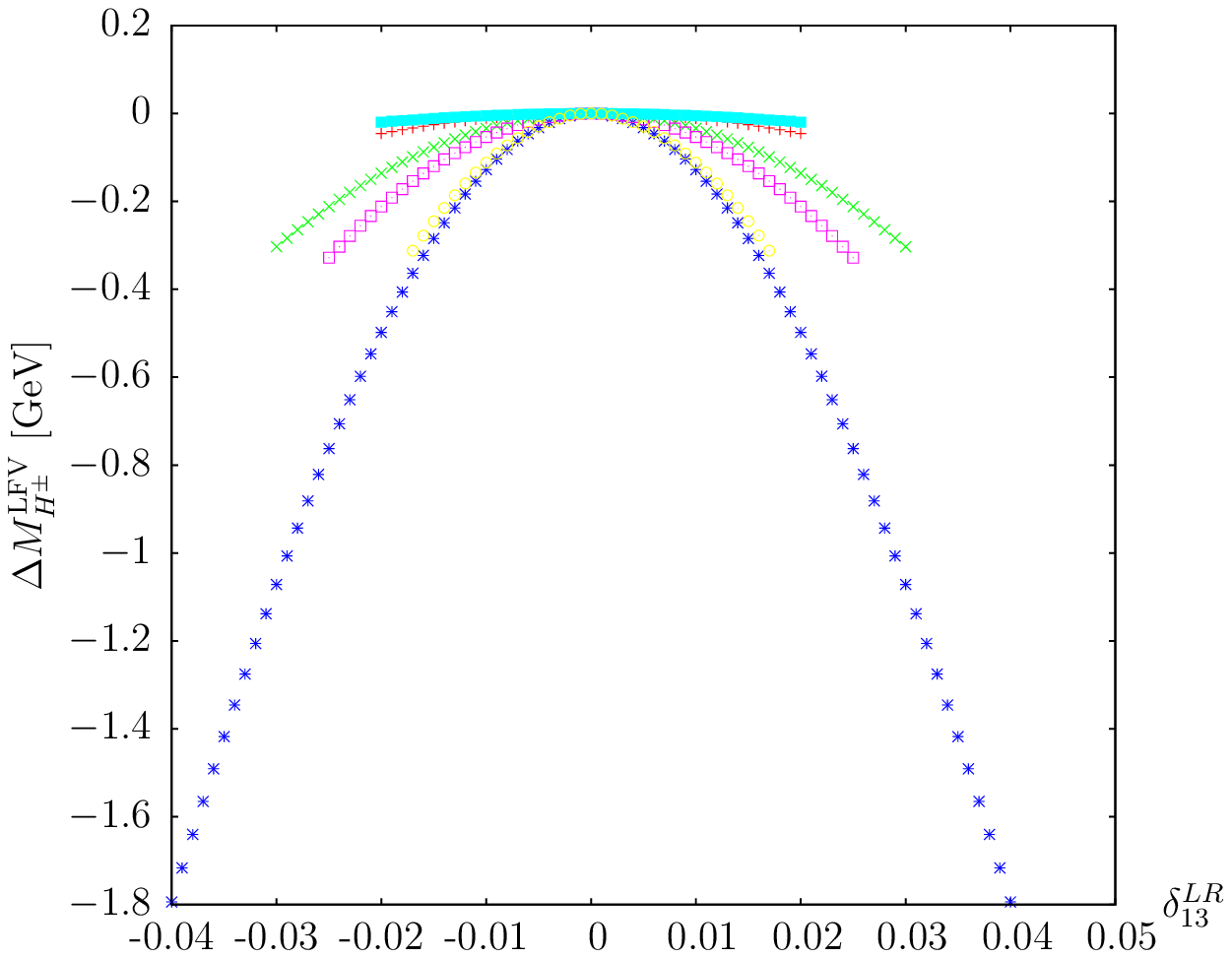  ,scale=0.57,angle=0,clip=}\\

\end{center}
\caption{EWPO and Higgs masses as a function of slepton
  mixing $\delta^{LR}_{13}$ for the six points defined in the \refta{tab:spectra}.}  
\label{figdLR13}
\end{figure}

\begin{figure}[ht!]
\begin{center}
\psfig{file=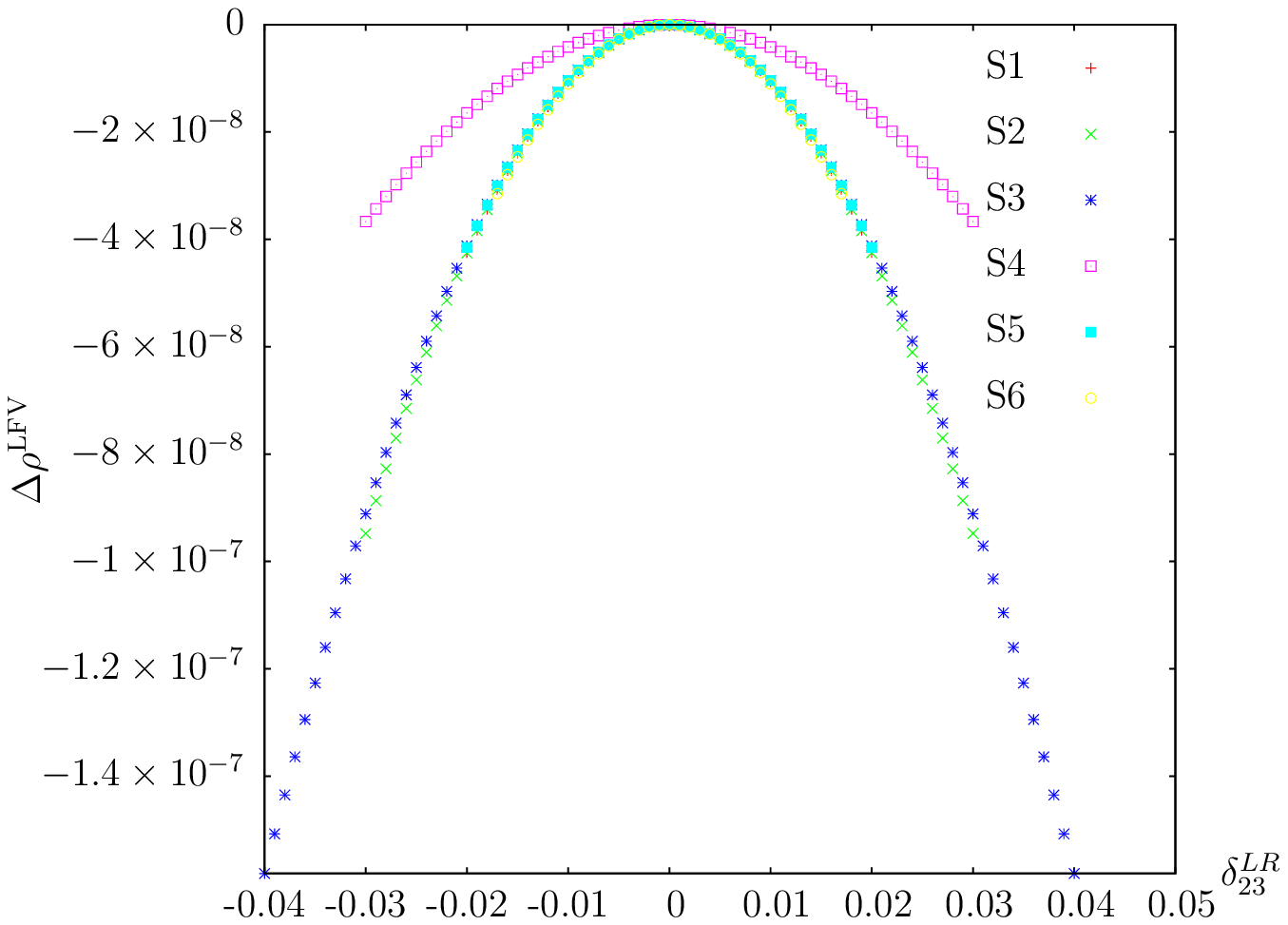  ,scale=0.57,angle=0,clip=}
\psfig{file=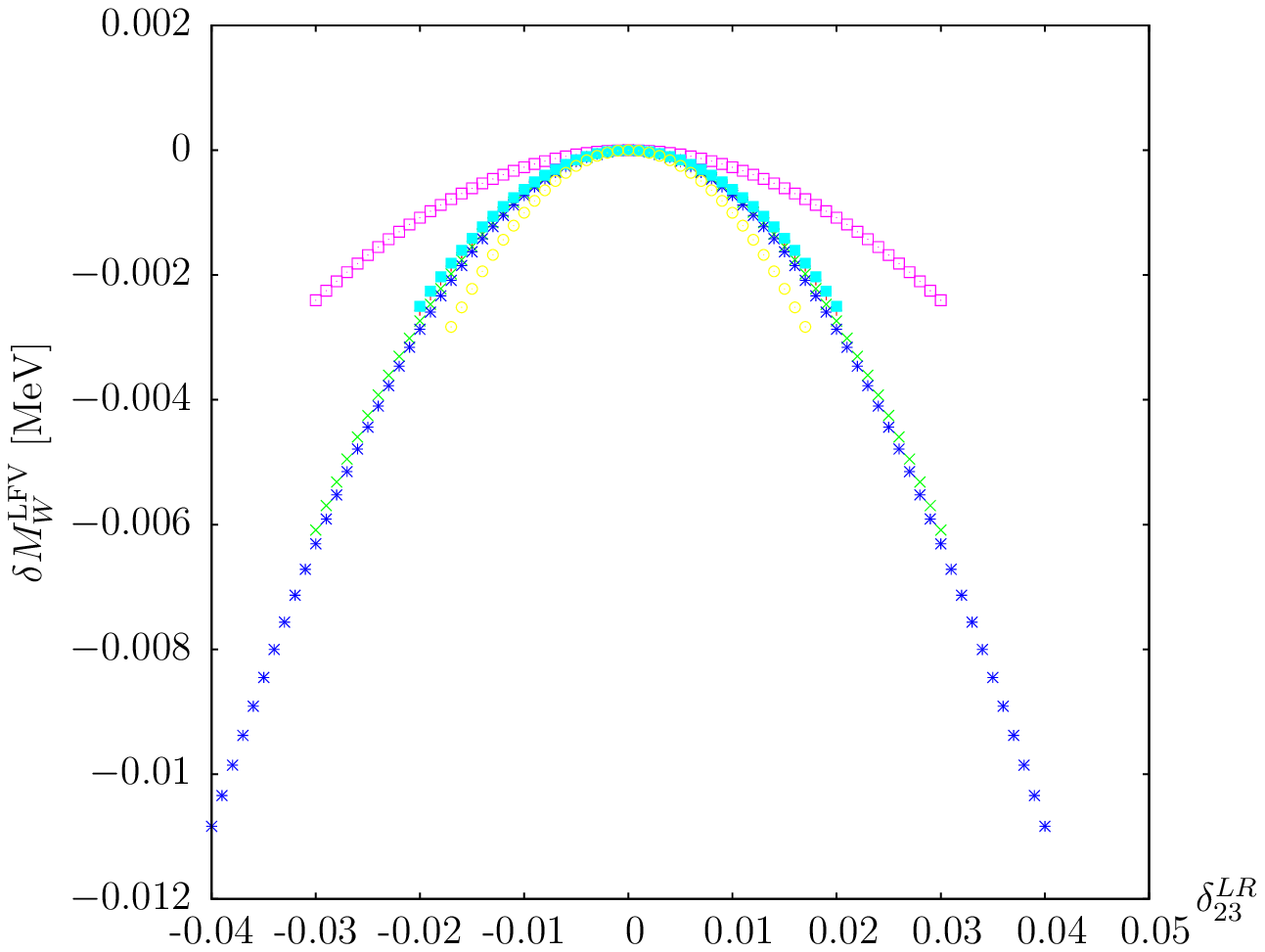  ,scale=0.57,angle=0,clip=}\\
\vspace{0.5cm}
\psfig{file=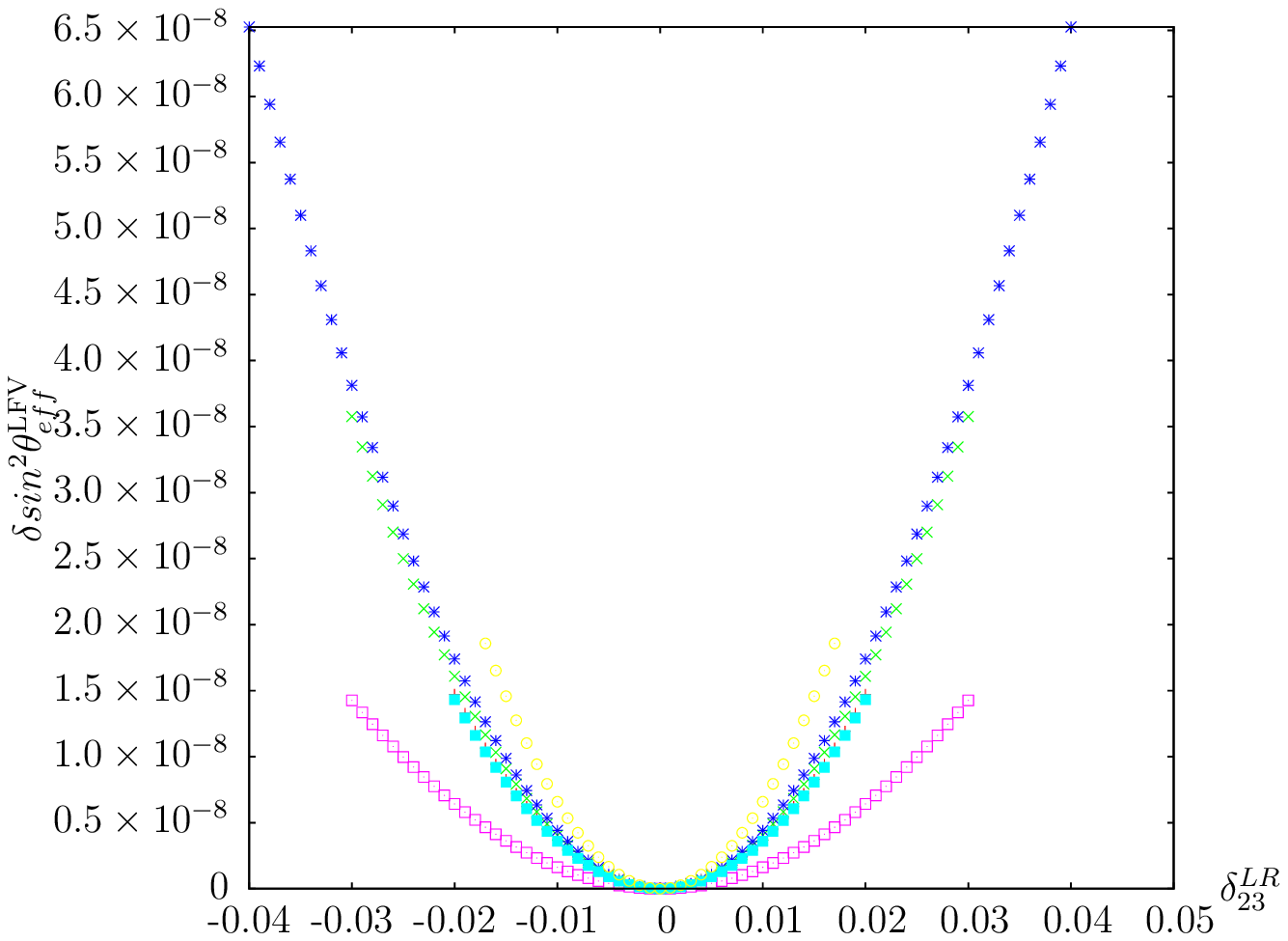 ,scale=0.57,angle=0,clip=}
\psfig{file=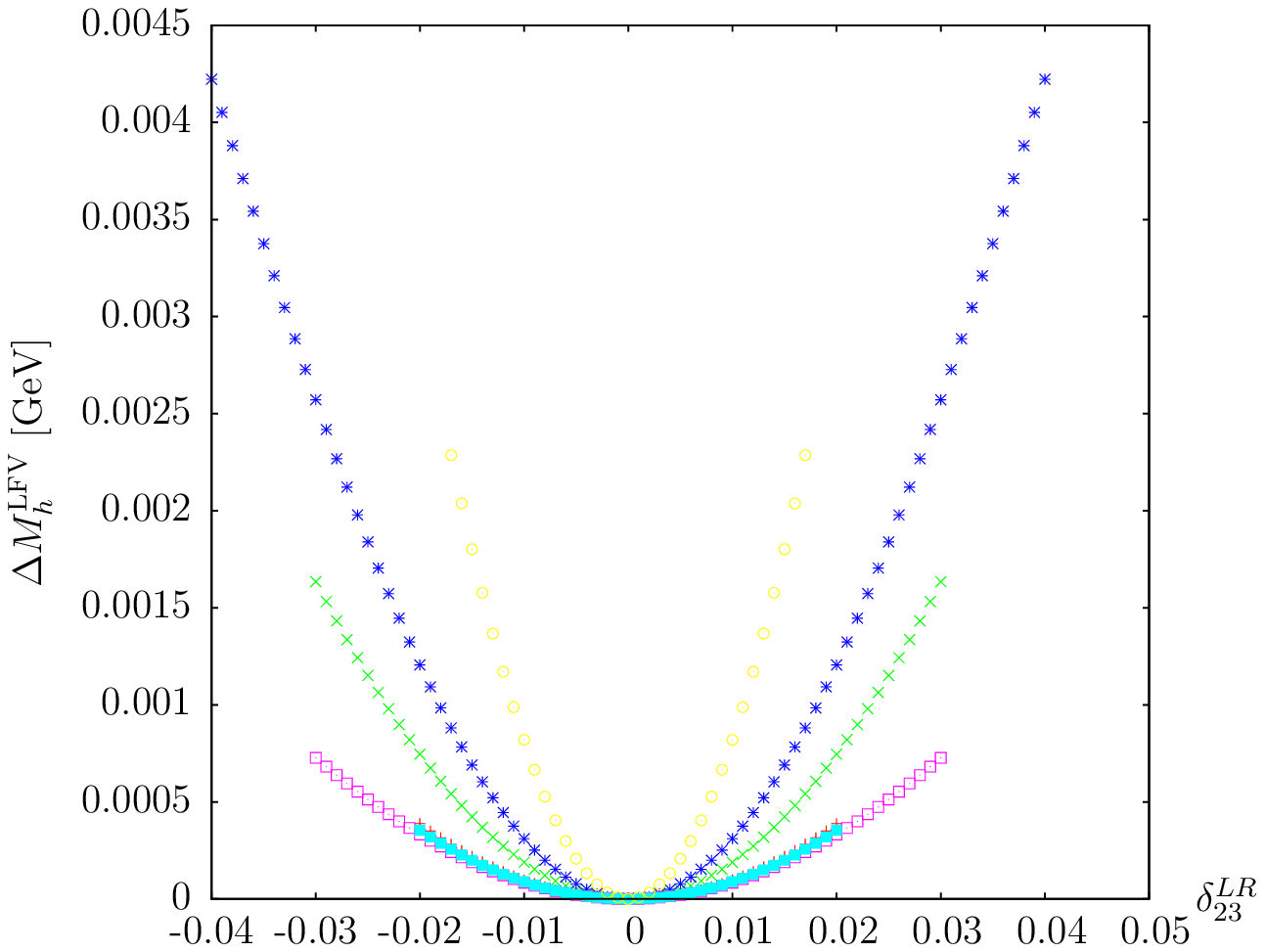   ,scale=0.57,angle=0,clip=}\\
\vspace{0.5cm}
\psfig{file=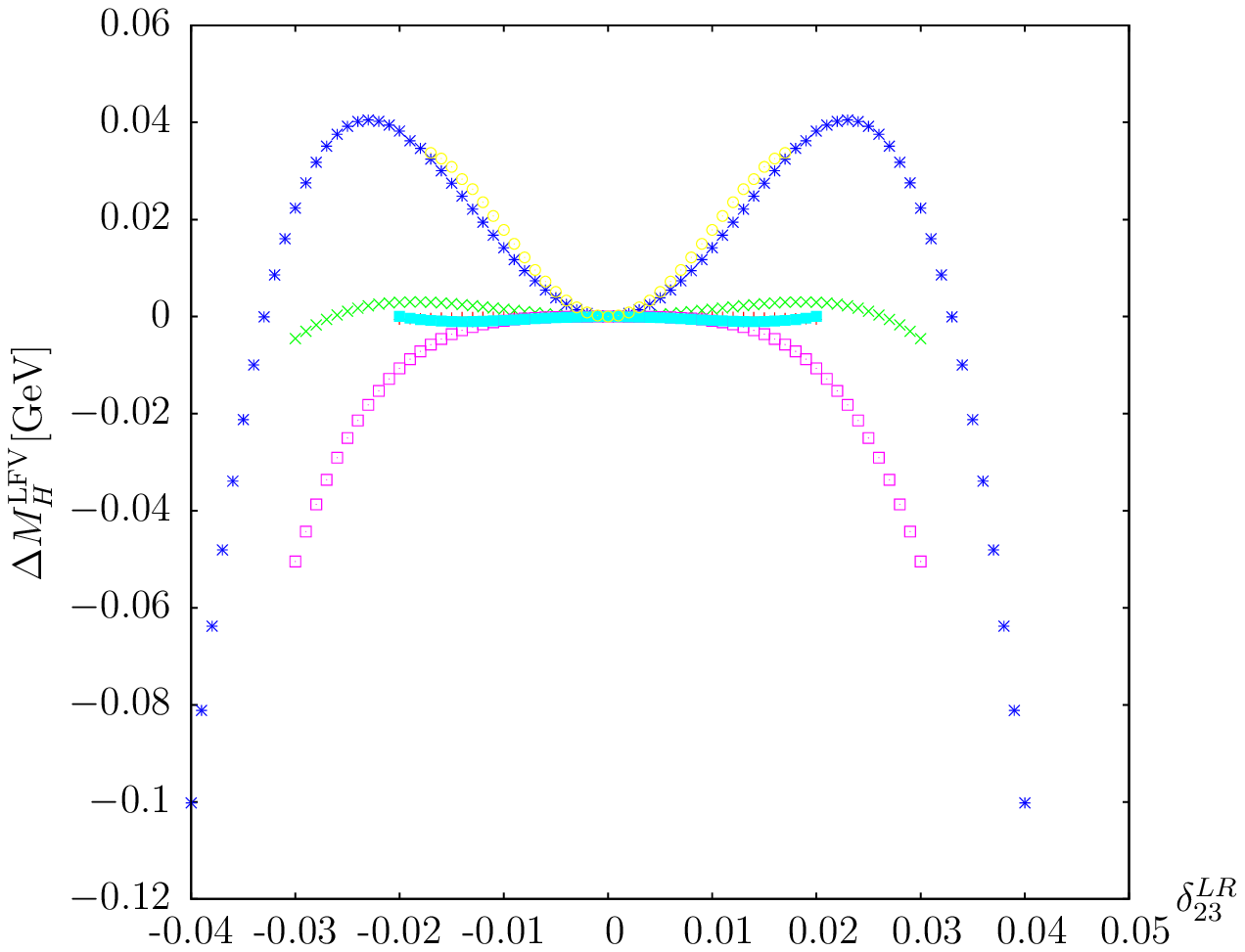  ,scale=0.57,angle=0,clip=}
\psfig{file=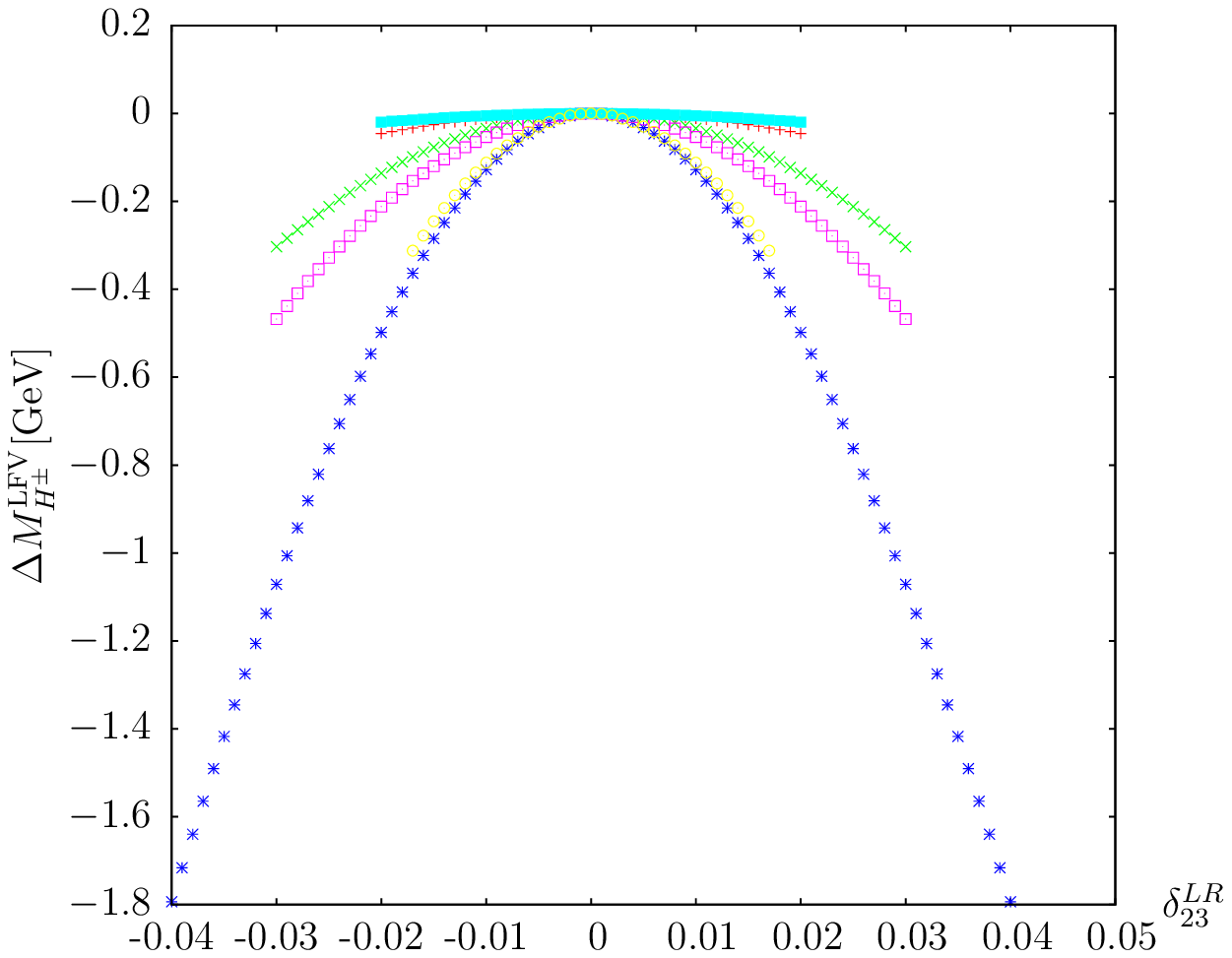  ,scale=0.57,angle=0,clip=}\\

\end{center}
\caption{EWPO and Higgs masses as a function of slepton
  mixing $\delta^{LR}_{23}$ for the six points defined in the \refta{tab:spectra}.}  
\label{figdLR23}
\end{figure}

\begin{figure}[ht!]
\begin{center}
\psfig{file=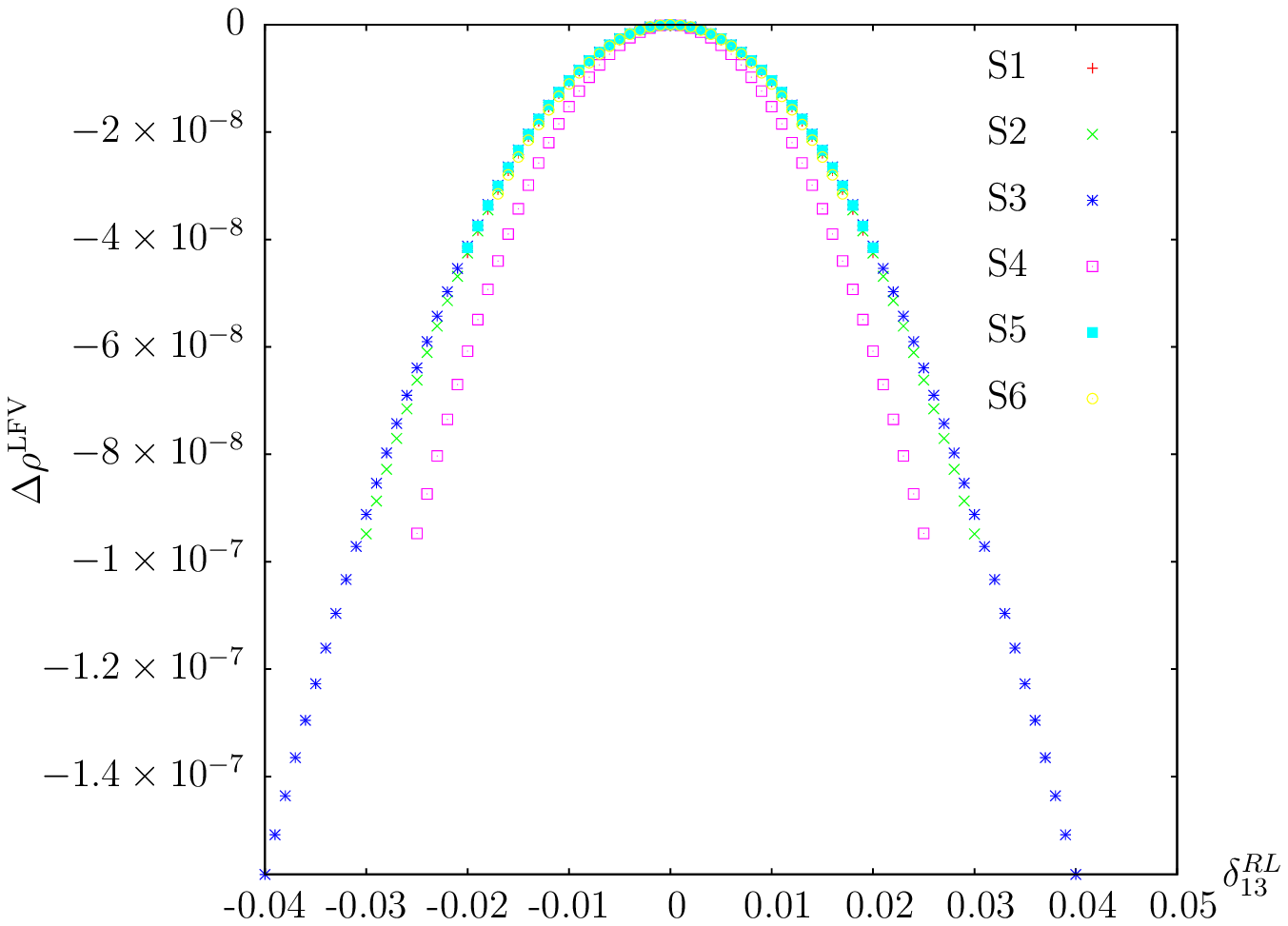  ,scale=0.57,angle=0,clip=}
\psfig{file=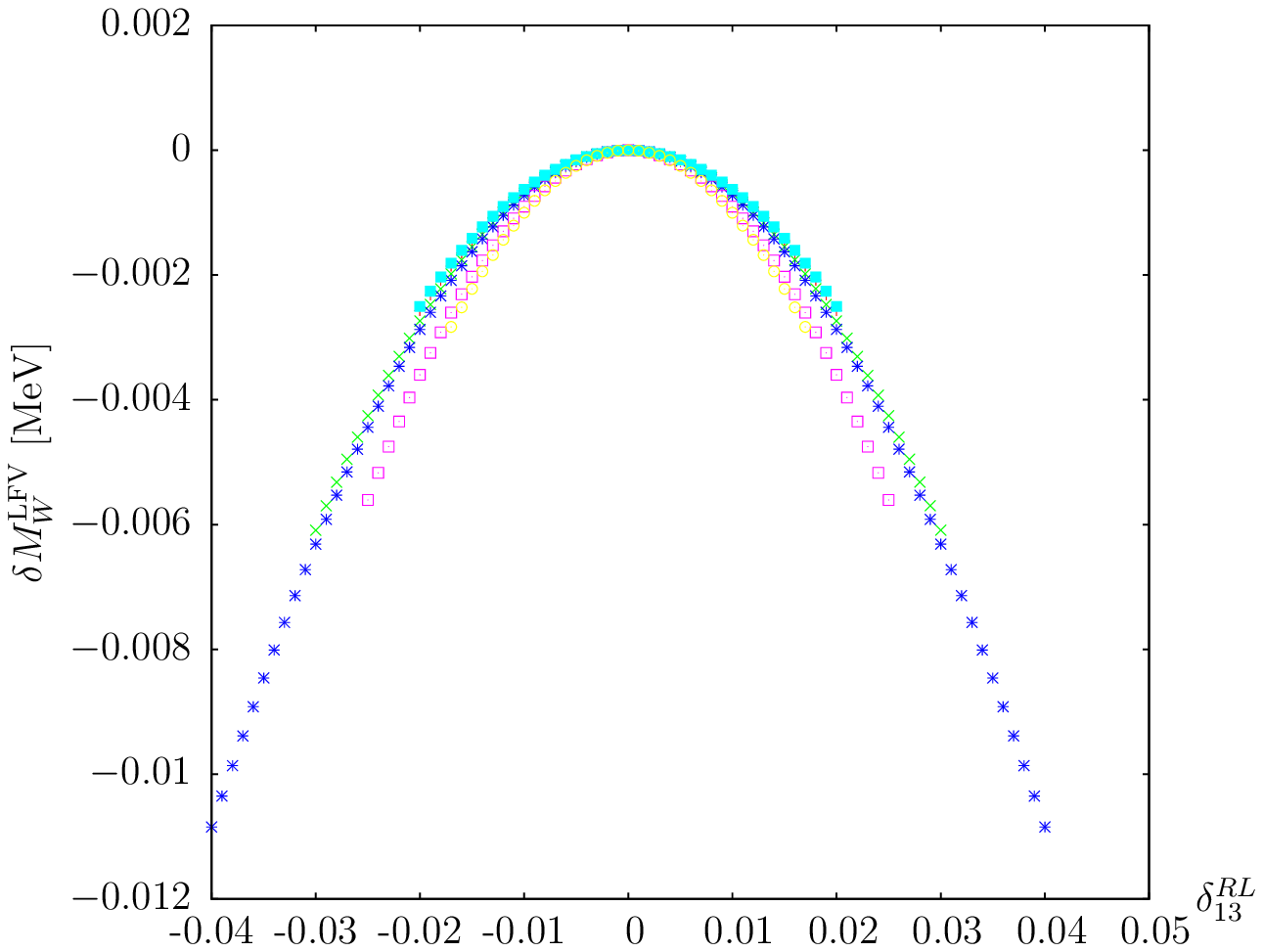  ,scale=0.57,angle=0,clip=}\\
\vspace{0.5cm}
\psfig{file=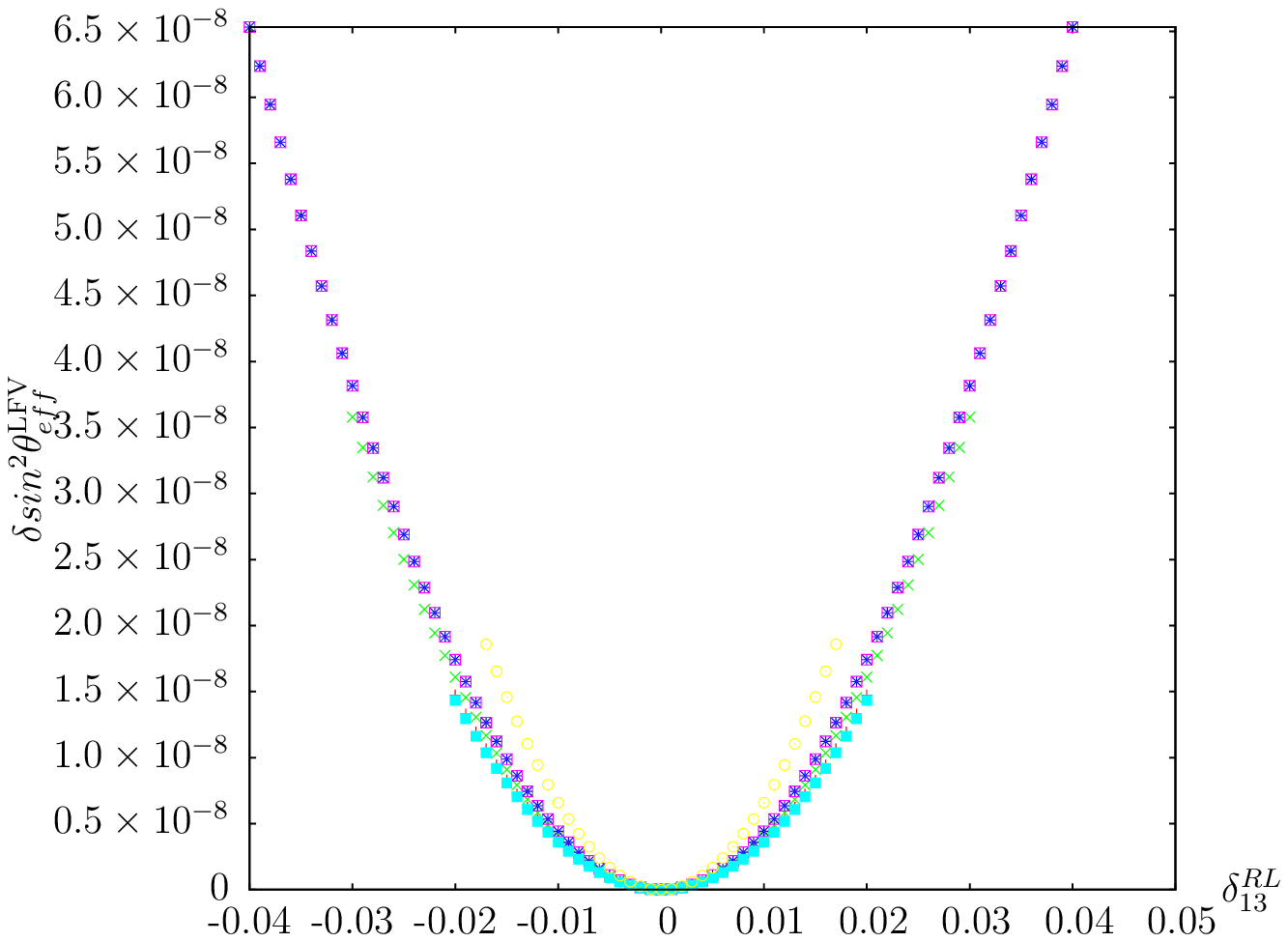 ,scale=0.57,angle=0,clip=}
\psfig{file=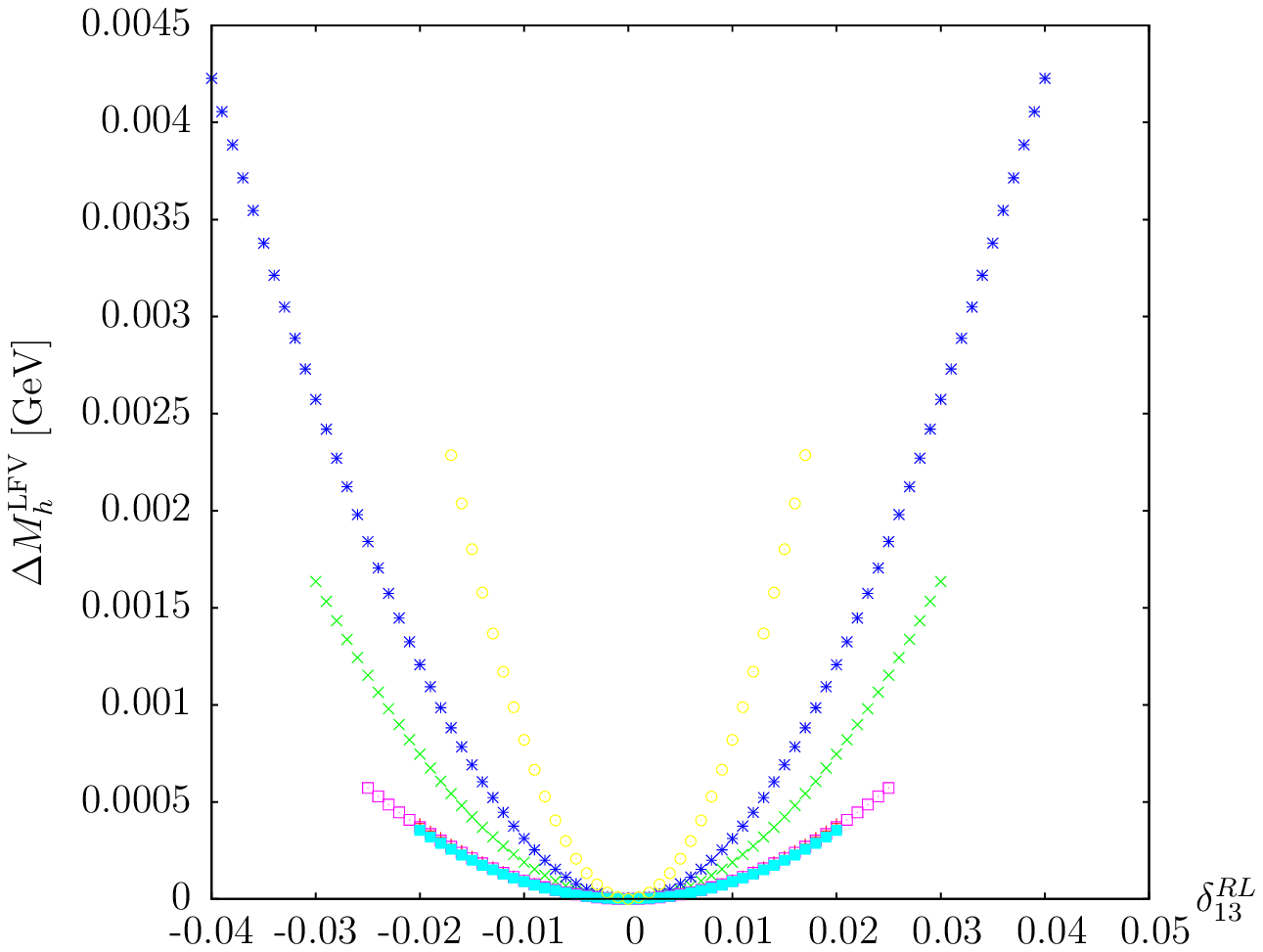   ,scale=0.57,angle=0,clip=}\\
\vspace{0.5cm}
\psfig{file=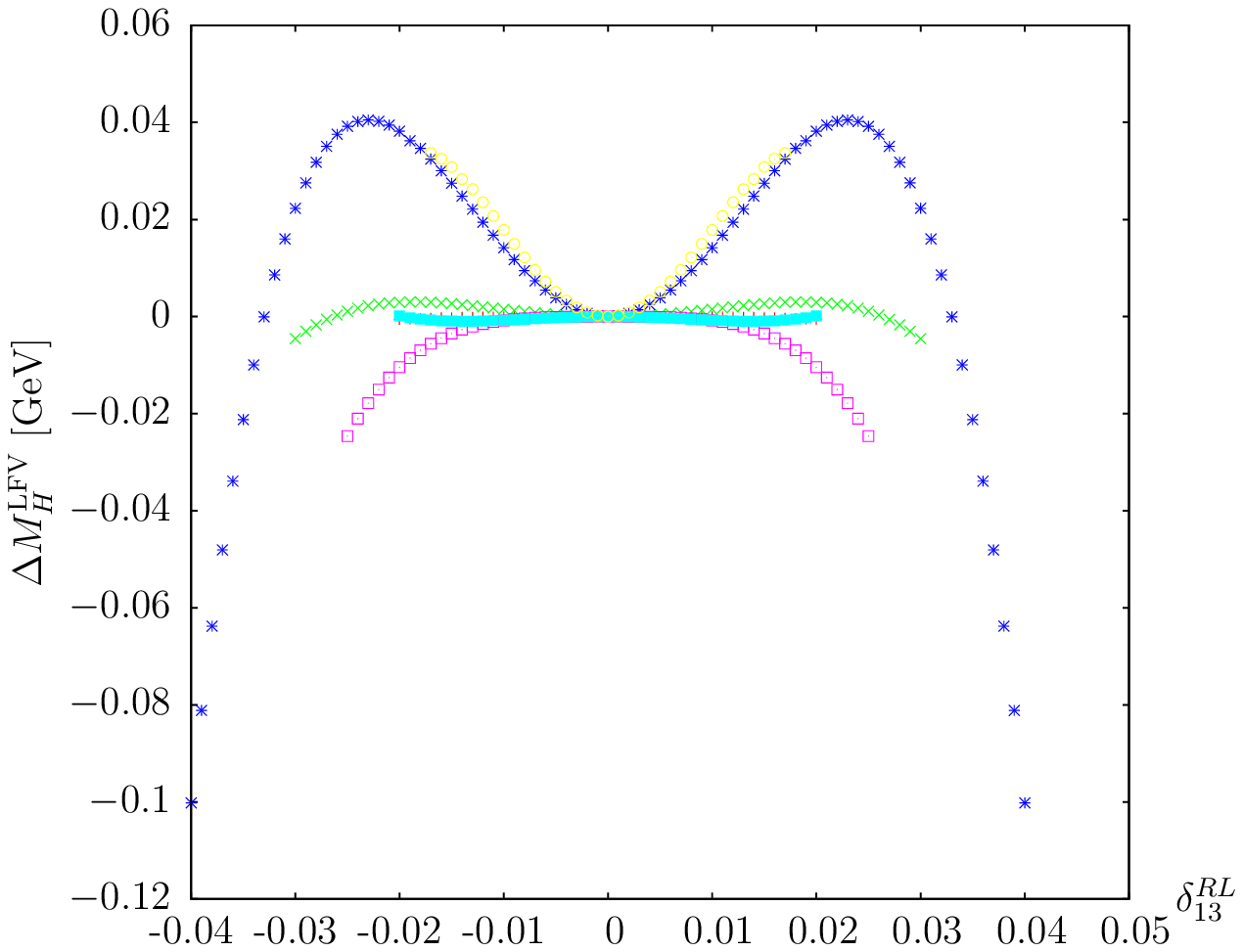  ,scale=0.57,angle=0,clip=}
\psfig{file=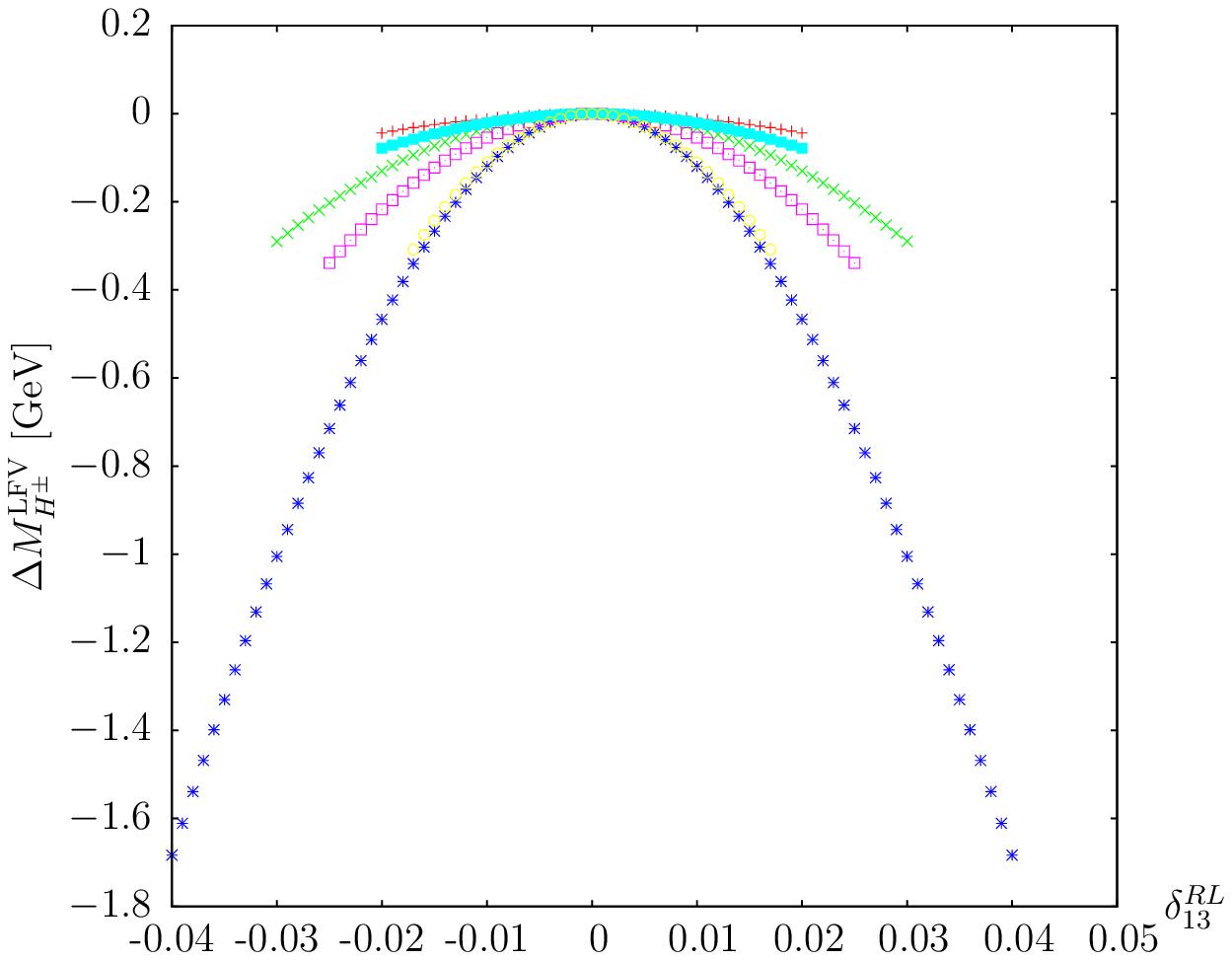  ,scale=0.57,angle=0,clip=}\\

\end{center}
\caption{EWPO and Higgs masses as a function of slepton
  mixing $\delta^{RL}_{13}$ for the six points defined in the \refta{tab:spectra}.}  
\label{figdRL13}
\end{figure}

\begin{figure}[ht!]
\begin{center}
\psfig{file=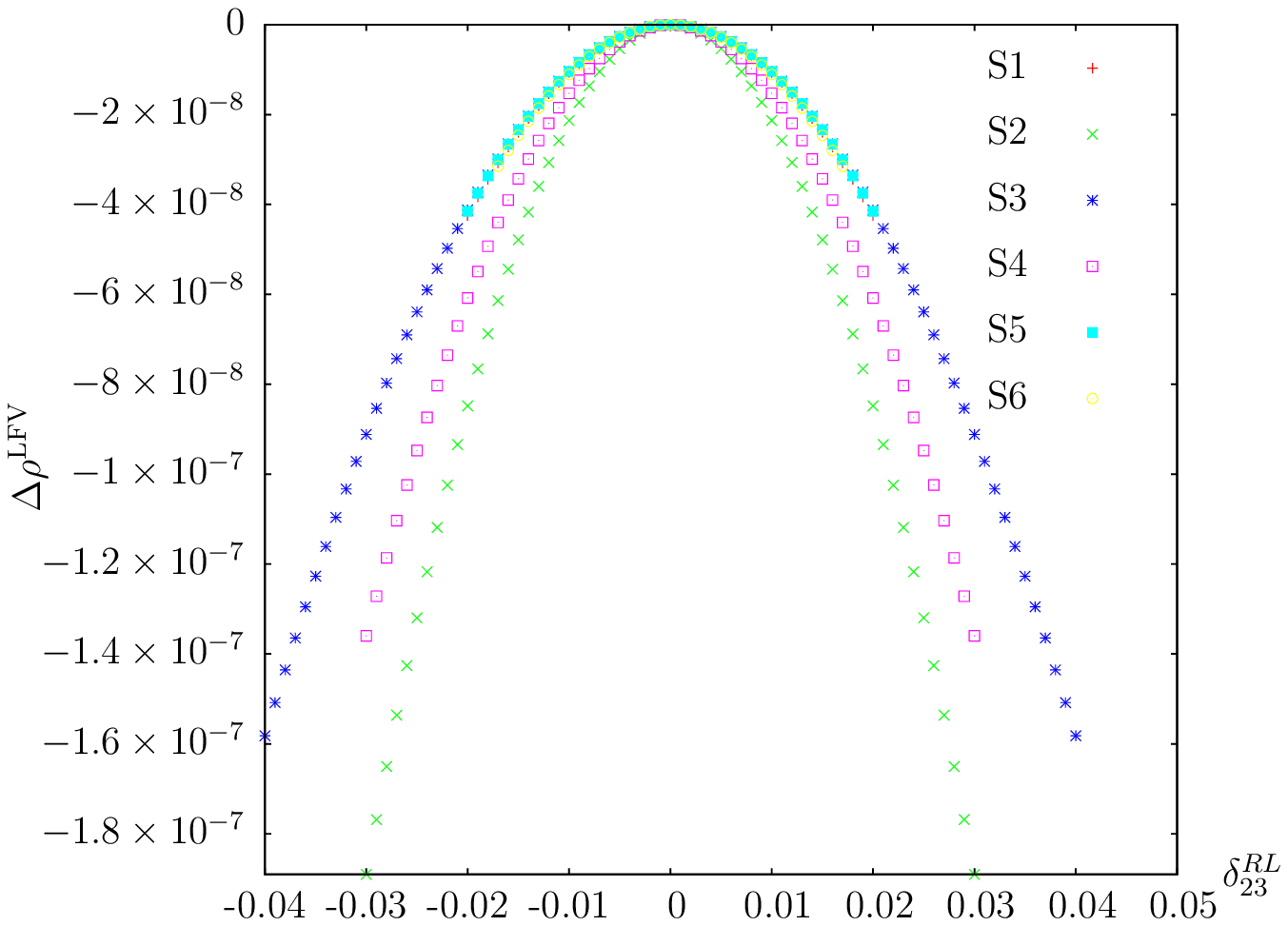  ,scale=0.57,angle=0,clip=}
\psfig{file=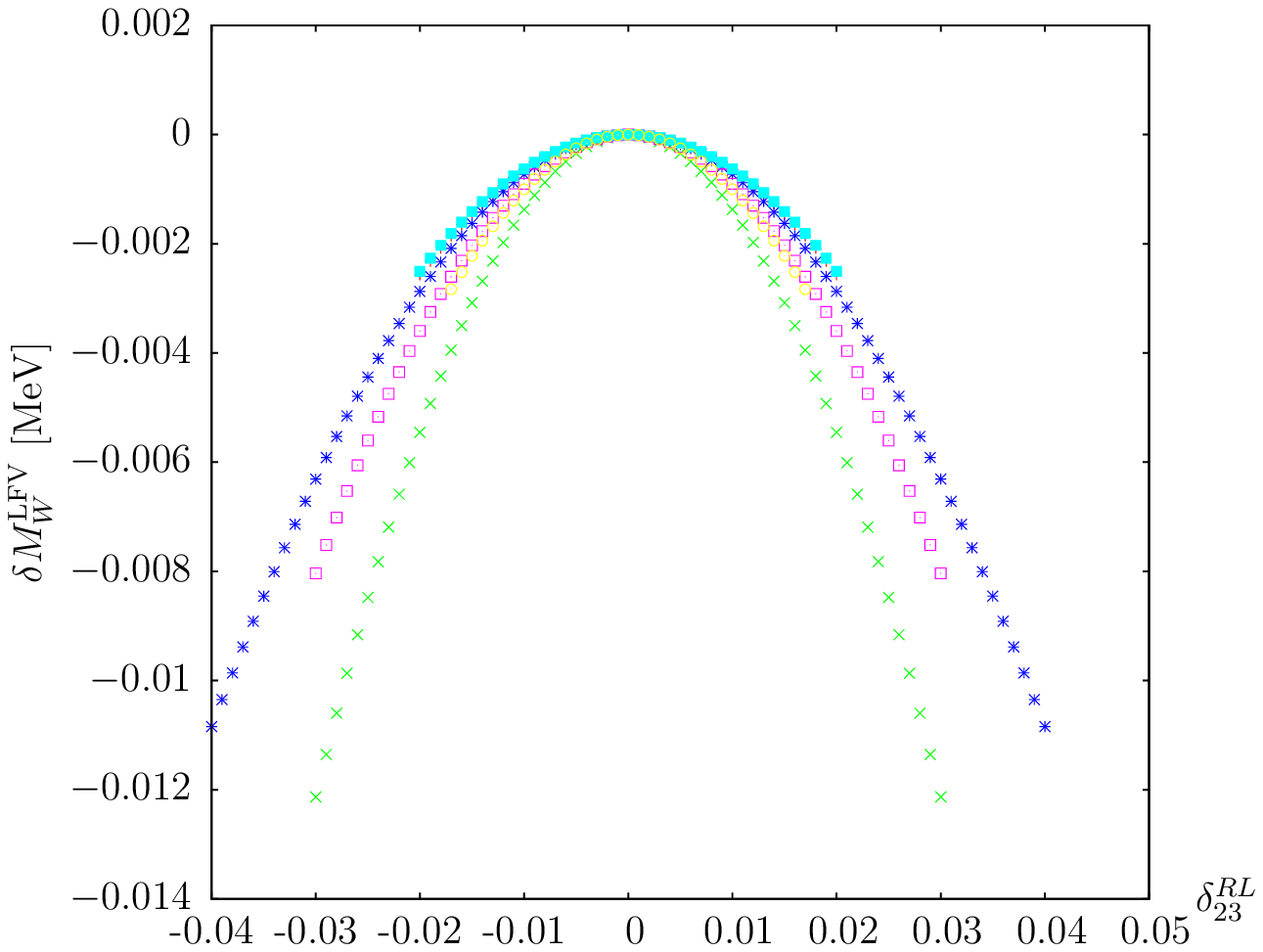  ,scale=0.57,angle=0,clip=}\\
\vspace{0.5cm}
\psfig{file=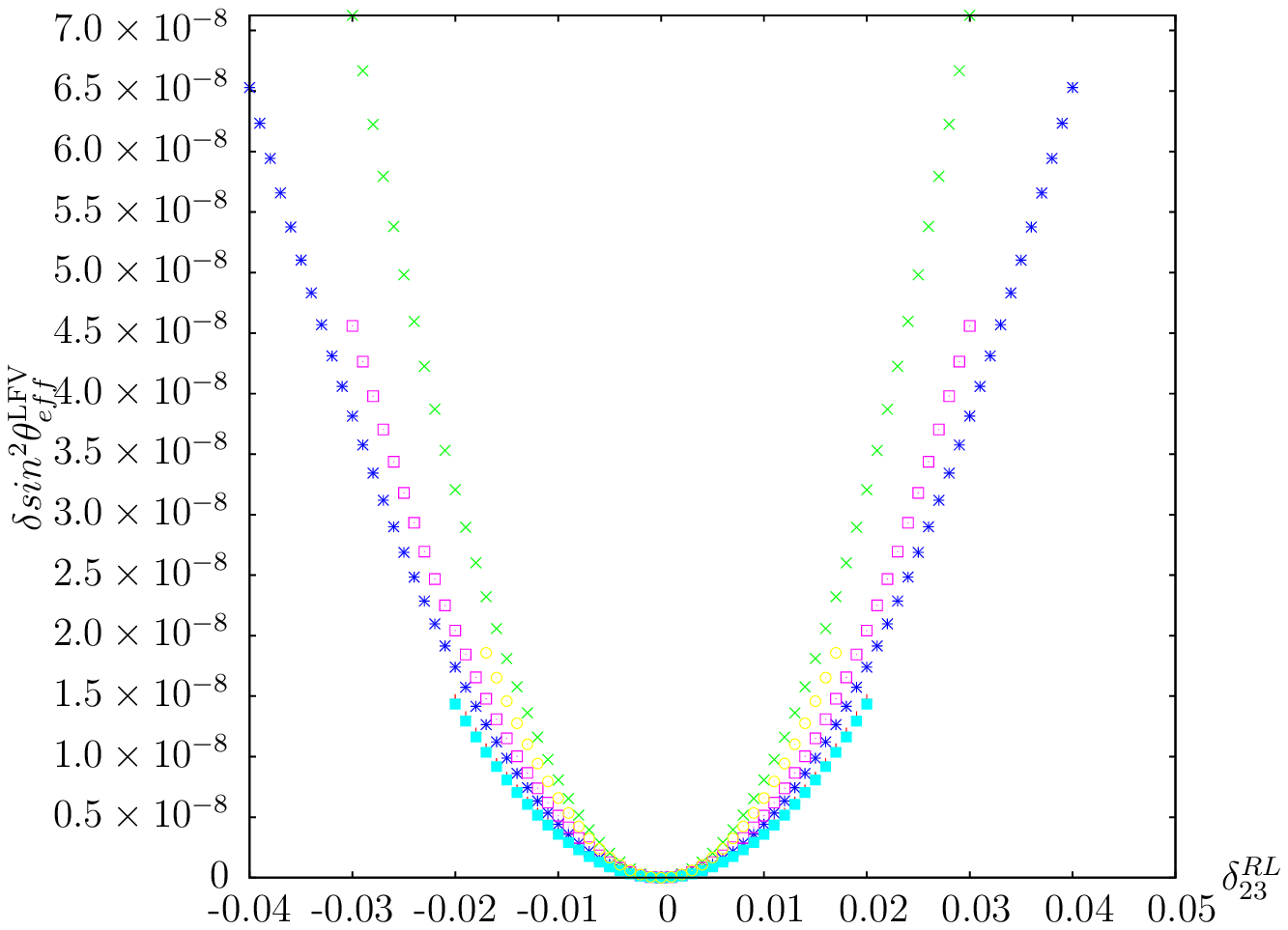 ,scale=0.57,angle=0,clip=}
\psfig{file=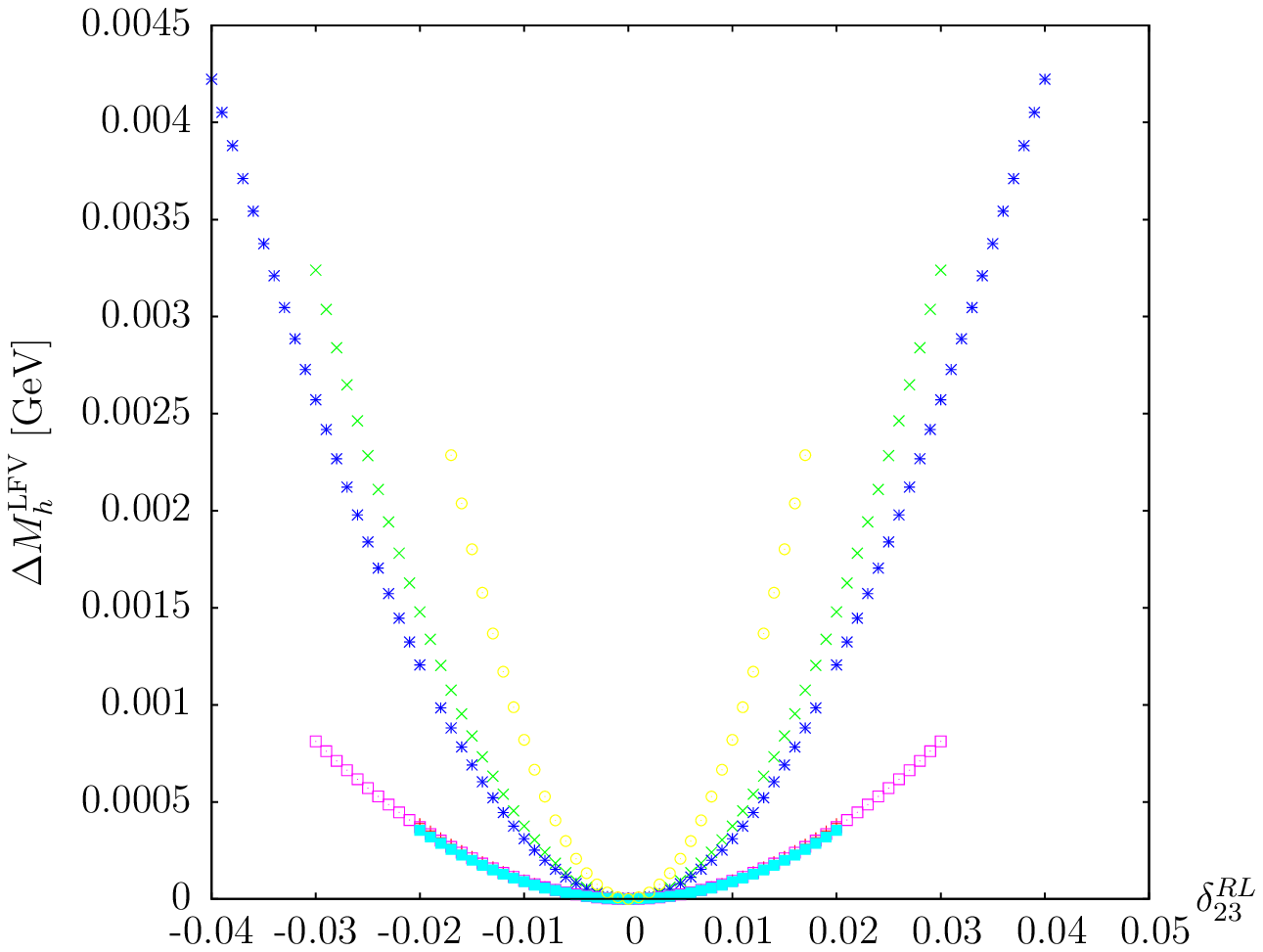   ,scale=0.57,angle=0,clip=}\\
\vspace{0.5cm}
\psfig{file=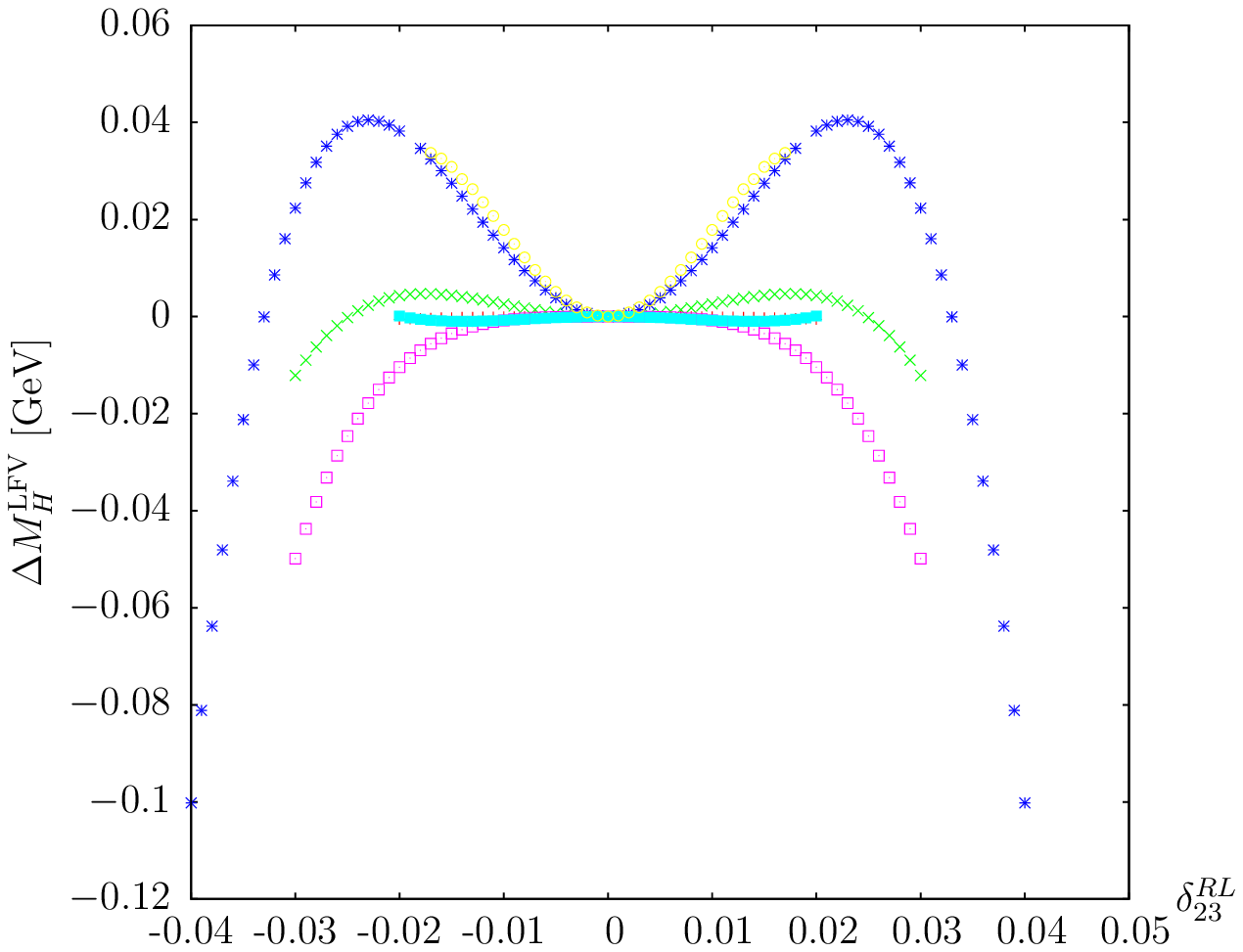  ,scale=0.57,angle=0,clip=}
\psfig{file=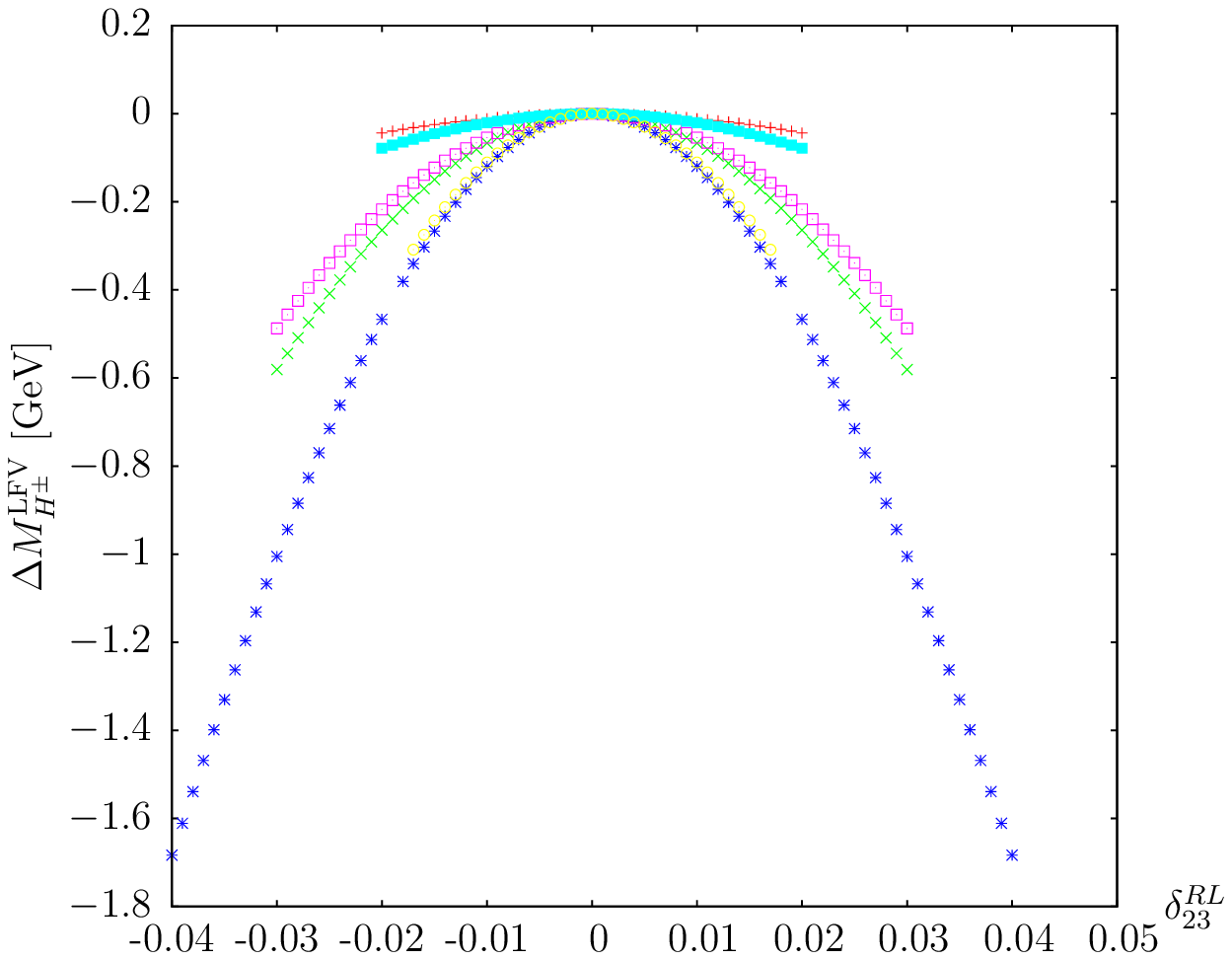  ,scale=0.57,angle=0,clip=}\\

\end{center}
\caption{EWPO and Higgs masses as a function of slepton
  mixing $\delta^{RL}_{23}$ for the six points defined in the \refta{tab:spectra}.}  
\label{figdRL23}
\end{figure}

Turning to the heavy Higgs bosons, the contributions to $\MH$ (most 
sizable again for $\del{LR,RL}{23,13}$) do not exceed \order{100 \mev} 
and are thus effectively negligible.  Substantially larger corrections 
are found, in agreement with the expectations from \citere{arana} for 
the charged Higgs-boson mass.  They can reach the level of nearly $-2 
\gev$, see \reffis{figdLR13}--\ref{figdRL23}.  For the chosen values of 
$\MA$ (or $\MHp$) this stays below the level of~1\%.  The absolute size 
of the corrections is not connected to the value of $\MHp$ in S1\dots 
S6, however.  Choosing starting values of $\MA$ somewhat smaller 
(requiring a new evaluation of the corresponding bounds on the LFV 
$\deABij$), could yield relative corrections to $\MHp$ at the level 
of~1\%.  Furthermore, as in the case of the light Higgs-boson mass, the 
explicit calculation of the LFV effects eliminates the theory 
uncertainty associated to these effects, thus improving the theoretical 
accuracy.

\newpage

\clearpage


\section{Conclusions}
\label{sec:conclusions}

We extended Lepton Flavor Violation in the MSSM into the setup of 
FeynArts and FormCalc; the corresponding model file is part of the 
latest release of these programs.

The LFV effects are parameterized in a complete set of $\deABij$ ($A,B = 
L,R$; $i,j = 1,2,3$) without any assumption on the physics at the GUT 
scale.  The inclusion of LFV into FeynArts/FormCalc allowed us to 
calculate the one-loop LFV effects on electroweak precision observables 
(via the calculation of gauge-boson self-energies) as well on the 
Higgs-boson masses of the MSSM (via the calculation of the Higgs-boson 
self-energies).  The corresponding results have been included in the 
code FeynHiggs and are publicly available from version 2.10.2 on.

The numerical analysis was performed on the basis of six benchmark 
points defined in \citere{Arana-Catania:2013nha}.  These benchmark 
points represent different combinations of parameters in the sfermion 
sector.  The restrictions on the various $\deABij$ in these six 
scenarios, provided by experimental limits on LFV processes (such as 
$\mu\to e\ga$) have been taken from \citere{Arana-Catania:2013nha}, and 
the effects on EWPO and Higgs-boson masses have been evaluated in the 
experimentally allowed ranges.  In this way we provide a general 
overview about the possible size of LFV effects and potential new 
restrictions on the $\deABij$ from EWPO and Higgs-boson masses.

The LFV effects in the EWPO turned out to be sizable for $\del{LL}{23}$
but (at least in the scenarios under investigation) negligible for the 
other $\deABij$.  The effects of varying $\del{LL}{23}$ in the 
experimentally allowed ranges turned out to exceed the current 
experimental uncertainties of $\MW$ and $\sweff$ in the case of heavy 
sleptons.  No new general bounds could be set on $\del{LL}{23}$, 
however, since the absolute values of $\MW$ and $\sweff$ strongly depend 
on the choices in the stop/sbottom sector, which is disconnected from 
the slepton sector presently under investigation.  Such bounds could be 
set on a point-by-point basis in the LFV MSSM parameter space, however.  
Looking at the future anticipated accuracies, also lighter sleptons 
yielded contributions exceeding that precision.  It may therefore be 
possible in the future to set bounds on $\del{LL}{23}$ from EWPO that 
are stronger than from direct LFV processes.

In the Higgs sector, based on evaluations for flavor violation in the 
squark sector, non-negligible corrections to the light $\cp$-even Higgs 
mass as well as to the charged Higgs-boson mass could be expected.  The 
associated theoretical uncertainties exceeded the anticipated future 
precision for $\Mh$ and $\MHp$.  Taking the existing limits on the 
$\deABij$ from LFV processes into account, however, the corrections 
mostly turned out to be small.  For the light $\cp$-even Higgs mass they 
stay at the few-MeV level.  For the charged Higgs mass they can reach 
\order{2 \gev}, which, depending on the choice of the heavy Higgs-boson 
mass scale, could be at the level of the future experimental precision.  
More importantly, the theoretical uncertainty from LFV effects that 
previously existed for the evaluation of the MSSM Higgs-boson masses, 
has been reduced below the level of future experimental accuracy.


\vspace{-0.5em}
\subsection*{Acknowledgments}

We thank M.~Arana-Catania and M.J.~Herrero for helpful discussions.  
The work of S.H.\ was partially supported by CICYT (grant FPA 
2010--22163-C02-01).  S.H.\ and M.R.\ were supported by the Spanish 
MICINN's Consolider-Ingenio 2010 Programme under grant MultiDark 
CSD2009-00064.



\end{document}